\magnification=\magstep1
\overfullrule=0pt
\hoffset -0.5cm
\vsize=46pc
\setbox4=\hbox{{\cal W}}
\def\ww{{\cal W}\hskip-\wd4\hskip 0.2 true pt{\cal W}}
\def\w{{\cal W}}
\def\lb{\lbrack}
\def\rb{\rbrack}
\def\de{\partial}
\def\vac{\vert 0 \rangle\, }

\def\det{{\rm det}}
\def\q#1{\lb#1\rb}
\def\bibitem#1{\parindent=8mm\item{\hbox to 6 mm{$\q{#1}$\hfill}}}
\def\mn{\medskip\smallskip\noindent}
\def\sn{\smallskip\noindent}
\def\bn{\bigskip\noindent}
\def\BZT{{\rm Z{\hbox to 3pt{\hss\rm Z}}}}
\def\BZS{{\hbox{\sevenrm Z{\hbox to 2.3pt{\hss\sevenrm Z}}}}}
\def\BZSS{{\hbox{\fiverm Z{\hbox to 1.8pt{\hss\fiverm Z}}}}}
\def\BZ{{\mathchoice{\BZT}{\BZT}{\BZS}{\BZSS}}}
\def\BQT{\,\hbox{\hbox to -2.8pt{\vrule height 6.5pt width .2pt
    \hss}\rm Q}}
\def\BQS{\,\hbox{\hbox to -2.1pt{\vrule height 4.5pt width .2pt\hss}$
   \scriptstyle\rm Q$}}
\def\BQSS{\,\hbox{\hbox to -1.8pt{\vrule height 3pt width
   .2pt\hss}$\scriptscriptstyle \rm Q$}}

\def\BCT{\,\hbox{\hbox to -3pt{\vrule height 6.5pt width
     .2pt\hss}\rm C}}
\def\BCS{\,\hbox{\hbox to -2.2pt{\vrule height 4.5pt width .2pt\hss}$
   \scriptstyle\rm C$}}
\def\BCSS{\,\hbox{\hbox to -2pt{\vrule height 3.3pt width
   .2pt\hss}$\scriptscriptstyle \rm C$}}
\def\BC{{\mathchoice{\BCT}{\BCT}{\BCS}{\BCSS}}}
\def\BHT{{\rm I{\hbox to 5.3pt{\hss\rm H}}}}
\def\BHS{{\hbox{\sevenrm I{\hbox to 4.2pt{\hss\sevenrm H}}}}}
\def\BHSS{{\hbox{\fiverm I{\hbox to 3.5pt{\hss\fiverm H}}}}}

\def\BPT{{\rm I{\hbox to 5pt{\hss\rm P}}}}
\def\BPS{{\hbox{\sevenrm I{\hbox to 4pt{\hss\sevenrm P}}}}}
\def\BPSS{{\hbox{\fiverm I{\hbox to 3pt{\hss\fiverm P}}}}}

\def\BST{\;\hbox{\hbox to -4.5pt{\vrule height 3pt width .2pt\hss}
   \raise 4pt\hbox to -2pt{\vrule height 3pt width .2pt\hss}\rm S}}
\def\BSS{\;\hbox{\hbox to -4.2pt{\vrule height 2.3pt width .2pt\hss}
   \raise 2.5pt\hbox to -4.8pt{\vrule height 2.3pt width .2pt\hss}
   $\scriptstyle\rm S$}}
\def\BSSS{\;\hbox{\hbox to -4.2pt{\vrule height 1.5pt width .2pt\hss}
   \raise 1.8pt\hbox to -4.8pt{\vrule height 1.5pt width .2pt\hss}
   $\scriptscriptstyle\rm S$}}

\def\BFT{{\rm I{\hbox to 5pt{\hss\rm F}}}}
\def\BFS{{\hbox{\sevenrm I{\hbox to 4pt{\hss\sevenrm F}}}}}
\def\BFSS{{\hbox{\fiverm I{\hbox to 3pt{\hss\fiverm F}}}}}

\def\BRT{{\rm I{\hbox to 5.5pt{\hss\rm R}}}}
\def\BRS{{\hbox{\sevenrm I{\hbox to 4.3pt{\hss\sevenrm R}}}}}
\def\BRSS{{\hbox{\fiverm I{\hbox to 3.35pt{\hss\fiverm R}}}}}
\def\BR{{\mathchoice{\BRT}{\BRT}{\BRS}{\BRSS}}}
\def\BNT{{\rm I{\hbox to 5.5pt{\hss\rm N}}}}
\def\BNS{{\hbox{\sevenrm I{\hbox to 4.3pt{\hss\sevenrm N}}}}}
\def\BNSS{{\hbox{\fiverm I{\hbox to 3.35pt{\hss\fiverm N}}}}}
\def\BN{{\mathchoice{\BNT}{\BNT}{\BNS}{\BNSS}}}
\def\BAT{\hbox{\raise1.8pt\hbox{\sevenrm/}{\hbox to 4pt{\hss\rm A}}}}
\def\BAS{\hbox{\raise1.4pt\hbox{\fiverm/}
{\hbox to 3pt{\hss\sevenrm A}}}}
\def\BASS{\hbox{\raise1.4pt\hbox{\fiverm/}
{\hbox to 3pt{\hss\sevenrm A}}}}

\def\section#1{\vbox{\leftline{\bf #1}\vskip-7pt\line{\hrulefill}}}
\def\subs#1{\vbox{\rightline{\bf #1}\vskip-7pt\line{\hrulefill}}}

\def\n{{\cal N}}         
\def\Q{{\cal Q}}         
\def\GB{\overline{G}}
\def\Lh{\hat{L}}
\def\ch{\hat{c}}
\def\hwt{\mid \!h, \tau \rangle\, }

\def\esl{$\widehat{sl(2,\BR)}$}
\def\eslk{$\widehat{sl(2,\BR)}_\kappa$}
\def\sluk{$\widehat{sl(2,\BR)}_\kappa/\widehat{U(1)}$}
\def\slukk{$\widehat{sl(2,\BR)}_k/\widehat{U(1)}$}
\def\slsl{$\widehat{sl(2,\BR)}_\kappa\oplus
\widehat{sl(2,\BR)}_{-{1\over2}}/\widehat{sl(2,\BR)}_{\kappa-{1\over2}}$}
\def\slslk{$\widehat{sl(2,\BR)}_k\oplus
\widehat{sl(2,\BR)}_{-{1\over2}}/\widehat{sl(2,\BR)}_{k-{1\over2}}$}

\def\slslgen{$\widehat{sl(2,\BR)}_\kappa\oplus
\widehat{sl(2,\BR)}_{\mu}/\widehat{sl(2,\BR)}_{\kappa+\mu}$}
\def\slnsln{$\widehat{sl(k,\BR)}_\kappa\oplus
\widehat{sl(k,\BR)}_{\mu}/\widehat{sl(k,\BR)}_{\kappa+\mu}$}
\def\slnslndual{$\widehat{sl({\kappa\!+\!\mu},\BR)}_k/
(\widehat{sl(\kappa,\BR)}_k\oplus\widehat{U(\mu)_k}$)}
\def\LLc{$L_{\Lambda,\Lambda'}^{{\hat g}/{{\hat g}'}}$}
\def\LL{$L_{\Lambda}^{\hat g}$}
\def\LLp{$L_{\Lambda'}^{{\hat g}'}$}
\def\gg{${{\hat g}/{{\hat g}'}}$}
\def\La{\Lambda}
\def\eps{\epsilon}
\def\pa{\partial}
\def\uone{$\widehat{U(1)}$}          
\def\slu{$\widehat{sl(2,\BR)}/\widehat{U(1)}$}
\def\ber{{$\w^{sl(3)}_{2,1}$}}                  
\def\beru{\ber $\!$/\uone}                   
\def\berut{\ber/\uone}                       
\def\ntsvir{$SVIR(N=2)$}     
\def\sviru{\ntsvir/\uone}    
\def\wfuenf{$\w(2,3,4,5)$}   
\def\wsieben{$\w(2,3,4,5,6,7)$} 
\def\idi{I}                  
\def\Jz{J^o}         
\def\Jp{J^+}         
\def\Jm{J^-}         
\def\Jpm{J^\pm}      
\def\cslc{c}         
\def\ks{k}           
\def\Tsc{L}          
\def\wsd{W_3}        
\def\wsv{W_4}        
\def\wsf{W_5}        
\def\kb{k}           
\def\cb{c}           
\def\Tb{L}           
\def\Gp{G^+}         
\def\Gm{G^-}         
\def\Gpm{G^{\pm}}    
\def\Tbc{\hat L}     
\def\cbc{\hat c}     
\def\wbd{W_3}        
\def\wbv{W_4}        
\def\wbf{W_5}        
\def\wbs{W_6}        
\def\wbsi{W_7}       
\def\SC{C}
\def\SCu{\widetilde{\SC}}       
\def\pjz{P_2^{11}}              
\def\wspdds{P_6^{33}}           
\def\wspdvs{P_7^{34}}           
\def\wspdva{P_8^{34}}           
\def\wbpdds{P_6^{33}}           
\def\wbpdda{P_8^{33}}           
\def\wbpdvs{P_7^{34}}           
\def\wbpdva{P_8^{34}}           
\def\wbpdfa{P_8^{35}}           
\def\wbpvva{P_8^{44}}           
\def\wbpdfn{P_9^{35}}           
\def\frac#1,#2{{#1}\over{#2}}
\def\NN#1#2{#2\,#1}
\def\jpm{J^{(\pm)}}
\def\jmp{J^{(\mp)}}
\def\jp{J^{(+)}}
\def\jn{J^{(0)}}
\def\jm{J^{(-)}}
\def\kp{\kappa}

\def\japm{J^{(1,\pm)}}

\def\jan{J^{(1,0)}}

\def\jbpm{J^{(2,\pm)}}

\def\jbn{J^{(2,0)}}

\def\jpm{J^{(\pm)}}
\def\jmp{J^{(\mp)}}
\def\SS#1#2{S_{#2}(#1)}
\def\sltr{\widehat{sl(2,\BR)}}
\def\soh#1{\widehat{so(#1)}}

\def\dcn{h^\vee}
\def\orb#1{\hbox{Orb}\left(#1\right)}
\def\pl{\hbox{$+$}}
\def\mi{\hbox{$-$}}
\def\appF{D}
\def\appD{E}
\def\appE{F}
\def\bouschou{1}
\def\flrep{2}
\def\zam{3}
\def\diealten{4}
\def\kauwatts{5}
\def\wirrep{6}
\def\hornfeck{7}
\def\klausREP{8}
\def\ajl{9}
\def\FORT{10}
\def\deboer{11}
\def\laszlorep{12}
\def\bowwatts{13}
\def\fehort{14}
\def\bbss{15}
\def\howcl{16}
\def\letter{17}
\def\bogo{18}
\def\altschuler{19}
\def\fatzam{20}
\def\dminusn{21}
\def\dijkgraaf{22}
\def\twists{23}
\def\frfolding{24}
\def\commute{25}
\def\thielemans{26}
\def\poly{27}
\def\bersh{28}
\def\bakas{29}
\def\bakpriv{30}
\def\narganes{31}
\def\peiwang{32}
\def\bmcp{33}
\def\yuwu{34}
\def\yuwuC{35}
\def\toppan{36}
\def\ralphnpb{37}
\def\nemesch{38}
\def\qiu{39}
\def\klausWAN{40}
\def\delduc{41}
\def\gosch{42}
\def\ralph{43}
\def\bouwknegt{44}
\def\kacpeter{45}
\def\kacwaki{46}
\def\ahn{47}
\def\kiritsis{48}
\def\wolfgang{49}
\def\rwofus{50}
\def\gso{51}
\def\mfl{52}
\def\andrdipl{53}
\def\don{54}
\def\koos{55}
\def\bakinf{56}
\def\frenkel{57}
\def\bakaskiritsis{58}
\def\awata{59}
\def\kneu{60}
\def\nakan{61}
\def\ginsparg{62}
\def\kauschpriv{63}
\def\luky{64}
\def\kfw{65}
\def\god{66}
\def\horst{67}
\def\supwir{68}
%
%
\font\large=cmbx10 scaled \magstep3
\font\largew=cmsy10 scaled \magstep3
\font\bigf=cmr10 scaled \magstep2
\pageno=0
\def\folio{
\ifnum\pageno<1 \footline{\hfil} \else\number\pageno \fi}
\phantom{not-so-FUNNY}
\rightline{ DFTT--25/94\break}
\rightline{ BONN--TH--94--11\break}
\rightline{ hep-th/9406203\break}
\rightline{ June 1994\break}
\rightline{ revised March 1995\break}
\vskip 2.0truecm
\centerline{\large Coset Realization of Unifying {\largew W}-Algebras}
\vskip 1.0truecm
\centerline{\bigf R.\ Blumenhagen\raise5pt\hbox{\tenrm 1},
W.\ Eholzer\raise5pt\hbox{\tenrm 1},}
\medskip
\centerline{\bigf A.\ Honecker\raise5pt\hbox{\tenrm 1},
K.\ Hornfeck\raise5pt\hbox{\tenrm 2},
R.\ H\"ubel\raise5pt\hbox{\tenrm 1}
}
\bigskip\medskip
\centerline{\it ${}^{1}$ Physikalisches Institut der Universit\"at Bonn}
\centerline{\it Nu{\ss}allee 12, 53115 Bonn, Germany}
\medskip
\centerline{\it ${}^{2}$ INFN, Sezione di Torino}
\centerline{\it Via Pietro Giuria 1, 10125 Torino, Italy}
\vskip 1.1truecm
\centerline{\bf Abstract}
\vskip 0.2truecm
\noindent
We construct several quantum coset $\w$-algebras, e.g.~\slu\ and
$\sltr \oplus \sltr /$ $\sltr$, and argue that they are finitely
nonfreely generated. Furthermore, we discuss in detail their r\^ole as
unifying $\w$-algebras of Casimir $\w$-algebras. We show that
it is possible to give coset realizations of various
types of unifying $\w$-algebras, e.g.\ the diagonal cosets based on the
symplectic Lie algebras $sp(2n)$ realize the unifying $\w$-algebras which
have previously been introduced as `${\cal WD}_{-n}$'. In addition,
minimal models of ${\cal WD}_{-n}$ are studied. The coset realizations provide
a generalization of level-rank-duality of dual coset pairs.
As further examples of finitely nonfreely generated
quantum $\w$-algebras we discuss orbifolding of $\w$-algebras which on the
quantum level has different properties than in the classical case.
We demonstrate in some examples that the classical limit according to
Bowcock and Watts of these nonfreely {\it finitely} generated quantum
$\w$-algebras probably yields {\it infinitely} nonfreely generated classical
$\w$-algebras.
\vfill
\centerline{to appear in Int.\ Jour.\ of Mod.\ Phys.\ {\bf A}}
\eject
%
%
\centerline{\bf Contents}
\bn
\leftline{1.\ Introduction\hfill 2}
\sn
\leftline{1.1.\ Notation \hfill 4}
\sn
\leftline{2.\ Non-freely generated $\w$-algebras in cosets and orbifolds
              \hfill 5}
\sn
\leftline{2.1.\ Coset $\w$-algebras \hfill 5}
\sn
\leftline{2.1.1.\ The cosets \slu\ and \sviru \hfill 5}
\sn
\leftline{2.1.2.\ The coset \beru  \hfill 12}
\sn
\leftline{2.1.3.\ The coset \slslgen \hfill 16}
\sn
\leftline{2.2.\ Orbifolds of quantum $\w$-algebras \hfill 23}
\sn
\leftline{2.2.1.\ General remarks and results \hfill 24}
\sn
\leftline{2.2.2.\ The orbifold of $\w(2,3)$\hfill 28}
\sn
\leftline{2.2.3.\ Remarks on the orbifold of ${\cal WA}_{n-1}$\hfill 30}
\sn
\leftline{3.\ General structures in cosets and orbifolds\hfill 32}
\sn
\leftline{3.1.\ Vacuum preserving algebras (VPA) and classical limits
                  \hfill 32}
\sn
\leftline{3.2.\ Coset realization of unifying $\w$-algebras and
                  level-rank-duality\hfill 38}
\sn
\leftline{3.2.1.\ Unifying $\w$-algebras for
                  the ${\cal WA}_n$ Casimir algebras \hfill 38}
\sn
\leftline{3.2.2.\ Level-rank-duality for the cosets $\widehat{so(n)}_k\oplus
             \widehat{so(n)}_1/\widehat{so(n)}_{k+1}$ \hfill 42}
\sn
\leftline{3.2.3.\ Realization of ${\cal WD}_{-n}$ as diagonal
                  $sp(2n)$ cosets \hfill 44}
\sn
\leftline{3.2.4.\ Minimal models of ${\cal WD}_{-m}$\hfill 46}
\sn
\leftline{3.2.5.\ The coset ${\hat g}_k/g$ for a
                  simple Lie algebra $g$\hfill 48}
\sn
\leftline{4.\ Conclusion \hfill 49}
\sn\sn\sn
\leftline{Appendix A: The simple fields with spin 3, 4, 5 of \slu \hfill 52}
\sn
\leftline{Appendix B: Some structure constants of \beru \hfill 52}
\sn
\leftline{Appendix C: The primary spin 4 generator of  \slslgen \hfill 53}
\sn
\leftline{Appendix \appF: Minimal models of Casimir $\w$-algebras \hfill 54}
\sn
\leftline{Appendix \appD: The orbifold of the $N=1$ Super Virasoro algebra
                   \hfill 54}
\sn
\leftline{Appendix \appE: Generators and structure constants of the orbifold
                   of $\w(2,3)$  \hfill 56}
\sn
\leftline{References \hfill 57}
\vfill
\eject
%
%
\section{1.\ Introduction}
\sn
One of the most interesting questions in two dimensional conformal
invariant quantum field theory is the classification
of rational conformal field theories (RCFT). An important
tool in the investigation of this question is provided by
extended conformal algebras also called $\w$-algebras (see e.g.\
$\q{\bouschou-\zam}$). Although $\w$-algebras have been the object of
intense studies in the last few years, a complete satisfactory
classification of $\w$-algebras and their representations
has not been achieved yet. One has to distinguish between
algebras which exist only for fixed values of the Virasoro central charge
$c$, called nondeformable, and deformable (or generic) $\w$-algebras existing
for generic central charge $c$. For the latter class
the structure constants are continuous functions of $c$ apart from a
finite set of singularities. Intense studies revealed that it is possible
to explain most of the nondeformable $\w$-algebras as truncations or
extensions of generically existing ones (see e.g.\ $\q{\diealten-\klausREP}$).
\mn
It was noticed very recently that the class of deformable
$\w$-algebras contains at least two subclasses which have completely
different features of the classical counterparts~$\q{\ajl}$.
The first class consists of deformable $\w$-algebras
originating from quantum Drinfeld-Sokolov (DS) type hamiltonian reduction of
affine Kac-Moody algebras (see e.g.\ $\q{\FORT - \laszlorep}$).
These algebras are by far the best understood ones. The main property of
this class of algebras is that the classical counterparts are Poisson
bracket algebras based on finitely and freely generated rings of
differential polynomials. It is believed that these algebras can be
classified by $sl(2,\BR)$ embeddings into simple Lie algebras
$\q{\bowwatts,\FORT,\fehort}$. The so-called Casimir algebras $\q{\bbss}$
correspond to the principal embedding. In contrast hereto the second
class of generically existing quantum $\w$-algebras have as classical
counterparts infinitely generated rings of differential polynomials with
infinitely many relations, i.e.\ the ring is nonfreely generated~$\q{\ajl}$.
\mn
In this paper we discuss features of quantum $\w$-algebras
belonging to this new class. The first observation
is that although the classical Poisson bracket algebra is
infinitely generated the quantum $\w$-algebra is generated by a finite
number of simple fields in all cases studied up to now. This is due to
the fact that on the quantum level normal ordered versions of
the classical relations between the
generators eliminate all but a finite number of generators $\q{\ajl}$.
A very intriguing point is that the unexplained solutions of
$\w(2,4,6)$ $\q{\kauwatts,\howcl}$ and $\w(2,3,4,5)$ $\q{\hornfeck}$
with generic null fields obtained by direct construction are of this
type and can be explained in terms of coset constructions (see also
$\q{\ajl}$). In $\q{\ajl}$ it has been argued that all classical coset
algebras are infinitely generated. In the present paper we argue for certain
cases that the corresponding quantum $\w$-algebras are {\it finitely}
generated. In the case of \slu\ we give arguments that the commutant of
the \uone-current yields the unexplained solution
of $\w(2,3,4,5)$. We argue that this coset algebra is
isomorphic to the commutant of the \uone-current
in the $N=2$ Super Virasoro algebra $SVIR(N=2)$. We also present
supporting arguments for the realization $\q{\ajl}$ of the previously
unexplained solution of $\w(2,4,6)$ as the diagonal coset \slsl\ and
study its representation theory.
\mn
In $\q{\letter}$ it has been observed that $\w$-algebras of the
second class have the interesting property of being unifying algebras
for special series of minimal models of Casimir $\w$-algebras.
In this paper we give explicit coset realizations for
many of these unifying algebras which generalize level-rank-duality
of coset pairs $\q{\bogo,\altschuler}$. In special cases like the unitary
models of the ${\cal WA}_{k-1}$ algebras the unifying algebras can be
inferred directly from level-rank-duality. The coset \slukk\ for example
describes the first unitary model of the ${\cal WA}_{k-1}$ Casimir algebra,
the so-called $\BZ_k$ parafermions $\q{\fatzam}$. For the corresponding values
of the central charge $c = {2 (k - 1) \over k + 2 }$ the
algebras ${\cal WA}_{k-1}$ truncate to \wfuenf\ $\q{\letter}$.
Even in the cases where level-rank-duality can be exploited
our considerations go beyond the `$T$-equivalence' of
coset pairs $\q{\bouschou,\bogo,\altschuler}$.
We compute the spins of the finite generating set of
the unifying algebras and check in some examples the isomorphism
of the extended symmetry algebras.
\sn
The unifying algebras for ${\cal WC}_k$ minimal models are the
new algebras ${\cal WD}_{-n}$ $\q{\klausREP,\letter}$.
We argue that these algebras ${\cal WD}_{-n}$ can be realized
in terms of a diagonal $\widehat{sp(2n)}$ coset.
The first member ${\cal WD}_{-1}$ is the formerly unexplained $\w(2,4,6)$.
In fact, negative-dimensional groups have already been encountered
in the representation theory of the classical groups
(see e.g.\ $\q{\dminusn}$).  The existence of ${\cal WD}_{-n}$ can
be expected by  a deep connection of the negative-dimensional
orthogonal groups $SO(-2n)$ and the symplectic groups $Sp(2n)$.
\sn
Apart from the coset construction there is another construction which
leads to $\w$-algebras with infinitely nonfreely generated
classical counterparts: orbifolding of $\w$-algebras
(see e.g.\ $\q{\dijkgraaf-\frfolding}$). If a given
$\w$-algebra possesses an outer automorphism one considers the projection
onto the invariant subspace. A special case of this construction is the
bosonic projection of a $\w$-algebra containing fermionic fields.
We show that in contrast to the classical case $\q{\frfolding}$ the
orbifold of a quantum Casimir algebra does in general
not contain a Casimir algebra as a subalgebra for generic central charge.
\mn
The plan of the paper is as follows. In the first part of section 2 we
will work out explicitly the algebraic structure of some cosets. Our
starting point are the cosets \slu\ and \sviru\ which we argue
to lead to a finitely generated algebra of type \wfuenf. After some remarks
concerning the unifying properties of these cosets we proceed with the coset
\beru\ leading to a $\w(2,3,4,5,6,7)$ (different from ${\cal WA}_6$).
As a further important example we investigate the diagonal
coset $\widehat{sl(2,\BR)}_\kappa\oplus\widehat{sl(2,\BR)}_\mu/$
$\widehat{sl(2,\BR)}_{\kappa+\mu}$ which, in particular, for $\mu=-{1\over 2}$
explains the special solution of $\w(2,4,6)$.
In the second part of section 2 we study orbifolds
of quantum $\w$-algebras. We determine the spin content of several
orbifolds, the explicit form of the invariant fields and
compute the structure constants for some examples.
\sn
In section 3 we discuss first the vacuum preserving algebra (VPA) and
classical limits of some prominent examples of deformable $\w$-algebras
outside the Drinfeld-Sokolov class. It turns out that the VPA is actually
infinite dimensional in contrast to algebras in the DS class.
We continue with a detailed discussion of the realization of
unifying algebras for Casimir algebras by coset and orbifold constructions.
Using character techniques we compute the spin content of these algebras
and discuss the relation to level-rank-duality.
\sn
The calculations presented in this paper have been performed in parts with a
small special purpose computer algebra system discussed in $\q{\commute}$
which we used for calculations with $\w$-algebra modules.
For the more complicated OPE calculations we used
the Mathematica-package of ref.\ $\q{\thielemans}$.
\mn
\section{1.1.\ Notation}
\sn
We begin with a short account of the notations which will be used
in the following. For a detailed description we refer to the
review article $\q{\bouschou}$.
\sn
Denote the extended symmetry algebra of a local chiral
conformally invariant quantum field theory in two dimensions
by ${\cal F}$. We assume that ${\cal F}$ is equipped
with the following three important operations: The commutator,
the normal ordered product and the derivative $\partial$
of local quantum fields.
Alternatively, we can demand that ${\cal F}$ is closed with respect
to the short-distance operator product expansion (OPE) of two local
fields. The singular part of an OPE reproduces the commutator, whereas
the constant term yields the normal ordered product (NOP).
The `mode expansion' of a field $\phi$ is defined by
$$\phi(z) = \sum_{n-d(\phi)\in\BZ}\ \phi_n\ z^{n-d(\phi)}.
\eqno{\rm (1.1.1)}$$
$d(\phi)$ is the `conformal dimension' of $\phi$ and $\phi_n$
are the `modes' (we deviate here from the standard conventions because
negative modes annihilate the vacuum state).
The modes of the energy momentum tensor $L$
satisfy the Virasoro algebra:
$$[L_n,L_m] = (m-n) L_{n+m} + {c\over {12}}(m^3-m)\delta_{n+m,0} .
\eqno{\rm (1.1.2)}$$
\sn
The Lie bracket structure in ${\cal F}$ is fixed by
a general formula for the commutator of two chiral fields
of half integral conformal dimension which involves universal
polynomials and some structure constants given by two- and three point
functions $\q{\diealten,\fehort}$. For the singular part of the OPE of
two fields $\phi_i,\phi_j$ we will use the following shorthand notation:
$$\phi_i\star\phi_j=\sum_k C_{i \,j}^k \phi_k .
\eqno{\rm (1.1.3)}$$
The coefficients $C_{i j}^k$ describe the coupling of the conformal
families of $\phi_i,\phi_j$ and $\phi_k$.
In repeated normal ordered products we use the convention:
$\phi_k\dots\phi_2\phi_1=(\phi_k(\dots(\phi_2\phi_1)))$ unless stated
explicitly. In the form $\chi\,\phi$ the NOP occurs as the constant
term in the OPE of $\chi$ and $\phi$
but in general it is not a quasi-primary field.
In explicit calculations it is sometimes
useful to work with the quasi-primary projection $\q{\diealten}$ of
$\de^n \phi_i \phi_j$ defined by
\footnote{${}^1$)}{This formula is the same as in $\q{\diealten}$,
the only difference is the normal ordering convention, i.e.\
$\phi_i \phi_j = N(\phi_j, \phi_i)$.}
$$\eqalign{
\n(\phi_j,\partial^{n}\phi_i) :&= \sum_{r=0}^{n} (-)^r
{n\choose r}{2(d(i)\!+\!d(j)\!+\!n\!-\!1) \choose r}^{-1}
{2d(i)\!+\!n\!-\!1 \choose r} \de^r ((\de^{n-r}\phi_i)\phi_j) \cr
& -(-)^{n}\!\!\!\sum_{\lbrace k:h(ijk)\geq1\rbrace}
\!\!\! C^k_{ij}\, {h(ijk)\!+\!n\!-\!1\choose n}\!
{2(d(i)\!+\!d(j)\!+\!n\!-\!1) \choose n}^{-1}  \cr
&\quad \times{2d(i)\!+\!n\!-\!1 \choose h(ijk)\!+\!n}\!{\sigma(ijk)\!-\!1
\choose h(ijk)\!-\!1}^{-1}
{\de^{h(ijk)+n} \phi_k \over (\sigma(ijk)\!+\!n)
(h(ijk)\!-\!1)!} \cr} \eqno({\rm 1.1.4})$$
with $\,\sigma(ijk)=d(i)+d(j)+d(k)-1\,,\,h(ijk)=d(i)+d(j)-d(k)$.
\sn
{}From any finite set of fields the operations $\partial$ and $\n$
generate infinitely many fields. It is therefore convenient
to define `simple' fields which are noncomposite and nonderivative.
To be more precise simple fields are defined to have a vanishing
projection onto derivative and composite fields.
An algebra generated by the simple fields $\phi_1,\dots,\phi_n$
is called a $\w(d(\phi_1),\dots,d(\phi_n))$. Primary fields composed of
two simple fields and their derivatives are sometimes abbreviated
using their dimensions if they determine the simple
fields uniquely. For example we
write $P_{d(\phi_1)+d(\phi_2)+2}^{d(\phi_1)\,d(\phi_2)}$ for
the primary field appearing in the OPE $\phi_1(z)\phi_2(w)$
at the zero of order 2 with normalization
$C_{\phi_1\,\phi_2}^{P_{\bullet}^{\bullet}}=1$.
\sn
We denote the Casimir algebra corresponding to a simple Lie algebra
${\cal L}_n$ by ${\cal WL}_n$. Casimir algebras arise from quantum
Drinfeld-Sokolov reduction with the principal $sl(2)$ embedding into
${\cal L}_n$. It should be clear to the reader that we reserve the
notation ${\cal WB}_n$ for the purely bosonic algebras
${\cal WB}_n \cong \w(2,4,\ldots,2n)$. In earlier papers, the same
notation was used for $\w$-algebras of type $\w(2,4,\ldots,2n,n+{1 \over 2})$
that contain one fermionic field. We denote these algebras by
${\cal WB}(0,n)$ because they can be obtained from quantum DS reduction for
the Super Lie algebras ${\cal B}(0,n)$.
For a $\w$-algebra coming from DS reduction of a nonprincipal $sl(2)$
embedding into ${\cal L}_n$ we use the notation $\w^{{\cal L}_n}_{\cal S}$
where ${\cal S}$ denotes the embedding. For example, for ${\cal L}_n
= {\cal A}_n$, ${\cal S}$ is the $r$-tuple of the dimensions of the
irreducible $sl(2)$ representations which appear in the defining
representation of ${\cal L}_n$ thus determining the embedding. The
Polyakov-Bershadsky algebra $\q{\poly,\bersh}$ which is obtained by a DS
reduction of the nonprincipal $sl(2)$ embedding into $sl(3)$ is abbreviated
in this notation by $\w^{sl(3)}_{2,1}$.
\bn
\section{2.\ Non-freely generated $\w$-algebras in cosets and orbifolds}
\sn
\sn
\section{2.1.\ Coset $\w$-algebras}
\sn
In this section we present the results of explicit constructions of the
symmetry algebras of several cosets. In order to do so we have to define
the coset construction algebraically: The algebraic coset $\w/\hat{g}_k$
of a $\w$-algebra $\w$ with a Kac-Moody subalgebra $\hat{g}_k$
is defined as the commutant of $\hat{g}_k$ in $\w$. Note that Kac-Moody
algebras are also $\w$-algebras and therefore $\w$ can also be a
Kac-Moody algebra. Similarly, $\w / g$ refers to those fields in $\w$
that commute with the horizontal Lie subalgebra $g$ of $\hat{g}_k$, i.e.\
the $g$-singlets in $\w$.
If one is interested in rational models one may define the
coset algebra slightly different namely as the maximal extended symmetry
algebra which contains the algebraic coset algebra as a subalgebra at
the given value of the central charge.
Since we are mainly interested in cosets which exist for
generic values of the central charge we will use only algebraic cosets
and call them just cosets from now on.
\sn
\section{2.1.1.\ The cosets \slu\ and \sviru}
\sn
In this section we study the quantum versions
of the cosets \slu\ and \sviru. It has been shown in $\q{\ajl}$
that these cosets are infinitely generated with infinitely
many relations in the classical case.
However, we will argue that in the quantum case these
cosets have at least a finitely generated subalgebra \wfuenf\
(most probably the cosets are equal to this algebra) what is already
indicated by counting arguments of the invariants in the classical
coset. By explicit calculation of the first generators in the commutant
we find that no new generators with conformal dimension 6,7 or 8 appear.
This is contrary to the claim of $\q{\bakas}$
\footnote{${}^2$)}{It has been known to the authors of $\q{\bakas}$
shortly after its publication that there is no independent spin 6 field
$\q{\bakpriv}$.} that one needs a new generator at each integer conformal
dimension greater than one. Furthermore, by computing structure constants
explicitly we find agreement with the second solution for \wfuenf\
$\q{\hornfeck}$. In fact, a character argument indicates that the
cosets \slu\ and \sviru\ are isomorphic to \wfuenf.
\sn
\subs{The coset \slu}
\sn
The equivalence \slu$\cong$\wfuenf\ can be checked explicitly and we
present here the calculations, even if they agree in part with
those performed by other authors~$\q{\bakas,\narganes}$. We start from the
$\sltr$ Kac-Moody algebra
$$ \Jz \star \Jz = 2 \, \ks \,\idi  \quad\quad
\Jz \star \Jpm = \pm \,2\,\Jpm  \quad\quad
\Jp \star \Jm =  \ks \, \idi \, + \, \Jz.
  \eqno{\rm(2.1.1)}$$
It is easily verified that the Virasoro-operator
$$\Tsc \, = \, {1\over {2\ks\left(\ks+2\right)}}
   \left({2\,\ks\,\Jp\,\Jm - \Jz\,\Jz -  \ks\,\pa\Jz}\right)
\eqno{\rm(2.1.2)}$$
has central charge $\cslc={{2(\ks-1)}\over{\ks+2}}$
and commutes with $\Jz$, that is
$ \Tsc \star \Jz \, = \, 0 $.
We can now construct a spin-3 field $\wsd$ that
is primary with respect to $\Tsc$ and commutes with $\Jz$.
$\wsd$ is a linear combination of operators with
zero $U(1)$-charge and we find
$$\wsd = -6{\ks^2}\Jp\pa\Jm -
  12\ks\Jz\Jp\Jm +
  4\Jz\Jz\Jz +
  6{\ks^2}\pa\Jp\Jm +
  6\ks\pa\Jz\Jz + {\ks^2}\pa^2\Jz.
\eqno{\rm(2.1.3)}$$
Starting from $\wsd$ we find the complete algebra by
successively calculating the OPEs
$$\eqalign{
\wsd \star \wsd &= d_{3,3}  \idi  +  \wsv  \qquad
\wsd \star \wsv = {\frac{d_{4,4}},{d_{3,3}}}  \wsd  +  \wsf \qquad
\wsd \star \wsf = {\frac{d_{5,5}},{d_{4,4}}} \wsv  +
\SCu_{3\,5}^{\wspdds} \wspdds  \cr
\wsv \star \wsv &= d_{4,4} \, \idi \, +\, \SCu_{4\,4}^4 \, \wsv \, + \,
\SCu_{4\,4}^{\wspdds} \, \wspdds \cr
\wsv \star \wsf &= {\frac{d_{5,5}},{d_{4,4}}} \,\wsd \,+ \,\SCu_{4\,5}^5
 \,\wsf\,
+\,\SCu_{4\,5}^{\wspdvs} \, \wspdvs \, + \, \SCu_{4\,5}^{\wspdva} \,
\wspdva \cr
\wsf \star \wsf &= d_{5,5}\, \idi +
            {{d_{5,5}}\over{d_{4,4}}}\widetilde{C}_{4\,5}^5 W_4 +
            \widetilde{C}_{5\,5}^{P_6^{33}} P_6^{33} +
            \widetilde{C}_{5\,5}^{P_8^{33}} P_8^{33} +
            \widetilde{C}_{5\,5}^{P_8^{44}} P_8^{44} +
            \widetilde{C}_{5\,5}^{P_8^{35}} P_8^{35} . \cr
}\eqno{\rm(2.1.4)}$$
The primary spin-4 and spin-5 fields as well as some two- and
three-point functions are given in appendix A. The tilde for the structure
constants $\SCu_{i\,j}^k$ distinguishes them from the structure constants
$\SC_{i\,j}^k$ used when the fields are in standard normalization.
We make the following two important observations:
In eq.~(2.1.4) new generators of dimensions~$>5$ do not appear so that
the algebra closes within the fields $L$, $\wsd$, $\wsv$ and
$\wsf$. The algebra can be identified with the special \wfuenf\ by
comparing the structure constants with those  given in $\q{\hornfeck}$
for this algebra. For example, we obtain for the structure
constant $\SC_{3\,3}^4$ in standard normalization
$$(\SC_{3\,3}^4)^2 = {{128(\ks-3)(2\ks+1){{(2\ks+3)}^2}}\over
   {(\ks-2)(\ks+2)(3\ks+4)(16\ks+17)}}=
{\frac{16},{3}} {\frac{(\cslc+2)(\cslc+10)^2(5\cslc-4)},
{(\cslc+7)(2\cslc-1)(5\cslc+22)}}
  \eqno{\rm(2.1.5)}$$
in accordance with the \wfuenf\ of $\q{\hornfeck}$ different from
${\cal WA}_4$. The corresponding structure
constant in~$\q{\narganes,\bakas}$ and in $\q{\peiwang}$ (derived from
a construction using $SU(2)$ parafermions) agrees with eq.\ (2.1.5).
\sn
The identification of \slu\ with the special \wfuenf\ is further supported by
the following calculation of the vacuum character of \slu. To this end one
has to count the uncharged states in the module generated freely by $J^{\pm}$:
$${1 \over \prod_{n\ge 1} (1-z^2 q^n)(1-z^{-2} q^n)} =
{(1\!-\!z^2)\sum_{m\in \BZ} \phi_{m}(q)  z^{2m} \over \prod_{n\ge 1}(1-q^n)^2}
= {\sum_{m\in \BZ} (\phi_{m}(q)\!-\!\phi_{m-1}(q)) z^{2m}
        \over \prod_{n\ge 1}(1-q^n)^2}
\eqno{\rm(2.1.6)}$$
where the first manipulation follows from the well-known identity
$${1 \over \prod_{n\ge 0} (1-z^2 q^n)(1-z^{-2} q^{n+1})} =
{\sum_{m\in \BZ} \phi_{m}(q)  z^{2m} \over \prod_{n\ge 1}(1-q^n)^2}
\eqno{\rm(2.1.7a)}$$
(see e.g.\ $\q{\bmcp}$) with
$$\phi_m(q)=\sum_{r\ge 0} (-1)^r q^{ {r(r+1)\over 2}+mr},\qquad
    \phi_{-m}(q)=q^m \phi_m(q).
\eqno{\rm(2.1.7b)}$$
The vacuum character $B_0^0(q)$ of \slu\ is given by the $m=0$ term of the
r.h.s.\ of eq.\ (2.1.6) which is the generating function of the $U(1)$
singlets:
$$B_0^0(q)={\phi_0(q)-q\phi_1(q) \over {\prod_{n\ge 1}(1-q^n)}^2}.
\eqno{\rm(2.1.8)}$$
This agrees with the corresponding formula of $\q{\bakas}$ after correcting
the following misprint: the exponent
of $f(q)$ in formula (4.14) of $\q{\bakas}$ should read 2 not 3.
For the difference of $B_0^0(q)$ and the character $\chi_{2,3,4,5}$ of the
vacuum module freely generated by fields of dimension 2, 3, 4 and 5
we obtain
$$B_0^0(q)-\chi_{2,3,4,5} = -2q^8 - 4q^9 -9q^{10} + {\cal O}(q^{11}) .
\eqno{\rm(2.1.9)}$$
This supports the identification of the coset algebra \slu\ with the special
solution of \wfuenf\ because both algebras have two generic null fields at
conformal dimension~8~$\q{\hornfeck}$.
\sn
\subs{Commuting charges}
\sn
One might wonder if the confusion in the literature about the existence
of a finite set of generators for the quantum coset \slu\ has any impact
on the associated set of commuting charges $\q{\yuwu}$. Note that the
computation of $\q{\yuwuC}$ need not necessarily reflect the general case
because it strongly relies on $k=-1$, i.e.\ $c=-4$. For $c=-4$ the
existence of one charge per integer dimension is due to the fact that
$c=-4$ is exactly one of the two values where $\w_\infty$ truncates to the
$\w(2,3,4,5)$ under consideration $\q{\letter}$. Therefore, the conserved
charges might be specific to $c=-4$ and the general case could be different.
\sn
In this section
we will present a few explicit results indicating that the misinterpretation
of the generating set has no impact on the conserved charges. In particular,
it seems to be true that there is one conserved charge for each integer
dimension $\q{\yuwu}$ -- in contrast to the ${\cal WA}_n$ Casimir
algebras where some of these charges are missing. The generic null fields
(or relations) play a crucial r\^ole because
the commutator of two charges ${\cal P}_i$ and ${\cal P}_j$ does no
longer need to be exactly zero but can be proportional to the
integral over a null field, relaxing the constraints on the commuting
charges. Using the explicit form of the abstract \wfuenf-algebra
$\q{\hornfeck}$
\footnote{${}^3)$}{Note that the prefactor of
$\SC_{3\,5}^{P^{33}_6}$ in the appendix of
this reference has a misprint and must
therefore be multiplied by a factor of ${4\over 5}$ !},
we can indeed construct the first few commuting charges:
$$\eqalign{
{\cal P}_1 &= \oint L\quad\quad {\cal P}_2 = \oint W_3\quad\quad
{\cal P}_3 = \oint \left(W_4+{3\over{16(2+c)}} C_{3\,3}^4\ L L\right)\cr
{\cal P}_4 &= \oint \left(W_5 + {{64}\over{5(22+5c)}}
      {{C_{3\,4}^5}\over{C_{3\,3}^4}} L W_3 \right)\cr
{\cal P}_5 &=\oint\Biggl( \left(4+c\right) W_3 W_3 +
{{\left(5{c^3}-100{c^2}-1100c-32\right)}\over
    {2\left(7+c\right)\left(2c-1\right)
    \left(22+5c\right)}}\ L L L\ +\cr
  &{{\left(20320+10928c+1032{c^2}+34{c^3}+5{c^4}\right)}\over
    {24\left(1-2c\right)\left(7+c\right)
    \left(22+5c\right)}}\partial L\partial L -
  {3\over 2} C_{3\,3}^4 \ L  W_4\Biggr).\cr}
\eqno{\rm (2.1.10)}$$
The various null fields with dimensions $\geq 8$ appear in the commutators
of ${\cal P}_3$ with ${\cal P}_4$ and of ${\cal P}_2$, ${\cal P}_3$ and
${\cal P}_4$ with ${\cal P}_5$. The presence of the charge ${\cal P}_5$
is contrary to the case of the ${\cal WA}_4$ algebra with the same spin
content, where all charges of dimension $0\ (\hbox{mod}\ 5)$ are missing.
\sn
We have checked that specializing (2.1.10) to $c=-4$ ($k=-1$) one finds
agreement with eq.\ (27) of $\q{\yuwuC}$. One can also check that the
classical limit of (2.1.10) is consistent with the classical conserved
charges of $\q{\toppan}$. The computation of this classical limit which
has to be performed on the construction in terms of $\sltr$ is
straightforward but tedious because leading orders in $\hbar$ cancel
and the classical conserved charges arise from (2.1.10) as subleading orders
in $\hbar$.
\sn
\subs{The coset \sviru}
\sn
In the following we report an explicit construction of the generators
in the quantum coset \sviru.
The classical version of this coset was presented in $\q{\ajl}$
where also some of the results presented below were already mentioned.
\bn
For the $N=2$ Super Virasoro algebra we use the explicit form
of $\q{\ralphnpb}$ eq.\ (2.12).
The coset we intend to consider is defined as the commutant of the
$U(1)$-current $J$. For explicit calculations it is convenient
to use a Vertex-operator approach. First we note that
$$\lb J_m, X_n \rb = 0 \quad \forall m,n
\qquad \Longleftrightarrow \qquad
J_{-m} X_{d(X)} \vac = 0 \ , \quad m=0,1,\ldots,d(X)
\eqno( {\rm 2.1.11})$$
for any field $X$ with fixed scale-dimension $d(X)$.
For later purposes we also
note a similar result for primary fields. The equivalence we will use is
$$\eqalign{
\lb \Lh_{m}, X_n \rb &= (n-(d(X)-1)\ m) \ X_{m+n} \quad \forall m,n \cr
\Longleftrightarrow \qquad
\Lh_{-m} X_{d(X)} \vac &= 0 \ , \quad m=1,\ldots,d(X) \cr
}\eqno( {\rm 2.1.12})$$
for the primarity of a field $X$ with scale-dimension $d(X)$ with respect to
some energy momentum tensor $\hat{L}$.
\mn
One now proceeds as follows: First one makes the most general ansatz at a
given scale dimension for an invariant field (it is better to use standard
normal ordered products and derivatives). In our case we also have a
$J_0$-grading in addition to the $L_0$-grading. Therefore, in our ansatz we
may restrict to `uncharged' composite fields (fields with $J_0$-grade 0).
This automatically ensures that the condition $J_0 X_{d(X)} \vac = 0$
is satisfied. When extracting conditions for the ansatz from the vacuum
module it is important to use a basis in the space of fields which can most
conveniently be implemented by choosing a lexicographic ordering.
\sn
First we find a unique invariant field at scale dimension 2
$$\Lh = L - {3 \over 2 c} \NN{J}{J}.\eqno{\rm(2.1.13)}$$
$\Lh$ satisfies the Virasoro-algebra eq.\ (1.1.2) with shifted central charge
$\ch = c - 1.$
Exploiting the first condition eq.\ (2.1.11) we find two- respective four- and
six-dimensional invariant spaces of fields at conformal dimensions 3, 4 and 5.
Imposing additionally the primarity condition eq.\ (2.1.12) we find
{\it unique} primary invariant fields at scale
dimensions 3, 4 and 5. For brevity
we omit the lengthy expressions for the primary fields $W^{(4)},W^{(5)}$
and present only $W^{(3)}$:
$$W^{(3)} = \nu \left(
-{6 \over c^2} \ J J J
+{6 \over c} \ L J
+ \GB G
+{c-9 \over 3 c} \ \de^2 J -  \de L
\right).\eqno{\rm(2.1.14)}$$
The normalization constant $\nu$ is fixed to
$\nu^2 ={3 (\ch + 1) \over 2 (2 \ch - 1) (\ch + 7)}$
by imposing the normalization condition $d_{3,3} = {\ch \over 3}$.
Now a standard computation shows that the structure constants
connecting the primary fields $W^{(3)},W^{(4)},W^{(5)}$
are identical to those obtained for the special solution of $\w(2,3,4,5)$
involving null fields $\q{\hornfeck}$. This suggests that the coset
considered here coincides with this algebra.
\sn
Analogous to \slu\ this identification can also be inferred
from a character argument, i.e.\ from the vacuum character of \sviru.
According to the results of $\q{\ajl}$ we have to count the states in the
vacuum representation of the $N=2$ algebra which are charge neutral and
do not contain $J$.
\sn
We use the Jacobi-triple product identity
to write the $N=2$ vacuum character as
$$ {\prod_{n\ge 1}  (1+z^2 q^{n+{1\over 2} })
(1+z^{-2} q^{n+{1\over 2}})
  \over \prod_{n\ge 1}  (1- q^n)(1-q^{n+1}) }=
 {(1-q)\over {\prod_{n\ge 1}(1- q^n)}^3}{ \sum_{m\in \BZ} q^{m^2\over 2}
  z^{2m} \over (1+z^2 q^{1\over 2})(1+z^{-2} q^{1\over 2}) }.
   \eqno{\rm(2.1.15)}$$
Expanding the denominator into the geometric series
$$ {1\over (1+z^2 q^{1\over 2})(1+z^{-2} q^{1\over 2}) }=
   \sum_{\alpha,\beta\ge 0} (-1)^{\alpha+\beta} q^{\alpha+\beta\over 2}
    z^{2(\alpha-\beta)}\eqno{\rm(2.1.16)} $$
one obtains for
the vacuum character $A_0^0(q)$ of the \uone-commutant in $SVIR(N=2)$:
$$A_0^0(q)= {(1-q)\over {\prod_{n\ge 1}(1- q^n)}^2 }
   \sum_{\alpha,\beta\ge 0} (-1)^{\alpha+\beta} q^{\alpha+\beta\over 2}
    q^{(\alpha-\beta)^2\over 2}. \eqno{\rm (2.1.17)}$$
This is compatible with three additional generators of dimensions 3, 4, 5
respectively and two generic null fields at scale dimension 8 -- precisely
what has been found for the special $\w(2,3,4,5)$ in $\q{\hornfeck}$.
This supports the above identification quite convincingly.
\mn
The question if the coset algebras \slu\ and
\sviru\ are isomorphic is a natural question since we claimed that both
cosets are isomorphic to \wfuenf.
First we show that the vacuum characters of \slu\ eq.\ (2.1.8) and
\sviru\ eq.\ (2.1.17) are equal.
To this end we define
$$f(\alpha,\beta)=  (-1)^{\alpha+\beta} q^{\alpha+\beta\over 2}
q^{(\alpha-\beta)^2\over 2}.
\eqno{\rm(2.1.18)}$$
Since $f(\alpha+1,\beta+1)=qf(\alpha,\beta)$ most terms in eq.\ (2.1.17)
cancel so that only the two axes $\alpha=0$ and $\beta=0$ survive.
Hence we obtain
$$A_0^0(q)= {1\over {\prod_{n\ge 1}(1- q^n)}^2}
   \left( \sum_{\alpha\ge 0} (-1)^{\alpha} q^{(\alpha+1)\alpha\over 2}
  + \sum_{\beta\ge 0} (-1)^{\beta+1} q^{(\beta+1)(\beta+2)\over 2} \right).
\eqno{\rm(2.1.19)}$$
Using the definition of $\phi_m(q)$ (eq.\ (2.1.7b)) this coincides obviously
with the vacuum character eq.\ (2.1.8) of \slu.
\sn
However, it is not difficult to show that the commutants are
algebraically isomorphic. One uses the realizations
of \esl\ and the \ntsvir\ in terms of a free boson and $\BZ_k$ parafermions
(see e.g.\ $\q{\nemesch,\qiu}$). The $U(1)$-current $J$ is just given by
the derivative of the free boson $\phi$. The fields $J^{\pm}$ resp.\
$G,\GB$ are represented by two parafermionic currents $\psi^\pm$
with conformal dimension $1-{1\over k}$ dressed with a vertex operator
$e^{\pm i \alpha(k) \phi}$ with a suitably chosen $\alpha(k)$
(compare $\q{\nemesch,\qiu}$).
The parafermions $\psi^\pm$ themselves can be realized in terms of two
free bosons $\q{\nemesch}$. The \uone-commutant is generated by the
invariants quadratic in $J^\pm$ resp.\ $G,\GB$ implying that in
both cases the commutant is built out of the two parafermions.
Thus the two \uone-commutants are isomorphic.
Note that in the $k\to\infty$ limit the two parafermions
have conformal dimension 1 and the character of the commutant
is exactly the one calculated in eq.\ (2.1.8).
\mn
Until now we have not treated the cancellation of
the classical generators and relations in the quantum case.
Therefore, let us understand why in the quantum case we fail to get a new
generator at scale dimension 6 in the coset \sviru\
and compare it with the classical situation. The field $((\GB G)(\GB G))$
does not belong to the set of quantum generators -- classically it vanishes
identically because of the Pauli principle. However, due to the
nonassociativity of the normal ordered product it satisfies the
following equality (quantum):
$$\eqalign{
 &((\GB G)(\GB G)) = (c+9) \ \NN{G}{\de^3 \GB}
+ 2 \ \NN{\de G}{\de^2 \GB}
+ \NN{\de^2 G}{\de \GB}
-{\textstyle{c \over 9}} \ \NN{\de^3 G}{\GB}
+ \NN{(\NN{G}{\de \GB})}{\de J} \cr
&+ \NN{(\NN{\de G}{\GB})}{\de J}
+ 2 \ \NN{(\NN{G}{\de^2 \GB})}{J}
+ 2 \ \NN{(\NN{\de^2 G}{\GB})}{J}
+ 2 \ \NN{(\NN{G}{\GB})}{\de L}
+ 2 \ \NN{(\NN{G}{\de \GB})}{L}
- 2 \ \NN{(\NN{\de G}{\GB})}{L} \cr
&+ 2 \ \NN{L}{\de^2 L}
- {\textstyle{2 \over 3}} \ \NN{J}{\de^3 L}
- \NN{\de J}{\de^2 L}
- {\textstyle{1 \over 3}} \ \NN{\de^3 J}{L}
+ {\textstyle{1 \over 6}} \ \NN{J}{\de^4 J}
+ {\textstyle{1 \over 6}} \ \NN{\de J}{\de^3 J}
- {\textstyle{c - 4 \over 60}} \ \de^4 J
+ {\textstyle{c \over 18}} \ \de^3 L . \cr}
\eqno{\rm (2.1.20)}$$
{}From this expression one can see explicitly that $((\GB G)(\GB G))$ tends
to zero in the classical limit (compare section 3.1.).
This equality implies that the square of the
generator $W^{(3)}$ contains correction terms that cancel contributions which
might give rise to new generators. This mechanism guarantees that the quantum
coset \sviru\ has at least a finitely generated subalgebra of type
$\w(2,3,4,5)$. Such cancellations do
not occur classically. Therefore, the coset under consideration is
infinitely generated classically with a first relation at scale dimension 6.
\mn
It is also straightforward to derive the representations of the coset
algebra from those of the $N=2$ Super Virasoro algebra. Each highest weight
representation of the Super Virasoro algebra satisfying
$$L_0 \hwt = h \hwt \qquad J_0 \hwt = \tau \hwt
\eqno{\rm(2.1.21)}$$
gives rise to one (in general reducible) representation of the coset algebra
with the following eigenvalue equations for the highest weight
vector:
$$\Lh_0 \hwt = \hat{h} \hwt \quad\quad\
W^{(i)}_0 \hwt = w_i \hwt\quad\quad\ i=3,4,5.
\eqno{\rm(2.1.22)}$$
Obviously, the central charge is shifted by one: $\ch = c-1$.
$\hat{h}$ and $w_i$ can be expressed through $h$, $\tau$
and $c$ using the realization of the generators. One finds:
$$\hat{h} = h - {3 \tau^2 \over 2 c}\quad \quad
w_3 = \nu \left(h {6 \tau \over c}
  - {\tau ( c^2 + 18 \tau^2) \over 3 c^2} \right).
\eqno{\rm(2.1.23)}$$
In particular, all minimal models of the coset algebra can be derived from
those of the $N=2$ Super Virasoro algebra. Note that the $N=2$ Super Virasoro
algebra presumably has only {\it unitary} minimal models.
\mn
We know that the \wfuenf\ is the unifying algebra for the first unitary model
of ${\cal WA}_{k-1}$ $\q{\letter}$, i.e.\ that ${\cal WA}_{k-1}$ truncates for
$c(k) = 2 {{k-1}\over{k+2}}$ to \wfuenf. From the above coset realizations
$${{\widehat{sl(2,\BR)}}\over{\widehat{U(1)}}}\cong
{{SVIR(N=2)}\over{\widehat{U(1)}}}\cong{\cal W}(2,3,4,5)
\eqno{\rm (2.1.24)}$$
we expect that the first unitary model of ${\cal WA}_{k-1}$ agrees with the
rational models of \sviru\ coming from the unitary minimal models
of the $N=2$ Super Virasoro algebra. Indeed, using eq.\ (2.1.23) one can
compute the conformal dimensions and $w_i$ quantum numbers for \sviru\ and
one finds perfect agreement with the first unitary model of ${\cal WA}_{k-1}$.
\sn
For ${\cal WA}_{k-1}$ we are in the fortunate position
that at least some structure constants are known generally
\q{\klausWAN}:
$$\eqalign{
\left(C_{3 \, 3}^{4}\right)^2
  =& {64 (k-3) (c+2) (c (k+3)+2 (4 k+3) (k-1))
     \over (k-2) (5 c+22)(c (k+2)+(3 k+2) (k-1))} \cr
C_{3 \, 3}^{4} C_{4 \, 4}^{4}
  =& {48 (c^2 (k^2\!-\!19)+3 c (6 k^3\!-\!25 k^2\!+\!15)
      +2 (k\!-\!1) (6 k^2\!-\!41 k\!-\!41))
      \over(k-2) (5 c+22) (c (k+2)+(3 k+2) (k-1))} \ . \cr
} \eqno{(\rm 2.1.25a)}$$
We compute the so far unknown structure constants
$\left(C_{3 \, 4}^{5}\right)^2$ and $C_{4\, 5}^5$.
This can be done in two different ways.
Either one uses an ansatz involving a representation
theoretic argument $\q{\klausREP}$ or one evaluates special
Jacobi identities for ${\cal WA}_{k-1}$. Both methods yield the same
result:
$$\eqalign{
&\left(C_{3 \, 4}^{5}\right)^2 =
{25 (c k + 4 c + 15 k^2 - 3 k - 12) (5 c + 22) (k-4) \over
(c k + 2 c + 3 k^2 - k - 2) (7 c + 114) (k-2)}\cr
&C_{4\, 5}^5 = {\textstyle{15\over 8}}
 \bigl(\left( 2 + c \right)
\left( 114 + 7c \right) \left( k-3 \right)
\left( 8{k^2}+ ck- 2k+ 3c -6  \right)\bigr)^{-1} C_{3\, 3}^4 \cr
&\times\! (6756 \! + \! 4120c \! - \! 483{c^2}\! \! - \! 97{c^3}\! \! - \!
6972{k^2} \!\! - \! 5192c{k^2} \!\! - \! 467{c^2}{k^2}\!\! + \!
3{c^3}{k^2} \!\! + \! 216{k^3}\! \! + \! 856c{k^3} \!\! + \!
94{c^2}{k^3})\cr
}\eqno{(\rm 2.1.25b)}$$
where we have also presented the result for $C_{4\, 5}^5$ that was
obtained exclusively with the methods of $\q{\klausREP}$.
Inserting $c(k) = 2 {{k-1}\over{k+2}}$ into eq.\ (2.1.25)
reproduces eq.\ (2.1.5) verifying again the truncation of ${\cal WA}_{k-1}$
at $c(k)$ to \wfuenf\ given in $\q{\letter}$. For this particular value of
the central charge all generators with scale dimension 6 or
higher turn out to be null fields. This can be verified directly
for $k=6$ by inspecting the structure constants presented in
$\q{\hornfeck}$ for ${\cal WA}_5 \cong \w(2,3,4,5,6)$
(see also $\q{\letter}$).
\sn
This identification can be interpreted in a more general way if
one inspects the ``dual coset pairs'' given
in $\q{\bouschou,\bogo,\altschuler}$.
Here dual coset pairs possessing equivalent energy momentum tensors
(``$T$-equivalence'') have been presented,
e.g.\ \slnsln\ $\cong$\ \slnslndual.
Specializing to $\kappa=\mu=1$ we know that the left hand
side yields the first unitary model
of ${\cal WA}_{k-1}$ at $c(k)$ $\q{\bbss}$. The right hand side
reduces to the coset $\widehat{sl(2,\BR)}_k/\widehat{U(1)}$
and we get back the equivalence
considered above.
Our investigations go beyond $\q{\bogo,\altschuler}$: we have
verified that the maximally extended symmetry algebras are
isomorphic and not just the energy momentum tensors.
In section three we will discuss the relation of unifying $\w$-algebras
to level-rank duality in more general situations.
\mn
\section{2.1.2.\ The coset \beru}
\sn
Like the \ntsvir-algebra, the \ber-algebra of Polyakov and
Bershadsky $\q{\poly,\bersh}$ has a primary field of dimension 1 and
two of dimension ${3\over 2}$ in addition to the energy momentum tensor.
\ber\ is obtained by quantum hamiltonian reduction for the nonprincipal
embedding of $sl(2)$ in $sl(3)$.
In contrast to the \ntsvir-algebra the spin ${3\over 2}$ fields $G^{\pm}$
obey bosonic statistics. One important consequence is that the OPE of the
generators does not close linearly, but quadratically in $\Jz$.
\sn
In the normalization of Bershadsky $\q{\bersh}$ the algebra reads
$$\eqalign{
\Jz \star \Jz &= {\textstyle{1\over 3}}(2\kb +3) \,\idi
\quad \quad \Jz \star \Gpm = \pm\,\Gpm
  \quad \quad G^{\pm}\star G^{\pm} = 0 \cr
\Gp \star \Gm &= (\kb+1)\,(2\kb+3)\, \idi \,
+ \, 3\,(\kb+1)\, \Jz \,+\,3\,\pjz \cr}
\eqno{\rm(2.1.26)}$$
where the central charge $\cb$ of this algebra is connected to $\kb$ by
$$
\cb = -{(2\,\kb+3)\,(3\,\kb+1)}({\kb+3})^{-1}
\eqno{\rm(2.1.27)}$$
and the field $\pjz$ is the composite primary field:
$$\pjz  =  \Jz\Jz + {\textstyle{2 \over 3}} (\kb+3)(3\kb+1)^{-1} \ \Tb.
\eqno{\rm(2.1.28)}$$
The classical version of the coset \beru\ has been treated in
$\q{\delduc}$ and an infinite generating set of the commutant has
been given. A nonredundant (infinite) set of generators and the
full (infinite) generating set of relations between these generators
have been presented in $\q{\ajl}$. We discuss now the quantum version
of this coset model.
\sn
As for \slu, we construct \beru\ by computing a Virasoro-operator
$\Tbc$ and a primary $\wbd$-field both commuting with $\Jz$
\footnote{${}^4)$}{Since $\wbd$ commutes with $\Jz$ it is simultaneously
primary under $\Tb$ and $\Tbc$.}, and recursive evaluation of the OPEs
starting with $W_3\star W_3$. The bosonic statistics of the fields
$G^{\pm}$ implies that the classical limit of $\wbpdds$ whose leading
term is $((\Gp \, \Gm)\, (\Gp \, \Gm))$ does not vanish. Therefore,
in the coset-algebra a $\wbs$-field
appears, contrary to the coset \sviru. Analogously the only composite
field with scale dimension 7, $\wbpdvs$, does not vanish in the classical
limit. From counting the invariant fields (commuting with $\Jz$) up to
scale dimension~8 we expect that the quantum coset does not contain a
spin 8 generator which is confirmed by our explicit calculations below.
\sn
First we have to construct the modified energy momentum tensor and the
$\wbd$-field both commuting with the \uone-current $\Jz$.
These fields are given by:
\footnote{${}^5)$}{The difference of $\Tbc$ to the one of the
\sviru-coset is due to the different normalization of $\Jz$.}
$$\eqalignno{
\Tbc  &= \Tb  - {\textstyle{3 \over 2}}(3+2\kb)^{-1} \Jz\Jz &{\rm(2.1.29)} \cr
\wbd  &= {1 \over {2(3\! +\! 2k)^2}} \Bigl(
2(3\! +\! 2k)^2\Gp\!\Gm\! -\! 18(2\! +\! k)\Jz\!\Jz\!\Jz +
6(3\! +\! k)(3\! +\! 2k)L\Jz\! -  \cr
&\ \  6(3\! +\! 2k)^2\pa\Jz\! \Jz +
    (3\! +\! k)(3\! +\! 2k)^2\pa\Tb -
    2(3\! +\! 2k)(6\! +\! 4k\! +\! k^2)\pa^2\Jz \Bigr)& {\rm(2.1.30)}\cr
}$$
where the central charge is again shifted by one,
$$
\cbc  =  \cb - 1  =  -6{(1+k)^2}({3+k})^{-1}.
\eqno{\rm(2.1.31)}$$
We verified the following OPEs:
$$\eqalign{
\wbd \star \wbd &= d_{3,3}\idi  +  \wbv \quad\quad
\wbd \star \wbv = {{d_{4,4}}\over{d_{3,3}}}  \wbd  +  \wbf \quad\quad
\wbd \star \wbf = {{d_{5,5}}\over{d_{4,4}}} \wbv  +  \wbs  \cr
\wbd \star \wbs &= {{d_6}\over{d_{3,3}}}\wbd+
{{d_{6,6}}\over{d_{5,5}}} \wbf  +  \wbsi  +
\SCu_{3\,6}^{\wbpdva} \wbpdva \cr
\wbv \star \wbv &= d_{4,4}  \idi + \SCu_{4\,4}^4 \wbv  +
\SCu_{4\,4}^6\wbs +  \SCu_{4\,4}^{\wbpdds}\wbpdds \cr
\wbv \star \wbf &= {{d_{5,5}}\over{d_{3,3}}}\wbd+\SCu_{4\,5}^5\wbf+
\SCu_{4\,5}^7\wbsi+\SCu_{4\,5}^{\wbpdvs}\wbpdvs+
\SCu_{4\,5}^{\wbpdva} \wbpdva \cr
\wbv \star \wbs &= \SCu_{4\,6}^4  \wbv+ \SCu_{4\,6}^6\wbs+
\SCu_{4\,6}^{\wbpdds}\wbpdds+\SCu_{4\,6}^{\wbpdda}\wbpdda+
\SCu_{4\,6}^{\wbpdfa}\wbpdfa+\SCu_{4\,6}^{\wbpvva}\wbpvva+
\SCu_{4\,6}^{\wbpdfn}\wbpdfn.\cr
}\eqno{\rm(2.1.32)}$$
The fields $\wbv$ to $\wbsi$ are again defined recursively by these OPEs.
In order to save space we omit the lengthy expressions of these fields.
For our purpose (i.e.\ to show that there is no new spin-8 field)
it was not necessary to define the new fields $\wbs$ and $\wbsi$
to be orthogonal to the composite fields $\wbpdds$ and $\wbpdvs$, i.e.\
they are just defined by eq.\ (2.1.32).
The appearance of $\wbd$ in the OPE of $\wbd$ with $\wbs$
reflects this fact, where $d_6$ is the off-diagonal element
of the spin-6 fields:
$\wbs \star \wbpdds \,=\, d_6\,\idi\,+\,\ldots\ .$
The OPE of $\wbv$ with $\wbs$ does not show a new spin-8 field.
We expect therefore that the \berut-coset has a
subalgebra of type \wsieben\ (being different from the
${\cal WA}_6-$algebra). The OPEs eq.\ (2.1.32) do not exclude
completely the existence of fields with dimension $\geq 9$ in the coset,
but supporting counting arguments on the classical level $\q{\ajl}$ can
be given. Moreover, these counting arguments indicate two
generic null fields at dimension 10.
\sn
The structure constant $\SC_{3\,3}^4$ in standard normalization
for this algebra is given by
$$\left(\SC_{3\,3}^4\right)^2=
{{128{\kb^2}\left( 3 + \kb \right) \left( 5 + 3\kb \right)
     \left( 12 + 5\kb \right) }\over
   {\left( 1 + 2\kb \right) \left( 9 + 4\kb \right)
     \left( -18 + 19\kb + 15{\kb^2} \right) }}.
\eqno{\rm(2.1.33)}$$
Rewriting it in terms of the new central charge
$\cbc$ would lead to square roots in the expression.
Additional structure constants are given in appendix B.
\mn
The spin content of \beru\
suggests that this algebra is a unifying algebra
for a certain model of the ${\cal WA}_{n-1}$ Casimir algebras.
Indeed, with the relation $k = {{n-3}\over 2}$ we find that the
central charge of the \beru-coset eq.\ (2.1.31) coincides with the
central charge $c_{{\cal A}_{n-1}}(n+1,n+3)= -3{{(n-1)^2}\over{n+3}}$ of the
{\it nonunitary} model $p=n+1,\ q=p+2$ of ${\cal WA}_{n-1}$.
Furthermore, we find agreement of the structure constants
$(\SC_{3\, 3}^4)^2$ of ${\cal WA}_{n-1}$ eq.\ (2.1.25a) and
the \beru-coset eq.\ (2.1.33) for these particular choices of $k$ and
$c$. Thus we expect that the generators of ${\cal WA}_{n-1}$
with scale dimension $\geq 8$
are null fields for these values of the central charge so that
the Casimir algebras ${\cal WA}_{n-1}$ truncate to this \wsieben.
This is also supported by the study of Kac-determinants in $\q{\letter}$.
\sn
Unfortunately, we are not able to verify that the highest weight
representations of the corresponding models coincide since the
highest weights of the minimal models of the \ber-algebra are
not known. We calculated explicitly the highest weights of the
first few minimal models of \ber\ in the sector with periodic
boundary conditions for the fields with half-integer spin. The
induced representations of the coset lie all in the Kac-table of
the corresponding ${\cal WA}_{n-1}$ models. Our explicit
calculations show that the set of highest weights given
in $\q{\bersh}$ is certainly too big.
\sn
\subs{Generalizations}
\sn
Our aim is to generalize the result of the preceding section
to unifying $\w$-algebras of the nonunitary models
$c_{{\cal A}_{n-1}}(n+1,n+r)$ of ${\cal WA}_{n-1}$.
\sn
{}From the experience with the case $r=3$ (the Polyakov-Bershadsky algebra) we
expect that these unifying $\w$-algebras should arise as cosets of DS type
$\w$-algebras $\w^{sl(r)}_{r-1,1}$ which have a $\widehat{U(1)}$ Kac-Moody
subalgebra. Consider therefore the $r$-dimensional defining representation
of $sl(r)$. One embeds \hbox{$sl(r-1)$} $\oplus U(1)$ into $sl(r)$ by
separating the last node from the Dynkin diagram of $sl(r)$
(the first $r-2$ nodes realize $sl(r-1)$). The defining representation of
$sl(r)$ splits into the $r-1$ dimensional defining representation of the
$sl(r-1)$ subalgebra and  the trivial representation. Take now the principal
$sl(2)$ embedding into the $sl(r-1)$.
Quantum hamiltonian reduction yields an algebra of type
$\w(1,2,\dots,r-1,{r \over 2},{r \over 2})$ with all fields bosonic.
The spin ${r \over 2}$ fields have $U(1)$-charge $\pm 1$
whereas all other simple fields are uncharged. In the notation introduced
in section 1.1.\ this algebra is denoted by $\w^{sl(r)}_{r-1,1}$.
\sn
Following $\q{\ajl}$ we obtain for the coset:
$${\w^{sl(r)}_{r-1,1}\over{\widehat{U(1)}}}\cong \w(2,3,\dots,2r+1).
 \eqno(2.1.34) $$
Comparing with table 1 in $\q{\letter}$ we see that this algebra
has indeed the spin content of the unifying algebra for the models
$c_{{\cal A}_{n-1}}(n+1,n+r)$.
\sn
To get a confirmation let us treat the case $r=4$. We constructed
explicitly the OPEs for the algebra $\w^{sl(4)}_{3,1}\cong
\w(1,2,2,2,3)$. To check our conjecture
we calculated the structure constant $C_{3\, 3}^4$ in the coset
$\w(1,2,2,2,3)$/\uone\ $\cong\w(2,3,4,5,6,7,8,9)$ yielding
truncations of ${\cal WA}_{n-1}$ to $\w(2,3,4,5,6,7,8,9)$ for:
$$
c_1(n) = -4 { {(n\!-\!1)(2n\!-\!3)}\over{(n\!-\!3)(n\!-\!4)} } \quad\quad
c_2(n) = -4 { {(n\!-\!1)(2n\!-\!1)}\over{(n\!+\!4)} } \quad\quad
c_3(n) = -  { {(3n\!+\!1)(5n\!+\!3)}\over{(n\!+\!3)} }.
\eqno({\rm 2.1.35})$$
The first truncation corresponds to the $\w_\infty$
truncation of section 3 in $\q{\letter}$ with $r=4$.
The second truncation is exactly the truncation of the
nonunitary models $c_{{\cal A}_{n-1}}(n+1,n+4)$ of
${\cal WA}_{n-1}$. This confirms our claim from above.
\sn
For a final check we compute the central charge of the algebra
$\w_{r-1,1}^{sl(r)}$ as a function of the level $k$ of the underlying
Kac-Moody $\widehat{sl(r)}_k$. We use the formula
$c = N_t - {1 \over 2} N_s - 12 \vert \sqrt{k + h^\vee} \delta
- {1 \over \sqrt{k + h^\vee}} \rho \vert^2$ (see e.g.\ $\q{\laszlorep}$).
For the embedding under consideration one has $N_t = r - 1$, $N_s = 2$
for $r$ odd
and $N_t = r+1$, $N_s = 0$ for $r$ even. A matrix representation for
$\rho$ and $\delta$ is $\rho = {\rm diag}({r-1 \over 2}, {r-3 \over 2},
\ldots, -{r-1 \over 2})$, $\delta = {\rm diag}({r-2 \over 2}, \ldots,
{r \over 2} - {\lbrack r \rbrack \over 2}, 0 , -{r \over 2} +
{\lbrack r \rbrack \over 2}, \ldots, -{r-2 \over 2})$. The scalar product
of two matrices is the usual one: $a \cdot b = {\rm tr}(a b)$. With this
data it is straightforward to calculate $c$ as a function of $k$ for
$\w_{r-1,1}^{sl(r)}$:
$$c =  - {(k r^2 - 2 k r + r^3 - 3 r^2 + 1)(k r - k + r^2 - 2 r)\over k + r}.
\eqno(2.1.36)$$
{}From the fact that the coset (2.1.34) is a unifying algebra for
${\cal WA}_{n-1}$ at $c_{{\cal A}_{n-1}}(n+1,n+r)$ one must have
$c - 1 = c_{{\cal A}_{n-1}}(n+1,n+r)$. This is indeed satisfied for
$$k =  {n- r^2 + 2 r \over r - 1}.
\eqno(2.1.37)$$
Note that the relation (2.1.37) between the level $k$ and the rank of
${\cal WA}_{n-1}$ is linear with a denominator $r-1$. This simple relation
increases our confidence that the cosets (2.1.34) play indeed the r\^ole
of unifying $\w$-algebras.
\mn
\section{2.1.3.\ The coset \slslgen}
\sn
In this section we study the diagonal $\sltr$ coset.
Noting from $\q{\gosch,\bogo,\ajl}$ that
$$\lim_{\mu \to \infty} {\sltr_\kp \oplus \sltr_\mu \over \sltr_{\kp+\mu}}
 = {\sltr_{\kp} \over sl(2,\BR)}
  \eqno({\rm 2.1.38})$$
and that the l.h.s.\ at generic $\mu$ is a deformation of the r.h.s.\
we conclude that much information about the coset we are interested in
can already be obtained from the simpler coset $\sltr_{\kappa} / sl(2,\BR)$.
Therefore, we treat this coset first.
\sn
\subs{\eslk/$sl(2,\BR)$}
\sn
Upon a mode expansion of the OPE eq.\ (2.1.1) we obtain the following
commutation relations for the Kac-Moody algebra $\sltr_\kp$:
$$\eqalign{
\lb \jpm_{m}, \jpm_{n} \rb =&\ 0 , \quad \quad \quad \quad \quad \quad
\lb \jn_{m}, \jn_{n} \rb = 2 \kappa \delta_{m,-n} n , \cr
\lb \jn_{m}, \jpm_{n} \rb =& \pm 2 \jpm_{m+n} ,  \quad \quad
\lb \jp_{m}, \jm_{n} \rb = \jn_{m+n} + \kappa \delta_{m,-n} n . \cr
}  \eqno({\rm 2.1.39})$$
We want to construct the invariant subspace under the natural action
of $sl(2,\BR)$ on $\sltr$. This means that we have to construct all
fields commuting with $\{\jpm_0, \jn_0\}$. From the coefficient of
$\lb \jn_0, X_m \rb$ in front of $X_m$ one concludes first that $X$
must be $\jn$-uncharged.
\sn
It is straightforward to make the most general ansatz in uncharged
fields of scale dimension 2 and determine the field(s) that also
commute(s) with $\jpm_0$. It is no surprise to find a unique
invariant field (up to normalization) at dimension 2, which
is just given by the Sugawara construction:
$$\eqalign{
2 (\kp + 2)\ L :&=
\n(\jp,\jm) + {\textstyle{1 \over 2}}\n(\jn,\jn) + \n(\jm,\jp) \cr
&=\jp\, \jm + {\textstyle{1 \over 2}} \jn\, \jn + \jm\, \jp  \cr
}\eqno({\rm 2.1.40})$$
with $\n(\jn, \jn) = \jn\, \jn$,
$\n(\jpm, \jmp) = \jmp\, \jpm \pm {1 \over 2} \de \jn$.
It is straightforward to check that $L$ satisfies
the Virasoro algebra eq.\ (1.1.2) with central charge
$c = 3 \kappa (\kappa + 2)^{-1}.$
Note that the original currents are primary with respect to $L$.
Therefore, one can conveniently use quasi-primary normal
ordered products in the calculations which we shall do below.
\sn
One can proceed along the same lines to find the next independent
invariant field at scale dimension 4. However, this approach becomes
unfeasible for higher dimensions. Therefore, we use group
theoretic knowledge about the generators of this coset $\q{\ajl}$.
Let $g^{i j}$ be the metric on the simple Lie algebra and $f_{i j}^{k}$
be the structure constants. With the inverse of the metric $g_{i j}$ we
define $\epsilon_{i j k} := \sum_l g_{k l} f_{i j}^{l}$. Then, the
$SL(2, \BR)$ invariant generators are given by $\q{\ajl}$
$$\eqalignno{
S_{m,n} := &\sum_{i,j} g_{i j}\ \de^n J^{(j)} \de^m J^{(i)},
  &({\rm 2.1.41a})\cr
S_{m,n,k} := &\sum_{i,j,l} \epsilon_{i j l}
    \ \de^k J^{(l)}\de^n J^{(j)}\de^m J^{(i)}
  &({\rm 2.1.41b})\cr
= & \, \de^k \jm (\de^n \jn\de^m \jp)
   - \de^k \jm (\de^n \jp\de^m \jn)
   - \de^k \jn (\de^n \jm\de^m \jp) \cr
& + \de^k \jn (\de^n \jp\de^m \jm)
  + \de^k \jp (\de^n \jm\de^m \jn)
  - \de^k \jp (\de^n \jn\de^m \jm). \cr
}$$
Instead of the second order invariants $S_{m,n}$ we use their
quasi-primary projections:
$$W^{(n+2)} :=
\n(\jp,\de^n \jm) + {\textstyle{1 \over 2}}\n(\jn,\de^n \jn)
+ \n(\jm,\de^n \jp)
 \eqno({\rm 2.1.42})$$
for all {\it even} $n$. For the third order invariants it is
more complicated to obtain quasi-primary projections; we will
come back to this problem below.
Note that it can also be verified case by case that $W^{(n)}$
and $S_{m,n,k}$ indeed commute with $\jpm_0$ (they commute with $\jn_0$
by construction) using the fact that
$\lb \jpm_0, X\rb = 0$ is equivalent to $\jpm_0 X_{d(X)} \vac = 0$
which is in the spirit of (2.1.11).
\sn
In order to obtain a generating set for the quasi-primary projections of the
third order invariants we first note that the third order invariants are
classically completely antisymmetric and therefore the quantization
$S_{m,n,k}$ can be expressed in terms of lower order invariants if any of the
arguments coincide. Next, the derivative $\de$ maps third order invariants to
third order invariants. Keeping all this in mind we may choose as independent
generators
$$S_{0,1,2}, \ 
 S_{0,1,4}, \ 
 S_{0,1,5}, \ 
 S_{0,1,6}, \ 
 S_{0,1,7}, \ 
 S_{0,1,8} 
 \eqno({\rm 2.1.43})$$
up to dimension 12. The quasi-primary projection of these fields can now be
calculated by orthogonalization with respect to all derivatives -- using again
the vacuum representation. One obtains the following
quasi-primary third order invariants:
$$\eqalign{
\hat{S}_{0,1,2} := & S_{0,1,2} - {\textstyle{1 \over 35}} \de^4 \hat{L} -
    {\textstyle{1 \over 18}} \de^2 W^{(4)} , \cr
\hat{S}_{0,1,4} := & S_{0,1,4} -  {\textstyle{1 \over 126}} \de^6 \hat{L}
    + {\textstyle{19 \over 39}} \de^4 W^{(4)} -
      {\textstyle{10 \over 13}}  \de^2 \hat{S}_{0,1,2} , \cr
\hat{S}_{0,1,5} := & S_{0,1,5} - {\textstyle{1 \over 210}} \de^7 \hat{L} -
      {\textstyle{1 \over 198}} \de^5  W^{(4)}
    + {\textstyle{45 \over 91}} \de^3 W^{(6)} -
      {\textstyle{50 \over 91}}  \de^3 \hat{S}_{0,1,2}
    + {\textstyle{1 \over 4}} \de W^{(8)}
    - {\textstyle{15 \over 8}} \de \hat{S}_{0,1,4} , \cr
\hat{S}_{0,1,6} := & S_{0,1,6} - {\textstyle{1 \over 330}} \de^8 \hat{L} -
      {\textstyle{7 \over 858}} \de^6  W^{(4)}
    + {\textstyle{17 \over 39}} \de^4 W^{(6)} -
      {\textstyle{5 \over 13}}  \de^4 \hat{S}_{0,1,2}
    + {\textstyle{41 \over 68}} \de^2 W^{(8)} \cr
   &- {\textstyle{315 \over 136}} \de^2 \hat{S}_{0,1,4}
    - {\textstyle{7 \over 3}} \de \hat{S}_{0,1,5} , \cr
\hat{S}_{0,1,7} := & S_{0,1,7} - {\textstyle{1 \over 495}} \de^9 \hat{L} -
      {\textstyle{4 \over 429}} \de^7  W^{(4)}
    + {\textstyle{14 \over 39}} \de^5 W^{(6)} -
      {\textstyle{7 \over 26}}  \de^5 \hat{S}_{0,1,2}
    + {\textstyle{140 \over 153}} \de^3 W^{(8)} \cr
   &- {\textstyle{245 \over 102}} \de^3 \hat{S}_{0,1,4}
    - {\textstyle{196 \over 57}} \de^2 \hat{S}_{0,1,5} +
      {\textstyle{1 \over 5}} \de W^{(10)}
    - {\textstyle{14 \over 5}}\de \hat{S}_{0,1,6} . \cr
} \eqno({\rm 2.1.44})$$
Using (2.1.44) we have computed determinants of all quasi-primary fields
up to scale dimension 9. From the zeroes of the determinants one can
read off those values of $\kp$ where truncations of the coset algebra
occur. All determinants have a singularity at $\kp = -2$ which
reflects the fact that the $c(\kappa)=3\kappa(\kappa+2)^{-1} \to \infty$
limit of this coset algebra is not well defined (see also section 3.1.).
Furthermore, all fields are null fields at $\kp = c =0$.
The remaining exceptional values of $\kp$
are listed in table 1. For these values of $\kp$ we have further determined
the kernels of the $d$-matrices giving us precise information which generators
drop out. The resulting generating sets are also presented in table 1.
\mn
\centerline{\vbox{
\hbox{\vrule \hskip 1pt
\vbox{ \offinterlineskip
\def\tablespace{height2pt&\omit&&\omit&&\omit&\cr}
\def\tablerule{ \tablespace\noalign{\hrule}\tablespace }
\hrule\halign{&\vrule#&\strut\hskip 4pt\hfil#\hfil\hskip 4pt\cr
\tablespace\tablespace
& $\kp$ && $c$   && {\it algebra} &\cr
\tablerule
& {\it generic} && {\it generic} && $\w(2,4,6,6,8,8,9,10,10,12)$&\cr\tablerule
& $1$ && $1$ && $\w(2)$ &\cr\tablerule
& $2$ && ${3 \over 2}$ && $\w(2,4,6)$ &\cr\tablerule
& $3$ && ${9 \over 5}$ && $\w(2,4,6,6,8,9)$ &\cr\tablerule
& $4$ && $2$ && $\w(2,4,6,6,8,8,9,10)$ &\cr\tablerule
& $5$ && ${15 \over 7}$ && $\w(2,4,6,6,8,8,9,10,10)$ &\cr\tablerule
& $-{1 \over 2}$ && $-1$ && $\w(2,4,6)$ &\cr\tablerule
& $-{4 \over 3}$ && $-6$ && $\w(2,6,8,10,12)$ &\cr\tablerule
& $-{8 \over 5}$ && $-12$ && $\w(2,4,6,8,9,10,12)$ &\cr\tablerule
& $-{12 \over 7}$ && $-18$ && $\w(2,4,6,6,8,9,10)$ &\cr
\tablespace
}
\hrule}\hskip 1pt \vrule}
\hbox{Table 1: Truncations of the $\sltr_\kp / sl(2,\BR)$-coset algebra}}
}
\sn
The information beyond scale dimension 9 for positive integer
$\kp$ in table 1 is taken from the character arguments presented in
$\q{\ralph}$ (see also $\q{\bouschou}$). For the remaining
cases information about dimensions higher than 9 is obtained from
character arguments which will be discussed at the end of this section.
The algebras of type $\w(2,4,6)$ appearing in table 1 are not identical;
the case $\kp = -{1 \over 2}$ was discussed in detail in $\q{\ajl}$ and
corresponds to the solution which was unexplained for some time
$\q{\kauwatts,\howcl}$, the case $\kp=2$ is the bosonic projection of
the $N=1$ Super Virasoro algebra.
\sn
It is now straightforward to obtain a {\it primary} set of generators
by orthogonalization of (2.1.42) and (2.1.44) with respect to normal ordered
products ${\cal N}$. Up to scale dimension~6 one can choose for example the
following primary generators in addition to $L$:
$$\eqalign{
\Phi^{(4)} =& 24 (\kappa + 2)^2 \n(L,L) - (37 \kappa + 44) W^{(4)} , \cr
\Phi^{({\rm 6a})} =& 1920 (\kp + 2)^3 \n(\n(L,L), L)
- 40 (209 \kp + 328) (\kp + 2) \n(W^{(4)}, L) \cr
& + 18 (31 \kp + 24) (9 \kp + 16) W^{(6)}
+ 45 (9 \kp + 16) (\kp + 8) \hat{S}_{0,1,2} , \cr
\Phi^{({\rm 6b})} =& 3840 (675 \kp^3\! +\! 75 \kp^2 \!-\! 1403 \kp \!+\! 4328)
 (\kp \!+\! 2)^3\n(\n(L,L), L)+\! {\cal M} (\kp \!+\! 2)^2 \n(L, \de^2 L) \cr
&- 360 (89 \kp + 136) (15 \kp - 8) (5 \kp + 13) (5 \kp -
2) (\kp + 2) \n(W^{(4)}, L) \cr
&+ 5 (3645 \kp^3 + 19947 \kp^2 - 200 \kp - 40192) (89 \kp + 136) (5
\kp - 2) W^{(6)} \cr
&- 30 (645 \kp^2 - 928 \kp - 3392) (89 \kp + 136) (5 \kp - 2)
 \hat{S}_{0,1,2} , \cr
} \eqno({\rm 2.1.45})$$
where we have used the abbreviation
$${\cal M} := 36 (18225 \kp^4 + 485475 \kp^3 + 637424 \kp^2 - 1738048 \kp -
2584576).$$
The two point functions (or central terms) read
$$\eqalign{
d_{4, 4} =& 120 (37 \kappa + 44) (3 \kappa + 4) (\kappa - 1) \kappa ,
\quad\quad\quad d_{\rm 6a, 6a} = 8 \, {\cal M} \,
 (9 \kp + 16) (\kp - 1) \kp , \cr
d_{\rm 6b, 6b} =& 252000 \, {\cal M} \,
 (89 \kp + 136) (5 \kp + 8) (5 \kp - 2) (3 \kp + 4)
(2 \kp + 1) (\kp - 1) (\kp - 2) \kp . \cr
} \eqno({\rm 2.1.46})$$
Note that we have chosen $d_{\rm 6a, 6b} = d_{\rm 6b, 6a} = 0$.
The normalization constants (2.1.46) indeed vanish for the
corresponding values of $\kp$ of table 1.
The additional zeroes belong to Virasoro-minimal values of $c$
(where a correction term becomes a null field and thus the primary
projection fails).
\sn
Finally, we present the first nontrivial structure constant of this
coset algebra in standard normalization ($\hat{d}_{4, 4} = {c \over 4}$):
$$\left(C_{4 \, 4}^4\right)^2 =
{2 (99 \kp^2 - 25 \kp - 236)^2 \over
5 (37 \kp + 44) (3 \kp + 4) (\kp + 2) (\kp - 1)} =
{(35 c^2 +211 c -354)^2 \over
15 (5 c + 22) (c + 6) (c - 1)}.
 \eqno({\rm 2.1.47})$$
Further structure constants would strongly depend on the choice
of basis and therefore we omit them.
\sn
\subs{\slslgen}
\sn
Next, we consider two commuting copies of the Kac-Moody algebra
$J^{(i,\pm,0)}$ ($i=1,2$) based on $sl(2,\BR)$. The
$J^{(i,\pm,0)}$ each satisfy eq.\ (2.1.39) independently, the
mixed commutators being zero. We denote the first level
by $\kappa$ and the second level by $\mu$.
All the currents are primary spin 1 fields with respect
to some energy momentum tensor.
We are however not going to use any information about the
nature of this energy momentum tensor or its central charge.
\sn
We want to construct the invariant subspace under the
action of the diagonally embedded $\sltr$. This
means that we have to construct all fields commuting with
$$\jpm := \japm + \jbpm , \quad\qquad
\jn := \jan + \jbn .
  \eqno({\rm 2.1.48})$$
This task is simplified by the observation that this $\sltr$ is generated
by the horizontal $sl(2,\BR)$ subalgebra of (2.1.48) and all modes of the
current $\jn$. So the commutant consists of those $sl(2,\BR)$-invariant
polynomials which also commute with $\jn$. The $sl(2,\BR)$-invariant
fields are generated by (2.1.41) where we have now the freedom to
insert the currents $J^{(1)}$ and $J^{(2)}$ for any of the arguments.
The second order invariants coming from (2.1.41a) are
$$\SS{r,s}{m,n} := \de^n J^{(s,-)}\de^m J^{(r,+)}
+ {\textstyle{1 \over 2}} \de^n J^{(s,0)}\de^m J^{(r,0)}
+ \de^n J^{(s,+)}\de^m J^{(r,-)}.
 \eqno({\rm 2.1.49})$$
The third order invariants $\SS{r,s,t}{m,n,k}$ are obtained from (2.1.41b)
in the same manner.
\sn
For explicit calculations we use again the Vertex-operator approach of
section 2.1.1. First, condition (2.1.11) is used to find the invariant
fields. Then, the primary generators can be computed from (2.1.12).
\mn
The first invariant we find with the ansatz (2.1.49) and the condition
(2.1.11) is the coset energy momentum tensor
$$L =  {1 \over \kappa + \mu + 2} \left(
{\mu \over 2 (\kappa + 2)} \SS{1,1}{0,0}
- \SS{1,2}{0,0} + {\kappa \over 2 (\mu + 2)} \SS{2,2}{0,0} \right).
\eqno({\rm 2.1.50})$$
The central charge $c$ of the coset theory is found immediately to be
$$c =
{3 \mu \over \mu + 2}
\left(1 - 2 {\mu + 2 \over (\kappa + \mu + 2) (\kappa + 2)}\right).
\eqno({\rm 2.1.51})$$
Note that for the diagonal coset the original currents are {\it not}
primary with respect to the coset energy momentum tensor (2.1.50).
This is the reason why it is not advantageous to use quasi-primary
normal ordered products $\n$.
\sn
The next primary generator can be constructed at scale dimension 4 using
eqs.\ (2.1.49), (2.1.11), (2.1.12). The explicit form can be found
in appendix C. From this realization one obtains the central term
(or normalization constant) which is omitted here. A tedious calculation
yields the self-coupling constant of the dimension 4 generator. In standard
normalization it reads
$$\eqalign{
&\left(C_{4\,4}^4\right)^2 \!=
   18 \Bigl(99 (\kp^2 \mu^4 \!+\! \kp^4 \mu^2)
   \!-\! 25 (\kp \mu^4 \!+\! \kp^4 \mu)
   \!-\! 236 (\mu^4 \!+\! \kp^4) \!+\! 198 \kp^3 \mu^3
   \!+\! 742 (\kp^2 \mu^3 \!+\! \kp^3 \mu^2)\cr
 & \!-\! 672 (\kp \mu^3 \!+\! \kp^3 \mu)
   \!-\! 1888 (\mu^3 \!+\! \kp^3) \!+\! 576 \kp^2 \mu^2
   \!-\! 4088 (\kp \mu^2 \!+\! \kp^2 \mu) \!-\! 5168 (\mu^2 \!+\! \kp^2)
   \!-\! 8592 \kp \mu \cr
 & \!-\! 5568 (\kp \!+\! \mu) \!-\! 1792 \Bigr)^2
   \Bigl(5 (\kp - 1) (\mu - 1) (2 + \kp) (2 + \mu) (4 + 3 \kp)
   (4 + 3 \mu) (2 + \kp + \mu) \cr
& (5 \!+\! \kp \!+\! \mu)(8 \!+\! 3 (\kp \!+ \!\mu))
   (176 (1\! +\! \kp \!+\! \mu) + 44(\kp^2\!+\!\mu^2)
    + 192 \kp \mu + 37 (\kp^2 \mu + \kp \mu^2))\Bigr)^{-1}. \cr
} \eqno({\rm 2.1.52})$$
A few remarks about the structure constant eq.\ (2.1.52) are in place.
It is symmetric in the levels $\kp$ and $\mu$ reflecting the
symmetry of the construction. Furthermore, eq.\ (2.1.47) is recovered from
eq.\ (2.1.52) in the limit $\mu \to \infty$, as it should be. Also the
singularities for one of the levels $\kp$ or $\mu$ equal to
$1$ or $-{4 \over 3}$ are expected because in these cases the field
$\Phi^{(4)}$ should be a null field. For $\mu = 2$ one recovers the structure
constants of the bosonic projection of the $N=1$ Super Virasoro algebra
presented in $\q{\kauwatts,\commute}$ (see also appendix \appD). For
$\mu = -{1 \over 2}$ the structure constant of the $\w(2,4,6)$ in
$\q{\kauwatts,\howcl}$ is reproduced. This last identity confirms
the identification in $\q{\ajl}$ of the last unexplained solution
$\w(2,4,6)$ with the above $sl(2,\BR)$ coset at $\mu=-{1 \over 2}$.
\sn
One should mention that one of the levels can be replaced by the central
charge $c$. The other level still occurs as a parameter in the structure
constants. Thus, the resulting $\w$-algebra can be regarded as a 1-parameter
family of algebras of type $\w(2,4,6,6,8,8,9,10,10,12)$ for generic $c$.
\sn
\subs{Representation theory of \slsl}
\sn
We conclude this section with a discussion of the representation theory of the
coset \slsl. For coset algebras of affine Kac-Moody algebras one has a natural
approach to representation theory, i.e.\ to the set of irreducible highest
weight modules (see e.g.\ $\q{\bouschou,\bouwknegt}$). One obtains a highest
weight module \LLc\ of the coset algebra \gg\ by the decomposition of a
highest weight module \LL\ of $\hat g$ under the ${\hat g}'$ action
$\hbox{\LL}=\bigoplus_{\La'} \hbox{\LLc}\otimes\hbox{\LLp}$
where $\La'$ runs over the weights of ${\hat g}'$ ($k'=jk$, where $j$
is the Dynkin index of the embedding $g' \hookrightarrow g$). The
corresponding formula for the characters is given by
$$\chi_{L_{\La}}(q) = \sum_{\La'} b_{\La'}^{\La}(q) \chi_{L_{\La'}}(q)
\eqno({\rm 2.1.53a})$$
with the so-called branching functions $b_{\La'}^{\La}(q)$. Similarly,
for the cosets $\hat{g} / g'$ one obtains a decomposition
$$\chi_{L_{\La}}(q) = \sum_{j} b_{j}^{\La}(q) \chi_{j}
\eqno({\rm 2.1.53b})$$
where the $\chi_j$ are now characters of irreducible representations of the
Lie algebra $g'$.
\sn
Let us focus on the diagonal cosets
$\hat{g}_{k_1} \oplus \hat{g}_{k_2} / \hat{g}_{k_1 + k_2}$ where the modules
$L_{\La,\La'}$ are believed to be irreducible. In this case, the branching
functions are equal to the characters of the coset up to some prefactor.
The representations of
$\hat{g}_{k_1} \oplus \hat{g}_{k_2}$ are now labeled by two weights
$\La_1$ and $\La_2$ instead of a single weight $\La$. One has the
following formula for the branching functions for integrable
weights $\La_i,\ i=1,2$ at level $k_i$ (see e.g.\ $\q{\bmcp}$):
$$b_{\La_3}^{\La_1,\La_2}=\sum_{w\in\widehat W}\eps(w)
c_{\La_3-w\star\La_2}^{\La_1}(q)\,\,
q^{{\vert w(\La_2+\rho)(k_1+k_2+h^\vee)-(\La_3+\rho)(k_2+h^\vee)\vert^2}
\over{2k_1(k_2+h^\vee)(k_1+k_2+h^\vee)}}
\eqno({\rm 2.1.54})$$
where $\La_3$ is an integrable weight at level $k_1+k_2$ such that
$\La_1+\La_2-\La_3$ belongs to the long root lattice of $\hat g$,
$\widehat{W}$ denotes the Weyl group of $\hat g$ and $w\star\cdot$
denotes the shifted action of the Weyl group element $w$.
The $c^{\La'}_{\La}(q)$ are the Kac-Peterson string functions
$\q{\kacpeter}$ which are defined via the identity
$b^{\La'}_{\La}(q) = \eta(q)^l c^{\La'}_{\La}(q)$ where $b^{\La'}_{\La}(q)$
are the branching functions of the coset $\hat{g}/\hat{h}$ with
$h$ the Cartan subalgebra of $g$, $l$ the rank of $h$ and $\eta(q)$
is Dedekind's eta-function.
\sn
The limit $k_2 \to \infty$ of the coset
$\hat{g}_{k} \oplus \hat{g}_{k_2} / \hat{g}_{k + k_2}$ is the
coset $\hat{g}_k / g$. The vacuum character for $\hat{g}_k / g$ is
obtained from the limit $k_2 \to \infty$ of (2.1.54) with $\La_1=k\La_0,\
\La_2=k_2\La_0,\ \La_3=(k+k_2)\La_0$ ($\La_0$ is the first fundamental
weight of $\hat g$):
$$b_0^{k\La_0}(q)=\sum_{w\in W} \eps(w) c_{w\star\rho+k\La_0}^{k\La_0}(q)
q^{{1\over{2k}}\vert w\star\rho\vert^2}
\eqno({\rm 2.1.55})$$
where $W$ is the Weyl group of $g$.
With the explicit form of the Kac-Peterson string functions
$\q{\kacpeter}$ it is relatively easy to calculate
the vacuum characters for $g=sl(2,\BR)$ and the integer levels specified
in table 1 above yielding the spins of the generating fields of
the coset.
\sn
However, we are interested in the generalization of eq.\
(2.1.54) to the case where one of the levels is fractional
so that the so-called {\it admissible} representations
of the Kac-Moody algebra carrying a representation of
the modular group enter the game $\q{\kacwaki}$.
In particular, we would like to calculate the branching
functions of the coset \slslk.
If the level $k=k_1$ is an integer we are allowed to calculate
the branching functions of the coset involving admissible representations:
$$\chi_{L_\La}(q) =
\chi_{\La_1}^{k}(q) \chi_{\La_2}^{-{1\over 2}}(q) = \sum_{\La_3}
   b^{\La_1,\La_2}_{\La_3}(q) \chi_{\La_3}^{k-{1\over 2}}(q),
\eqno({\rm 2.1.56})$$
where the sum runs over all admissible representations (we denoted
the levels of $\La_i$ explicitly). Eq.\ (2.1.54) may still applied to
one fractional and one integer level if one modifies the range of summation
such that the powers of $q$ remain integer spaced. Using the parametrization
of $\q{\kacwaki}$ for the admissible representations
$k_2=-{1\over 2}$, $p=2k+4$, $q=p-1$, $l = \sqrt{2} \La_1$ and
$j=\sqrt2 \La_2+1$, $j'=\sqrt 2\Lambda_3+1$ one obtains from eq.\ (2.1.54):
$$b^{l,j}_{j'}(q) = q^{-{c\over{24}}}\!\!\sum_{m\in\BZ}\!\left(
    c^{l}_{j'-j-mp}(q) q^{-{ (mpq+jq-j'p)^2-(p-q)^2 \over 2pq}} -
    c^{l}_{j'+j-mp}(q) q^{-{ (mpq-jq-j'p)^2-(p-q)^2 \over 2pq}}\right)
\eqno({\rm 2.1.57})$$
where $0<j<{p\over 2}$, $0<j'<q$, $j-j' \in 2 \BZ$,
$l\in\lbrace0,1\rbrace$ and $c=-1+{12(p-q)^2\over pq}$.
By definition, the string functions satisfy
$$c^{l}_{m}(q) ={B^{l}_m(q)\over \eta(q)}
\eqno({\rm 2.1.58})$$
where the $B^{l}_m(q)$ are the branching functions of the coset
$\sltr_{-{1 \over 2}} / \widehat{U(1)}$. One can read off these branching
functions from (2.1.7a) observing that the modules of $\sltr_{-{1 \over 2}}$
are freely generated in terms of a $\beta-\gamma$ system (see $\q{\ajl}$).
Substituting $z^2 \to z^2 q^{1 \over 2}$ in (2.1.7a) leads to the character
of $\sltr_{-{1 \over 2}} / \widehat{U(1)}$. Multiplication with the correct
prefactor gives the result
$$B^{l}_m = q^{1\over{12}}{q^{m(m+1)\over 2} \phi_m(q) \over
  \prod_{n\ge 1}(1- q^{n}) } \quad\quad \hbox{for } l \equiv m \hbox{ mod } 2
  \eqno({\rm 2.1.59})$$
where $\phi_m(q)$ is defined in eq.\ (2.1.7b).
Note that the following symmetry holds
$$b^{0,j}_{j'} = b^{1,j}_{q-j'}.
\eqno({\rm 2.1.60})$$
Thus we get the following Kac-table for the conformal dimensions of the
branching functions:
$$\eqalign{
 h(j,j') &= {\vert j-j' \vert(\vert j-j' \vert +1)\over 2}-
     {(jq-j'p)^2-(p-q)^2 \over 2pq}
     \quad\quad\quad \quad {\rm for\ } j'<j<{p\over 2} \cr
 h(j,j') &= h(j,q-j') \quad\quad {\rm for\ } j\le j'<q = p -1,
\ j<{p\over 2}
\qquad \qquad \vert j-j' \vert \equiv 0 \hbox{ (mod 2)}. \cr
}\eqno({\rm 2.1.61})$$
The second line of (2.1.61) does not reflect a symmetry but means
that the conformal dimension for $j' \ge j$ is obtained from the
first line evaluated at $h(j,q-j')$.
\mn
For the rational models of $\w(2,4,6)$ calculated in $\q{\howcl}$
we get perfect agreement with the set of highest weights given
by eq.\ (2.1.61). Taking the limit $p\to\infty$ one obtains
$b^{0,1}_1(q)=c^0_0(q)-qc^0_2(q)$ from which one can read off
the generators of the underlying $\w-$algebra yielding a $\w(2,4,6)$.
For general fractional level the vacuum character can be computed
using the results of $\q{\ahn}$ for string functions
\footnote{${}^{6})$}{Note that the example in table 3 of $\q{\ahn}$
arising from a coset with two fractional levels $k_1$, $k_2$ cannot be
a RCFT in contrast to the claim of $\q{\ahn}$. Firstly, it is known that
there is no RCFT with positive conformal dimensions and $c = {1 \over 5}$
(see e.g.\ $\q{\kiritsis}$). Secondly, using the ideas of $\q{\wolfgang}$
one can check that there is no representation of the modular group inducing
a `good' fusion algebra with conformal dimensions as in $\q{\ahn}$.}.
\mn
Eq.\ (2.1.61) agrees with the conjecture of $\q{\klausREP}$
for the minimal series of this $\w(2,4,6)$ which was obtained by
{\it formal} extrapolation to ${\cal WD}_{-1}$. We will continue this
formal extrapolation to the algebras ${\cal WD}_{-m}$ in section 3.2.4.
\sn
Using formula (\appF.2) of appendix {\appF} it is easy to show that the
conformal dimensions for the ${\cal WC}_{m}$ minimal models with
central charge $c_{{\cal C}_m}(m+2,2m+3)$ agree exactly with (2.1.61).
This supports that ${\cal WD}_{-1}\cong \w(2,4,6)$ is the unifying algebra
for the minimal models of the Casimir algebras ${\cal WC}_m$
at $c_{{\cal C}_m}(m+2,2m+3)$ as indicated in $\q{\letter}$.
\mn
%
%
\section{2.2.\ Orbifolds of quantum $\w$-algebras}
\mn
In the previous section we discussed cosets. Coset constructions can be
viewed as projections onto the subspace invariant under an {\it inner}
symmetry (realized by a subalgebra) of a $\w$-algebra. Orbifolds are
projections onto subspaces invariant under {\it outer} automorphisms
(leaving the algebraic structure invariant) and behave therefore similarly to
coset constructions. It was shown in $\q{\ajl}$ that these two constructions
lead in general to nonfreely generated $\w$-algebras. We will
further comment on the similar behaviour of these two constructions
in section 3.1.
\sn
There are two further questions, both turning up in the
general context of classification of $\w$-symmetries, which motivate the
study of orbifold constructions of $\w$-algebras. The first motivation
is that recently projections of the generators onto invariant
subspaces were calculated for the classical Casimir algebras ${\cal WA}_{n-1}$
in $\q{\frfolding}$ and were shown to give rise to other Casimir algebras.
In all known cases of quantum Casimir $\w$-algebras one can read off from
the structure constants that such an identity is not true on the quantum
level. In this section we will show that indeed such an identity is never
true for the quantum Casimir algebras ${\cal WA}_{n-1}$.
\sn
The second motivation is the observation that instead of considering
symmetry algebras containing fermions -- in particular Super $\w$-algebras --
it might be simpler to use their bosonic projections $\q{\rwofus}$.
In particular, a classification program might turn out to be simpler
for the bosonic case. From this point of view it is certainly important
to study orbifolds of $\w$-algebras.
\mn
Orbifolds also turn up in various applications of CFT. For example
they occur as the chiral algebras of the GSO projected models
used in superstring theory $\q{\gso}$. Orbifolds are also useful
for applications to statistical mechanics.
Spin models on the cylinder, torus or other higher genus surfaces
are difficult to realize in experiments. From this point of view
the most natural boundary conditions for spin systems are {\it free}
boundary conditions. At conformally invariant second order phase
transitions of such two dimensional statistical systems the spectrum
generating algebra should be the {\it orbifold} of the underlying
$\w$-algebra. This expectation comes from the observation
in $\q{\twists}$ that generally boundary conditions of
$\BZ_n$ quantum spin systems 
are in one to one correspondence with boundary conditions of
their spectrum generating algebras ${\cal WA}_{n-1}$.
\mn
In this section we will take an algebraic approach to orbifolds
which is different from the usual orbifolds that deal with the
complete CFT (see e.g.\ $\q{\dijkgraaf}$). If a $\w$-algebra has a
nontrivial outer automorphism group one can consider the projection
onto the invariant subspace. We will denote this projection by
`orbifold'. Available results on outer automorphisms of $\w$-algebras
have been collected in $\q{\twists}$. In all known cases the group of
outer automorphisms is discrete, in many cases the automorphism
group is just $\BZ_2$. For simplicity, we will restrict to
the case of $\BZ_2$ automorphisms. Note that this covers in particular the
bosonic projections of fermionic $\w$-algebras. Here the automorphism
maps any fermion $\psi$ to $-\psi$ and the invariant subspace is
precisely given by the space of the bosonic fields.
\mn
\section{2.2.1.\ General remarks and results }
\sn
By definition, the energy momentum tensor $L$ of any $\w$-algebra is
invariant under all outer automorphisms. Therefore, orbifold constructions
never change the energy momentum tensor $L$ (in contrast to
coset constructions). This observation will be exploited
using a quasi-primary basis from the very beginning which
simplifies the transition to primary generators. For $\BZ_2$
automorphisms a minimal generating set of the {\it classical}
orbifolds was presented in $\q{\ajl}$. The normal ordered
versions of these generators are also a (redundant) generating
set for the quantum orbifolds. From this we obtain in the
case of $\BZ_2$ orbifolds that the nonzero quasi-primary
normal ordered products
$$\n(\phi^{(1)}, \de^n \phi^{(2)}) \eqno({\rm 2.2.1})$$
of any two generators $\phi^{(j)}$ transforming under
the automorphism as $\phi^{(j)} \mapsto -\phi^{(j)}$
constitute a generating set for the orbifold together with
the invariant generators. One can further restrict to
those normal ordered products eq.\ (2.2.1) where the fields
$\phi^{(j)}$ appear in a certain order.
\sn
The field content of an orbifold can be easily predicted
using a character argument which also enables one (at least
in principle) to determine the representations of the
orbifold. Let $V$ be an irreducible highest weight module of
the underlying $\w$-algebra. Then we define a character $\chi(q,z)$
which also encodes the $\BZ_2$ automorphism by
$$\chi(q,z) := {\rm tr}_V^{} \left(q^{(L_0 - {c \over 24})} z^P\right)
 \eqno({\rm 2.2.2})$$
where $P$ denotes the parity of a state in $V$ with respect to the
automorphism. The parity of invariant states is zero ($P=0$).
The states transforming with a sign under the automorphism
group are defined to have parity $P=1$. The subspace of $V$
invariant under the automorphism as well as the subspace
of states transforming with a sign both provide irreducible
representations of the orbifold $\w$-algebra. From this we
conclude that the decomposition
$$\chi(q,z) = \chi^{(0)}(q) + z \, \chi^{(1)}(q)
\eqno({\rm 2.2.3})$$
yields the two characters $\chi^{(i)}(q)$ for the
two representations of the orbifold obtained from $V$.
In particular, the constant part $\chi_0^{(0)}(q)$ in $z$ of the
vacuum character $\chi_0(q,z)$ of the underlying $\w$-algebra
is the vacuum character of the orbifold. It is straightforward
to compute this character for a $\w$-algebra without null fields
(which applies to Drinfeld-Sokolov $\w$-algebras at generic $c$)
up to a finite order. The field content of the orbifold can now
be read off by determining the minimal set of fields whose
free vacuum module is at least as large.
\sn
If $\chi_i(q,z)$ are the characters of a rational model of the
underlying $\w$-algebra, the characters obtained from (2.2.3)
are the characters of the associated rational model of the
orbifold. Note, however, that for this statement to be valid
one has to take the representations of all sectors of the original
algebra into account, in particular for bosonic algebras also the
twisted sector $\q{\twists}$. Then the statement about rational
models follows from the identity (3.6) in $\q{\twists}$
for the partition functions.
\sn
Note that instead of determining the field content of the
orbifold using this character argument it can also be
computed by applying a basis algorithm to those simple fields
that will drop out. The invariant normal ordered
products must be contained in the orbifold. Either
they can be considered as composite or must be
added to the generating set. This approach is slightly
more involved but has the small advantage that one can trace
invariant fields and up to the cancellation of correction terms in relations
and generators (see $\q{\ajl}$) ensure closure of the orbifold $\w$-algebra.
\sn
The {\it primary} generators in the orbifold can be efficiently
determined using the definition of {\it simple} fields (see section 1.1.).
Instead of taking the quasi-primary normal ordered
products (2.2.1) we orthogonalize them with respect to all
quasi-primary normal ordered products in the orbifold because
simple fields are orthogonal to derivatives and quasi-primary
normal ordered products.
\sn
Finally, the structure constants are determined by a standard procedure:
Two- and three-point functions of the simple fields are evaluated
in the vacuum representation and the coupling constants arise as solutions
of a linear system of equations (see section 1.1.).
\sn
Below we will use the following notations: $W^{(\delta)}$ denotes a simple
field of dimension $\delta$ in the original algebra, $\Phi^{(n)}$ denotes a
simple field of dimension $n$ in the orbifold. $d_{n,n}$ is the
central term in the commutator of $\Phi^{(n)}$ with itself which
equals its norm squared. The structure constants in the
original algebra are denoted by $C_{W^{(\kappa)}\, W^{(\lambda)}}^{W^{(\mu)}}$
whereas for the orbifold we just write the dimensions, i.e.\
$C_{k \,\, l}^m$. Note that the structure constants for the orbifold
are given in the standard normalization $\hat{d}_{n,n} = {c \over n}$
and {\it not} in the induced one given by $d_{n,n}$.
\sn
%
%
\subs{Throwing out one field}
\sn
Before presenting a collection of results for $\BZ_2$ orbifolds
we will first consider the simplest case where we project out a single
field $W^{(\delta)}$ using an automorphism $W^{(\delta)} \mapsto
 - W^{(\delta)}$. In order to determine the field content of the
orbifold let us look at all invariant fields that can be built out of
$W^{(\delta)}$ alone. This argumentation will rely on the absence of
null fields, i.e.\ it will apply to generic $c$ only.
\sn
First, we discuss the bosonic case where $\delta$ is an integer
$\delta = n$. According to eq.\ (2.2.1) all fields
$\n(W^{(n)}, \de^{2 k} W^{(n)})$ ($k \ge 0$) are invariant under the
automorphism. Their primary projections $\Phi^{(2(n+k))}$ cannot be
composite for $k < n$. At dimension $4 n$ we have invariant fields of the form
$\n(W^{(n)}, \de^{2 n} W^{(n)})$ and
$\n(\n(\n(W^{(n)},W^{(n)}),W^{(n)}),W^{(n)})$. These two fields are equivalent
to $\Phi^{(4 n)}$ and $\n(\Phi^{(2 n)}, \Phi^{(2 n)})$. Clearly, there
is precisely one additional generator $\Phi^{(4 n)}$ at scale
dimension $4 n$. At dimension $4 n + 2$ the invariant fields are
$\n(W^{(n)}, \de^{2 n + 2} W^{(n)})$ and
$\n(\n(\n(W^{(n)},W^{(n)}),W^{(n)}),\de^2 W^{(n)})$. Assuming the
cancellation procedure described in $\q{\ajl}$ to be valid in general,
this space is also spanned by $\n(\Phi^{(2 n)}, \de^2 \Phi^{(2 n)})$ and
$\n(\Phi^{(2 n+2)}, \Phi^{(2 n)})$.
Finally, at dimension $4 n + 4$ there are two fourth order and one
second order invariant fields (three altogether). At this dimension we have
in the orbifold $\n(\Phi^{(2 n)}, \de^4 \Phi^{(2 n)})$,
$\n(\Phi^{(2 n+2)},\de^2 \Phi^{(2 n)})$, $\n(\Phi^{(2 n+2)},\Phi^{(2 n + 2)})$
and $\n(\Phi^{(2 n+4)}, \Phi^{(2 n)})$. Thus, there must be a generic null
field at dimension $4n +4$. In summary, we have generators $\Phi^{(2 k)}$,
$n \le k \le 2 n$ in the orbifold and a first generic null field
at dimension $4 n + 4$.
\sn
The same procedure can be applied to a fermionic field, i.e.\ $\delta$
half-integer. Again, we obtain generators of the orbifold
$\Phi^{(2 \delta + 2 k +1)}$ from the primary projection of
$\n(W^{(\delta)}, \de^{2 k + 1} W^{(\delta)})$ for
$0 \le k \le \delta - {1 \over 2}$ and the first generic
null field at dimension $4 \delta + 4$.
\sn
Note that the Casimir algebras ${\cal WD}_n$ and ${\cal WB}(0,n)$ have
this property. ${\cal WD}_n$ ($n>4$) contains
precisely one simple bosonic field which
can be projected out $\q{\twists}$. This field has dimension $n$.
${\cal WB}(0,n)$ has precisely one simple fermionic field.
The bosonic subalgebra can therefore be obtained
by projecting out the fermion with dimension $n+{1 \over 2}$.
It was noted in $\q{\bouschou}$ that this bosonic subspace can be realized
in terms of the diagonal coset
$(\hat{\cal B}_n)_k \oplus (\hat{\cal B}_n)_1 /(\hat{\cal B}_n)_{k+1}$.
Eq.\ (7.19) of $\q{\bouschou}$ is a closed formula for
the vacuum character of this coset.
\sn
\subs{Results}
\sn
{}From this result and the character argument explained above one can
predict the generating set for many orbifolds. Table 2 contains a
collection of results
\footnote{${}^7)$}{For the field content of the orbifolds of
${\cal WA}_2$ and ${\cal WA}_3$ see also $\q{\bouschou}$.}.
Strictly speaking, one would have to check in each case separately
that the cancellation mechanism of $\q{\ajl}$ indeed takes place.
To determine the orbifold of the cosets \slslgen\ and \sluk\ one has to use
the explicit knowledge of the invariants (and relations) generating the
coset because of the presence of generic null fields.
In the case of \sluk\ the outer automorphism of the coset comes from
the inner automorphism of the $\sltr$ Kac-Moody algebra (2.1.1) that
maps $\Jp \leftrightarrow \Jm$, $\Jz \mapsto -\Jz$. This automorphism
leaves the even dimensional generators of the coset invariant and changes
the sign of the odd dimensional ones.
\mn
The automorphism of the other coset \slslgen\ is induced by the map
$J^{(1,\cdot)} \leftrightarrow J^{(2,\cdot)}$. The effect is that for
$\kp = \mu$ the quadratic fields (2.1.41a) are left invariant
and the third order invariants (2.1.41b) change their sign. With this
information the field content of the orbifold can easily be inferred
from $\q{\ajl}$ by dropping the third order invariants from the
generating set and the relations.
\mn
\centerline{\vbox{
\hbox{\vrule \hskip 1pt
\vbox{ \offinterlineskip
\def\tablespace{height2pt&\omit&&\omit&&\omit&\cr}
\def\tablerule{ \tablespace\noalign{\hrule}\tablespace}
\hrule\halign{&\vrule#&\strut\hskip 4pt\hfil#\hfil\hskip 4pt\cr
\tablespace\tablespace
& {\it algebra} && {\it projection}   && {\it dimension of first} &\cr
\tablespace
& \omit && {\it for generic $c$}   && {\it generic null field} &\cr
\tablerule
& $\w(2,1)$&& $\w(2,2,4)$   && 8 & \cr \tablerule
&$\w(2,{3\over2})$&&$\w(2,4,6)$&& 10 & \cr \tablerule
& $\w(2,2)$&& $\w(2,4,6,8)$ && 12 & \cr \tablerule
& ${\cal WA}_2 \cong \w(2,3)$&& $\w(2,6,8,10,12)$ && 16  & \cr \tablerule
& ${\cal WA}_3 \cong \w(2,3,4)$&& $\w(2,4,6,8,10,12)$ && 16  & \cr \tablerule
& ${\cal WA}_4 \cong \w(2,3,4,5)$
&&$\w(2,\!4,\!6,\!8^2\!,\!9,\!10^3\!,\!11,\!12^3\!,\!13,\!14^2)$
          && 16 &\cr\tablerule
& ${\cal WA}_5 \cong \w(2,3,4,5,6)$
&&$\w(2,\!4,\!6^2\!,\!8^2\!,\!9,\!10^3\!,\!11,\!12^3\!,\!13,\!14^2)$
          && 16 &\cr\tablerule
& $\sltr_{\kp} / \widehat{U(1)} \cong \w(2,3,4,5)$
 && $\w(2,4,6,8,10)$ && $14$ & \cr \tablerule
& $\sltr_\kp \oplus \sltr_\kp / \sltr_{2 \kp}$
 && $\w(2,4,6,8,10,12,14,16,18)$ && $22$ & \cr \tablerule
& ${\cal WD}_n \cong \w(2,4,\ldots,2 n-2,n)$
      && $\w(2,4,\ldots,4n)$ &&  $4 n + 4$ & \cr \tablerule
& ${\cal WB}(0,\!n)\! \cong\! \w(2,\! 4,\ldots,\!2 n,\!n\! +\! {1 \over 2})$
      && $\w(2,4,\ldots,4n+2)$ && $4 n + 6$ & \cr \tablerule
& ${\cal SW}({3 \over 2},2)$ && $\w(2^2,4^2,5,6^3,7)$ && 9 & \cr \tablespace
}
\hrule}\hskip 1pt \vrule}
\hbox{\quad Table 2: Field content of some orbifolds of $\w$-algebras}}
}
\sn
In the simplest cases one can also determine some primary
generators of the orbifold and calculate the corresponding structure
constants. We briefly summarize results for
$\w(2,1)$ and $\w(2,2)$. The more interesting
cases of the $\w(2,3)$ and ${\cal WA}_{n-1}$ will be treated in later
subsections and the orbifold of the $N=1$ Super Virasoro algebra
$\w(2,{3 \over 2})$ can be found in appendix \appD.
Below, we present structure constants connecting additional simple
fields.
\sn
{\bf $\bf \ww(2,1)$:} The (Lie) algebra $\w(2,1)$ is the extension of
the Virasoro algebra $L$ by a primary $U(1)$ current $J$. The map
$J \mapsto -J$ is the unique nontrivial automorphism of the algebra,
providing us with one of the simplest examples of an orbifold
construction. The orbifold contains two additional primary fields of
dimensions 2 and 4. Both of them are null fields at $c=1$. At
$c=-{17 \over 5}$ the dimension 4 generator vanishes. The vanishing
of the additional dimension 2 generator at $c=1$ is
to be expected because the Sugawara energy momentum tensor
of the current is the unique energy momentum tensor for $c=1$, i.e.\
$L = {1 \over 2} \n(J,J) $ at $c=1$.
The structure constants connecting these two fields can be determined as:
$$\eqalign{
\left(C_{2 \, 2}^2\right)^2 &= 4 (c - 2)^2(c - 1)^{-1} \quad
\left(C_{2 \, 2}^4\right)^2 = 24 (5 c + 17) c^3
\bigl((25 c^2 + 180 c + 383) (c - 1)\bigr)^{-1} \cr
C_{2 \, 2}^2 C_{4 \, 4}^2 &= 4 (25 c^3 + 95 c^2 - 61 c - 383) (c - 2)
\bigl((25 c^2 + 180 c + 383) (c - 1)\bigr)^{-1} \cr
C_{4 \, 4}^4 C_{2 \, 2}^4 &= 36 (375 c^4 \!+\! 2400 c^3\!
 +\! 2090 c^2 \!-\! 9864 c \!-\! 11801) c^2
 \bigl((25 c^2 \!+\! 180 c \!+\! 383)^2 (c \!-\! 1)\bigr)^{-1}\!. \cr
} \eqno({\rm 2.2.4})$$
\sn
{\bf $\bf \ww(2,2)$:} The algebra $\w(2,2)$ admits a nontrivial
outer automorphism iff the self-coupling constant vanishes. In this
case, it can be realized in terms of two commuting
copies of the Virasoro algebra ($L_1$ and $L_2$) with
equal central charge. $W := L_1 - L_2$ is primary with
respect to $L := L_1 + L_2$. Furthermore, the map $L \mapsto L$,
$W \mapsto - W$ is an automorphism of this algebra. The generating
set of the orbifold was discussed in $\q{\ajl}$. It was also
verified in $\q{\ajl}$ that there is indeed no dimension 10
generator in the orbifold which supports the character argument
predicting a $\w(2,4,6,8)$.
\sn
We have further determined a basis of {\it primary} fields and
calculated the structure constants. Omitting those involving the
complicated dimension 8 generator one obtains
the following list:
$$\eqalign{
\left( C_{4 \, 4}^4 \right)^2 =& 2 (5 c^2 + 66 c - 176)^2
\bigl( (5 c + 44) (5 c + 22) c\bigr)^{-1} \cr
\left( C_{4 \, 4}^6 \right)^2 =& 8 (7 c + 136) (5 c + 22)^2 (c + 4)^2 (c - 1)
\bigl(3 (7 c + 68) (5 c + 44) (2 c - 1) (c + 24) c\bigr)^{-1} \cr
C_{4 \, 4}^4 C_{6 \, 6}^4 =&
  4 (5 c^2 + 66 c - 176) (7 c + 68) (5 c + 88) (2 c - 1)
\bigl(9 (5 c + 44) (5 c + 22) (c + 24) c\bigr)^{-1} \cr
C_{4 \, 4}^6 C_{6 \, 6}^6 =&
  {20 (1106 c^5 \!+\! 50845 c^4 \!+\! 705182 c^3 \!+\! 2270104 c^2
\!-\! 5361664 c \!-\! 1192448)}\times \cr
& (5 c + 22) (c + 4) \bigl(27 (7 c + 68) (5 c + 44) (2 c - 1)
  (c + 24)^2 c\bigr)^{-1}  \ . \cr
} \eqno({\rm 2.2.5})$$
It is interesting to notice that for $c \in \{-4, 1, -{136\over 7} \}$ the
coupling constant $C_{4\, 4}^6$ vanishes whereas $\Phi^{(4)}$ is not a
null field (for $c\in \{1, -{136\over 7} \}$ the field $\Phi^{(6)}$ is a
null field). Thus, for these values of the central charge the orbifold of
$\w(2,2)$ reduces to $\w(2,4)$ or has at least a $\w(2,4)$ subalgebra.
\mn
\section{2.2.2.\ The orbifold of $\w(2,3)$}
\sn
Zamolodchikov's $\w(2,3)$ $\q{\zam}$ is not only one of the first
$\w$-algebras which appeared in the literature but also one of
the most frequently used ones. Therefore we also discuss
it in detail here.
The computations which will be reported below were
carried out with the $\w(2,3)$ as it was presented e.g.\ in
$\q{\twists}$. The field content of the orbifold for generic $c$
can be found e.g.\ in $\q{\bouschou}$.
\mn
First, we have computed determinants of the
{\it invariant} quasi-primary fields up to scale dimension 13.
The zeroes of the determinants tell us where null fields occur.
For these values of the central charge we have
further calculated the dimension
of the space of nonnull invariant fields. Define a counting function
$$\Pi(q) := \sum_{n=1}^{\infty} q^n (\hbox{\# quasi-primary fields with
   dimension $n$}) \ .
\eqno({\rm 2.2.6})$$
$\Pi(q)$ and the vacuum character are related by
$q^{c \over 24} \chi_0(q) = (1-q)^{-1} \Pi(q) + 1$.
These counting functions are presented up to order 13 for generic value of
$c$ and all exceptional values of the central charge in table 3.
Table 3 also contains the field content of a $\w$-algebra which
would give rise precisely to these counting functions.
\mn
\centerline{\vbox{\hbox{\vrule \hskip 1pt
\vbox{ \offinterlineskip
\def\tablespace{height2pt&\omit&&\omit&&\omit&&\omit&\cr}
\def\tablerule{ \tablespace\noalign{\hrule}\tablespace}
\hrule\halign{&\vrule#&\strut\hskip 4pt\hfil#\hfil\hskip 4pt\cr
\tablespace\tablespace
& $c$ && $\Pi(q)$   && {\it orbifold} && {\it first null field} &\cr
\tablerule
& {\it generic} && $\scriptstyle q^2+q^4+3 q^6 + 5 q^8 + 2 q^9 + 8 q^{10}
   + 5 q^{11} + 16 q^{12} + 10 q^{13} + {\cal O}(q^{14})$
      && $\w(2,6,8,10,12)$ && $16$ & \cr \tablerule
& $-2$ && $\scriptstyle q^2+q^4+2 q^6 + 3 q^8 + q^9 + 5 q^{10}
   + 2 q^{11} + 8 q^{12} + 4 q^{13} + {\cal O}(q^{14})$
      && $\w(2,10)$ && $20$ & \cr \tablerule
& $-23$ && $\scriptstyle q^2+q^4+2 q^6 + 4 q^8 + q^9 + 5 q^{10}
   + 3 q^{11} + 9 q^{12} + 5 q^{13} + {\cal O}(q^{14})$
      && $\w(2,8)$ && $16$ & \cr \tablerule
& ${6 \over 5}$ && $\scriptstyle q^2+q^4+2 q^6 + 5 q^8 + 2 q^9 + 8 q^{10}
   + 5 q^{11} + 15 q^{12} + 10 q^{13} + {\cal O}(q^{14})$
      && $\w(2,6,8,10)$ && $> 13$ & \cr \tablerule
& ${4 \over 5}$ && $\scriptstyle q^2+q^4+2 q^6 + 3 q^8 + q^9 + 4 q^{10}
   + 2 q^{11} + 7 q^{12} + 3 q^{13} + {\cal O}(q^{14})$
      && $\w(2)$ && $20$ & \cr \tablerule
& $-{98 \over 5}$ && $\scriptstyle q^2+q^4+3 q^6 + 5 q^8 + 2 q^9 + 8 q^{10}
   + 5 q^{11} + 15 q^{12} + 10 q^{13} + {\cal O}(q^{14})$
      && $\w(2,6,8,10)$ && $> 13$ & \cr \tablerule
& $-{186 \over 5}$ && $\scriptstyle q^2+q^4+3 q^6 + 4 q^8 + 2 q^9 + 7 q^{10}
   + 4 q^{11} + 13 q^{12} + 8 q^{13} + {\cal O}(q^{14})$
      && $\w(2,6,10)$ && $> 13$ & \cr \tablerule
& $-{40 \over 7}$ && $\scriptstyle q^2+q^4+3 q^6 + 5 q^8 + 2 q^9 + 8 q^{10}
   + 5 q^{11} + 14 q^{12} + 9 q^{13} + {\cal O}(q^{14})$
      && $\w(2,6,8,10)$ && $12$ & \cr \tablerule
& $-{114 \over 7}$ && $\scriptstyle q^2+q^4+2 q^6 + 3 q^8 + q^9 + 4 q^{10}
   + 2 q^{11} + 7 q^{12} + 3 q^{13} + {\cal O}(q^{14})$
      && $\w(2)$ && $26$ & \cr \tablerule
& $-{470 \over 7}$ && $\scriptstyle q^2+q^4+3 q^6 + 5 q^8 + 2 q^9 + 8 q^{10}
   + 5 q^{11} + 15 q^{12} + 10 q^{13} + {\cal O}(q^{14})$
      && $\w(2,6,8,10)$ && $> 13$ & \cr \tablerule
& $-{490 \over 11}$ && $\scriptstyle q^2+q^4+3 q^6 + 5 q^8 + 2 q^9 + 7 q^{10}
   + 5 q^{11} + 14 q^{12} + 9 q^{13} + {\cal O}(q^{14})$
      && $\w(2,6,8)$ && $> 13$ & \cr \tablerule
& $-{774 \over 13}$ && $\scriptstyle q^2+q^4+3 q^6 + 5 q^8 + 2 q^9 + 8 q^{10}
   + 5 q^{11} + 15 q^{12} + 10 q^{13} + {\cal O}(q^{14})$
      && $\w(2,6,8,10)$ && $> 13$ & \cr \tablespace
}
\hrule}\hskip 1pt \vrule}
\hbox{\quad Table 3:
            Orbifold of $\w(2,3)$ where generators become null fields}}
}
\mn
Using the procedure described in section 2.2.1.\ we calculated
the composite primary fields of dimension 6, 8 and 10
in the orbifold. The dimension 6 generator is given by
$$\eqalign{
\Phi^{(6)} &= 9 (43 c - 844) (5 c + 22) \n(L, \de^2 L)
+ 480 (191 c + 22) \n(\n(L,L),L) \cr
&  -90 (7 c + 68) (5 c + 22) (2 c - 1) \n(W^{(3)},W^{(3)}). \cr
}\eqno({\rm 2.2.7})$$
The two point function turns out to be
$$ d_{6,6} = 3600 (7 c + 114) (7 c + 68)
(5 c + 22) (5 c - 4) (2 c - 1)
(c + 23) (c + 2) c.
\eqno({\rm 2.2.8})$$
After rescaling to standard normalization one obtains the following
structure constant:
$$
\left( C_{6 \, 6}^6 \right)^2 = {50 (14 c^3 + 915 c^2 + 14758 c - 22344)^2
   (c + 2) \over 3 (7 c + 114)
(7 c + 68) (5 c + 22) (5 c - 4) (2 c - 1) (c + 23)} .
\eqno({\rm 2.2.9})$$
The explicit form of the spin 8 and 10 generators, other two-point functions
and additional structure constants can be found in appendix~\appE.
\mn
At $c=-23$ the field $\Phi^{(6)}$ is a null field whereas $\Phi^{(8)}$ is
nonzero.  This is to be expected because one knows that for this value of the
central charge $\w(2,3)$ has a $\w(2,8)$ subalgebra $\q{\mfl}$. At $c=-2$ both
fields $\Phi^{(6)}$ and $\Phi^{(8)}$ turn out to be null fields. However
$\Phi^{(10)}$ is nonzero and $\w(2,3)$ has a $\w(2,10)$ subalgebra at $c=-2$.
This agrees with the results of $\q{\howcl}$ and in particular confirms that
the field of dimension 10 is quadratic in $W^{(3)}$. Furthermore,
$c={4 \over 5}$ and $c=-{114 \over 7}$ are Virasoro minimal and therefore the
orbifold must be just the Virasoro algebra. These statements about
$c = -2, -23, {4 \over 5}, -{114 \over 7}$ are confirmed by the dimensional
arguments in table 3 and we can use known facts (see e.g.\ $\q{\mfl,\twists,
\howcl}$) about representations of these algebras to predict the dimension of
the first null field. For the remaining cases one has to be more careful
because we have not checked all structure constants. For some values of the
central charge it might turn out that null fields actually make the orbifold
inconsistent. Note, however, that the induced normalizations eq.\ (2.2.8),
(\appE.2) for the generators of the orbifold are consistent with the field
content predicted in table 3.
\mn
\section{2.2.3.\ Remarks on the orbifold of ${\cal WA}_{n-1}$}
\sn
It has been shown in $\q{\frfolding}$ that
the orbifolds of the classical ${\cal WA}_{n-1}$ possess other classical
Casimir algebras as subalgebras. In this subsection we will show that such
a relation does not hold true for the quantum orbifolds.
\sn
For ${\cal WA}_{n-1}\cong \w(2,\ldots,n)$
some structure constants (2.1.25) are known generally.
The first primary composite field in the $\BZ_2$ orbifold (the $\BZ_2$
automorphism changes the sign of the odd dimensional simple fields)
can be calculated for all $n \ge 4$:
$$\eqalign{
\Phi^{(6)} =& 27 (43 c - 844) (5 c + 22) (c + 24) \n(L, \de^2 L)
+ 1440 (191 c + 22) (c + 24) \n(\n(L,L),L) \cr
&+ 1980 (7 c + 68) (5 c + 22) (2 c - 1) C_{W^{(3)} \, W^{(3)}}^{W^{(4)}}
  \n(W^{(4)},L) \cr
&- 270 (7 c + 68) (5 c + 22) (2 c - 1) (c + 24) \n(W^{(3)},W^{(3)}).
} \eqno{(\rm 2.2.10)}$$
Using (2.1.25a) this induces the following normalization
$$\eqalign{
d_{6,6} =& \Bigl(32400 c \bigl(7 (n + 2) (n - 2) c^4
 + (21 n^3 + 380 n^2 - 1800) c^3 + (1399 n^3 + 1585 n^2 \cr
&\ - 7700) c^2
+ 4 (1179 n^3 + 4375 n^2 - 1770) c
- 32 (473 n^2 - 81 n - 81) (n - 1) \bigr) \cr
&\ (7 c + 68) (5 c + 22)^2 (2 c - 1) (c + 24) (c + 2)\Bigr)
\Bigl((c n + 2 c + 3 n^2 - n - 2) (n - 2) \Bigr)^{-1} \ .
} \eqno{(\rm 2.2.11)}$$
The simplest nontrivial structure constant
of the orbifold reads in the standard normalization
($\hat{d}_{6,6} = {c \over 6}$)
$$\eqalign{
\left(C_{4 \, 4}^6\right)^2 =&
6 \Bigl(29 c^3 n^2 - 284 c^3 + 255 c^2 n^3 - 427 c^2 n^2
 - 368 c^2 + 540 c n^3 + 3016 c n^2 - 1636 c \cr
&\ - 1920 n^3 + 2032 n^2 - 112\Bigr)^2 (5 c + 22)^2 \cr
&\Biggl(
\Bigl(7 c^4 n^2 - 28 c^4 + 21 c^3 n^3 + 380 c^3 n^2 - 1800 c^3 + 1399 c^2
n^3 + 1585 c^2 n^2 - 7700 c^2 \cr
&\ + 4716 c n^3 + 17500 c n^2
- 7080 c - 15136 n^3 + 17728 n^2 - 2592 \Bigr) \cr
& \
(c n + 2 c + 3 n^2 - n - 2) (7 c + 68) (2 c - 1) (c + 24) (c + 2) (n - 2)
\Biggr)^{-1} \cr
} \eqno{(\rm 2.2.12)}$$
where we have used both (2.1.25a) and (2.1.25b).
Fortunately (2.1.25b) vanishes for $n=4$ such that we can apply it to
$\w(2,3,4)$ as well. Note that the structure constant (2.2.12) is nonzero
for any $n$ and the central charge $c$ generic. This means that the
invariant original generators do not close among themselves and one is
forced to include the dimension 6 generator (2.2.10) into the orbifold.
In particular, for $n \ge 6$ the orbifold contains two simple fields of scale
dimension 6. This has to be contrasted with the classical situation where
it has been shown in $\q{\frfolding}$ that already the invariant original
generators generate a closed subalgebra of the complete orbifold. The fact
that (2.2.12) vanishes in the limit $c \to \infty$ is consistent with this
statement because this limit should correspond to a classical limit
(see section 3.1.).
\mn
Specializing (2.2.11) to $n=4$ we obtain for the induced normalization of
the first composite field in the orbifold of $\w(2,3,4)$:
$$d_{6,6} = 10800 (7 c \!+\! 114) (7 c \!+\! 68) (5 c \!+\! 22)^2
(3 c \!+\! 116) (2 c \!-\! 1) (c \! +\! 24)
(c \!+\! 13) (c \!+\! 2) (c \!-\! 1) c  (c\! +\! 7)^{-1}.
\eqno{(\rm 2.2.13)}$$
Of course, it is also straightforward to specialize (2.1.25) and
(2.2.12) to $n=4$ to obtain the first structure constants. From
(2.2.13) we read off some interesting values of the central charge $c$
for the orbifold of $\w(2,3,4)$. For $c \in \{-13,1,-{116 \over 3}\}$ the
field $\Phi^{(6)}$ is a null field but does not make $\w(2,3,4)$ inconsistent
(like it happens e.g.\ for $c=-24$). Thus, for these values of the central
charge the orbifold of $\w(2,3,4)$ is a $\w(2,4)$ or at least has a $\w(2,4)$
subalgebra. In particular, $\w(2,3,4)$ itself has a $\w(2,4)$ subalgebra for
$c \in \{-13,1,-{116 \over 3}\}$. For $\w(2,3,4)$ there are 8 quasi-primary
invariant fields at scale dimension 8. Their determinant reads (up to a
nonzero constant of proportionality which depends on the choice of basis):
$${\rm det}_8\!\sim (11 c \!+\! 702) (7 c \!+\! 114) (7 c \!+\! 27)
(5 c \!+\! 22)^4 (3 c \!+\! 116)^2 (c \!+\! 51)(c \!+\! 13)^2
(c\! +\! 2) (c \!-\! 1)^3 c^8 (c \!+\! 7)^{-3}.
\eqno({\rm 2.2.14})$$
{}From this we observe that the additional scale dimension 8 generator
in the orbifold vanishes for $c \in \{-13,1,-{116 \over 3}, -51,
-{27 \over 7}, -{702 \over 11}\}$ which includes in particular the
three values of $c$ where already the scale dimension 6 composite
generator vanishes. Whereas $c=-51$ and $c=-{27\over 7}$ do not belong
to the minimal series of the ${\cal WB/C}$ Casimir algebras the value
$c=-{702 \over 11}$ lies in the minimal series of ${\cal WB}_3$
($p=11,q=7$). Comparison of the set of highest weights of the minimal
models and structure constants indicates that the orbifold of
$\w(2,3,4)$ is a ${\cal WB}_3$ for $c=-{702 \over 11}$ (compare
appendix \appF).
\sn
Coupling constants connecting two simple fields with primary
normal ordered products have been determined for $\w(2,3,4,5)$
and $\w(2,3,4,5,6)$ before in $\q{\hornfeck}$. However,
in this work the coefficient of $\n(W^{(3)}, W^{(3)})$ was
chosen independent of $c$ such that one cannot read off the values of the
central charge where $\Phi^{(6)}$ becomes a null field.
\mn
For the orbifold of $\w(2,3,4,5)$ we obtain from the specialization
of (2.2.11) to $n=5$:
$$d_{6,6} = 10800 (7 c \!+\! 68) (7 c \!-\! 8) (5 c \!+\! 22)^2
(3 c \!+\! 116) (2 c \!-\! 1)(c \!+\! 24) (c \!+\! 23) (c \!+\! 2) c.
\eqno{(\rm 2.2.15)}$$
It is remarkable that for $c={8 \over 7}$ the field $\Phi^{(6)}$ turns
out to be a null field. Thus, the orbifold of $\w(2,3,4,5)$ has a $\w(2,4)$
subalgebra at $c={8 \over 7}$ or probably even reduces to $\w(2,4)$.
On the one hand this is the first unitary minimal
model of $\w(2,3,4,5)$ -- the $\BZ_5$ parafermions $\q{\fatzam}$.
On the other hand this is presumably the only nontrivial unitary minimal
model of $\w(2,4)$ $\q{\andrdipl}$. The orbifold construction explains
why precisely half of the representations of $\w(2,4)$ at $c={8 \over 7}$
are parafermionic representations $\q{\andrdipl}$ because each representation
of the original algebra splits into two representations when a $\BZ_2$
orbifolding procedure is applied.
We also re-encounter the value $c=-{116 \over 3}$ which is the only other
rational value of $c$ for which the dimension 6 generator vanishes and
$C_{4\, 4}^6 = 0$.
\sn
For $n=6$, i.e.\
the orbifold of $\w(2,3,4,5,6)$ one finds no rational value of the central
charge $c$ where the additional (composite) dimension 6 generator could drop
out.
\bn
\section{3.\ General structures in cosets and orbifolds}
\bn
\section{3.1.\ Vacuum preserving algebras (VPA) and classical limits}
\mn
Once a construction of a $\w$-algebra as a reduction of a Kac-Moody algebra or
a similar linear system is known, the $\w$-algebra is usually well under
control. In particular, one can easily discuss its classical counterpart, i.e.\
the analogous reduction of the corresponding classical linear system.
Note, however, that in general several constructions of the same quantum
$\w$-algebra are possible and can lead to different classical counterparts.
For a classification one needs more general methods which do not refer to
any particular construction. Two closely related ideas in this direction have
been put forward in $\q{\bowwatts}$ for deformable $\w$-algebras:
The vacuum preserving algebra (VPA) as well as a particular classical limit of
a $\w$-algebra. These methods work nicely for $\w$-algebras obtained
by Drinfeld-Sokolov reduction $\q{\bowwatts,\fehort}$. The $\w$-algebras
discussed in this paper are not in the DS class and therefore it is
interesting to see to what extent these methods work for them.
\mn
We start with a discussion of a general approach to the VPA. First, one
introduces the `vacuum preserving modes' of all quasi-primary fields. The
space spanned by them carries a Lie algebra structure. Next, one considers
the limit $c \to \infty$ of this algebraic structure. The VPA is the smallest
subalgebra of this algebra containing the vacuum preserving modes of
the simple fields. To be more precise, the vacuum preserving modes
of a quasi-primary field $\Phi$ are given by
$$\{\Phi_n \mid \vert n \vert < d(\Phi) \} .
   \eqno({\rm 3.1.1})$$
The vacuum preserving modes of all quasi-primary fields have the
important property $\q{\bowwatts,\fehort}$ that the commutator
closes among them and does not have any central term. Note that
the vacuum preserving modes $L_{\pm 1}, L_0$ of the energy momentum
tensor $L$ form an $sl(2)$ subalgebra of all vacuum preserving modes.
The space spanned by all vacuum preserving modes (3.1.1) is in general
infinite dimensional and the commutator still depends continuously on $c$.
In order to cure the second property one takes the limit $c \to \infty$.
In general, one will have to rescale the generators $W^{(i)}$ of the
finitely generated quantum $\w$-algebra in order to make sense of the
limit $c \to \infty$:
$$\hat{W}^{(i)} := c^{-\alpha_i} W^{(i)} .
   \eqno({\rm 3.1.2})$$
The exponents $\alpha_i$ have to be adapted in order to make all structure
constants connecting the fields $\hat{W}^{(i)}$ bounded and nontrivial for
$c \to \infty$. Even in this limit, the algebra of all vacuum preserving
modes is a very complicated object. Therefore, the VPA is defined as the
smallest subalgebra of the limit $c\to \infty$ of the algebra of all vacuum
preserving modes that contains the vacuum preserving modes of the simple
fields.
\mn
It was shown in $\q{\bowwatts}$ that this works nicely for
the algebras in the DS class: One can
associate to them a finite dimensional Lie algebra with an $sl(2)$ embedding
that encodes the spin content of these algebras.
For the $\w$-algebras in the DS class
one can set all $\alpha_i := 0$ and then the VPA is defined by
$${\cal V} := {\rm span}\{W^{(i)}_n \mid \vert n \vert < d(W^{(i)}) \}
   \eqno({\rm 3.1.3})$$
where $W^{(i)}$ are the simple fields of the $\w$-algebra.
On this space a Lie bracket is induced by taking the limit $c \to \infty$
of the commutator.
The crucial point is that the commutator linearizes, i.e.\ that the induced
Lie bracket closes in the space (3.1.3). The data (3.1.3) together with the
Lie bracket and the $sl(2)$ embedding is equivalent to the original data
used for the DS reduction $\q{\bowwatts,\fehort}$.
\mn
The situation is much less clear for $\w$-algebras outside the
DS class, i.e.\ those $\w$-algebras which do not have `nice'
asymptotic properties of the structure constants. We will first discuss
one example in detail: The bosonic projection of the $N=1$ Super
Virasoro algebra. Let us
for the moment forget about the construction of this $\w(2,4,6)$
and look what can be said about the VPA solely from inspection of the
structure constants (Set 1 in section 6.2 of $\q{\kauwatts}$).
Denote the generators of scale dimension 2, 4 and 6 by
$L$, $W^{(i)}$ ($i=4,6$). We observe from the structure
constants in $\q{\kauwatts}$ that we have to rescale
$$\hat{L} := L \, , \qquad
\hat{W}^{(i)} := {1 \over \sqrt{c}} W^{(i)}.
   \eqno({\rm 3.1.4})$$
Then, all structure constants connecting these three simple fields
are bounded and nonzero for $c \to \infty$. Unlike for $\w$-algebras
in the DS class the commutator does not linearize for this $\w(2,4,6)$,
i.e.\ the commutator does not close in the space (3.1.3).
The structure constants $C_{W^{(4)} \, W^{(6)}}^{(W^{(4)} W^{(4)})}$
and $C_{W^{(6)} \, W^{(6)}}^{(W^{(4)} W^{(4)})}$ are invariant
under the rescaling (3.1.4). Furthermore, these structure constants
tend to a nonzero constant in the limit $c \to \infty$. This means
that the vacuum preserving modes $\hat{W}^{(8)}_n$, $\vert n \vert < 8$
of the primary projection $\hat{W}^{(8)}$ of $(\hat{W}^{(4)} \hat{W}^{(4)})$
have to be included into the VPA. We have checked that the same happens
at scale dimension 10. At scale dimension 10 one finds the quadratic
fields $(\de^2 \hat{W}^{(4)} \hat{W}^{(4)})$ and
$(\hat{W}^{(4)} \hat{W}^{(6)})$.
Due to the presence of a generic null field at scale dimension 10
$\q{\bouwknegt}$ these two fields give rise to precisely one primary
field at scale dimension 10. Since both
$(\de^2 \hat{W}^{(4)} \hat{W}^{(4)})$ and $(\hat{W}^{(4)} \hat{W}^{(6)})$
do not decouple in the limit $c \to \infty$,
the vacuum preserving modes of the primary field with dimension 10
have to be included into the VPA. Thus, the VPA does definitely
not close on the vacuum preserving modes of the simple fields only
but one has to include the vacuum preserving modes of further primary fields
(at least up to scale dimension 10). We expect that one must indeed
include the vacuum preserving modes of infinitely many primary fields
because there is no reason to expect any of the crucial structure
constants to vanish. Therefore, we actually expect the VPA of this
$\w(2,4,6)$ to be infinite dimensional.
\mn
Similar reasoning applies to the second $\w(2,4,6)$ outside
the DS class (Set 2 in section 6.2 of $\q{\kauwatts}$ and eq.\
(6) in $\q{\howcl}$). The situation is slightly different at scale
dimension 10. Here, no null field is present and therefore we do
indeed have two primary fields which turn up in the commutator
of $\hat{W}^{(6)}$ with itself. However, up to that stage only a
particular linear combination of these fields appears in the commutators
and therefore it might be sufficient to include the vacuum preserving
modes of only one field at scale dimension 10 into the VPA.
As before the realization of this $\w(2,4,6)$ in terms of the coset
$\sltr_k \oplus \sltr_{-{1 \over 2}} / \sltr_{k-{1 \over 2}}$
$\q{\ajl}$ (see also section 2.1.3.) is not needed for the discussion
of the VPA.
\mn
Let us now turn to the $\w(2,3,4,5)$ which we discussed in
section 2.1.1.\ (see also $\q{\ajl}$). The structure constants
for this algebra were first calculated in $\q{\hornfeck}$ checking
associativity of the OPE without reference to the cosets
$\sltr_k / \widehat{U(1)}$ or $SVIR(N=2) / \widehat{U(1)}$.
Denote the generators of this $\w(2,3,4,5)$ by $L$ and $W^{(i)}$
($i=3,4,5$). Then we see from the structure constants in table
2 of ref.\ $\q{\hornfeck}$ that we have to rescale as in eq.\ (3.1.4).
Up to scale dimension 8 there are 6 further quadratic fields in terms
of the $W^{(i)}$, two of them at scale dimension 8 give rise to null fields
(see eq.\ (2.11) and (2.12) of ref.\ $\q{\hornfeck}$) leaving
us with 4 primary fields: One each at scale dimensions 6 and 7,
two at scale dimension 8. As in the previous cases, the structure
constants connecting the simple primary fields with these composite
primary fields are invariant under the rescaling (3.1.4) and tend to
a nonzero constant for $c \to \infty$ (compare table
2 of ref.\ $\q{\hornfeck}$). This means that the vacuum preserving modes
of the fields at scale dimensions 6 and 7 have to be included into the VPA.
Only a linear combination of the two fields at scale dimension 8
appears in the OPEs of the simple fields. Therefore, so far we have to include
the vacuum preserving modes of only one primary field at scale dimension 8
into the VPA. As before, we have not calculated structure constants
involving fields of higher dimension, but there is no reason to
expect those structure constants involving primary fields to vanish.
Thus, we expect also the VPA of this algebra to be infinite dimensional.
\bn
The above discussion can further be supported by looking at the classical
counterparts of these algebras that correspond to the above constructions
$\q{\ajl}$. These classical counterparts are infinitely generated. From
$\q{\bowwatts,\fehort}$ we know that in the classical case the VPA consists
of the vacuum preserving modes of all generators. In particular, for the
examples under consideration it is definitely infinite dimensional. Even
more, the composite primary fields which had to be added precisely correspond
to the additional generators of the classical counterparts. This means that
the VPA of a $\w$-algebra encodes some information of possibly underlying
constructions.
\mn
We have seen in three examples outside the DS class that their
VPA is an infinite dimensional Lie algebra with $sl(2)$ embedding
which decomposes into finite dimensional representations under
this $sl(2)$. This indicates that probably all $\w$-algebras
outside the DS class have infinite dimensional VPAs which makes
the VPA as a tool for classification unhandy. At least, for
any explicitly known algebra the construction of the VPA is
purely algorithmic but does unfortunately not necessarily stop after
finitely many steps. It should be noted that the impact of the realization
of composite fields contributing to the VPA in terms of finitely many simple
ones is not completely clear to us, neither is the impact of the relations
satisfied by the generators of these $\w$-algebras $\q{\ajl}$ on the VPA.
\bn
Next we consider classical limits of $\w$-algebras. A
classical Kac-Moody algebra is a Lie algebra and can therefore
be quantized according to Dirac's rule. In particular, the
r.h.s.\ of the commutator is multiplied by $\hbar$, and the canonical
classical limit of the quantum Kac-Moody algebra is the limit $\hbar \to 0$.
For a $\w$-algebra that arises as some reduction of a Kac-Moody algebra, this
classical limit of the Kac-Moody algebra induces a classical limit of the
$\w$-algebra. If we have a particular construction in mind we refer to the
induced classical limit as `the classical limit'.
In contrast hereto, the procedure of $\q{\bowwatts}$ is a set of rules
to re-institute $\hbar$'s in a $\w$-algebra without referring to any
particular reduction and afterwards define a classical limit $\hbar\to 0$.
We will refer to this classical limit as the `BW classical limit'.
\mn
To be more explicit, the BW classical limit is defined by the following
set of rules
$$W^{(i)} \mapsto {\hat{W}_\hbar^{(i)} \over \hbar^{1+ \alpha_i}} \, , \qquad
c \mapsto {\hat{c} \over \hbar}
   \eqno{(\rm 3.1.5)}$$
where $W^{(i)}$ are the simple generators of the $\w$-algebra. Then, the
quantum fields $W^{(i)}(z)$ are replaced by classical fields $w^{(i)}(z) :=
\lim_{\hbar \to 0} \hat{W}_\hbar^{(i)}(z)$ and Poisson brackets
as well as the ring structure are defined by the following identifications
$$\{w^{(i)}(z), w^{(j)}(w)\} = \lim_{\hbar \to 0} {1 \over \hbar}
   \lb \hat{W}_\hbar^{(i)}(z), \hat{W}_\hbar^{(j)}(w) \rb \, , \qquad
w^{(i)}(z) w^{(j)}(z) = \lim_{\hbar \to 0}
   (\hat{W}_\hbar^{(i)} \hat{W}_\hbar^{(j)})(z)
   \eqno{(\rm 3.1.6)}$$
where the choice of normal ordering prescription on the quantum level
is actually irrelevant. The crucial point is to choose the exponents
$\alpha_i$ in (3.1.5) such that the limits in (3.1.6) exist and are
nontrivial. The reader should note that in general the existence
of such a set of exponents $\alpha_i$ is not guaranteed.
For $\w$-algebras in the DS class one can choose all
$\alpha_i = 0$ $\q{\bowwatts}$ and then (3.1.5) and (3.1.6) are
one-to-one maps between the one-parameter families of
$\w$-algebras on the quantum level and on the classical level where
the parameter is the central charge. It should be noted that the BW classical
limit according to (3.1.5) and (3.1.6) is not necessarily the same
as the one induced by a reduction. Close inspection shows that
even in the DS class the central charges of these two classical
limits are indeed different, and therefore the classical limits
are equivalent only when looking at one-parameter families of $\w$-algebras.
Note that also nondeformable $\w$-algebras can have a classical limit
(compare e.g.\ the $\beta-\gamma$ system of $\q{\ajl}$). Thus, deformability
and existence of a classical limit should not be confused. However, only for
deformable $\w$-algebras the rules (3.1.5) can be applied.
\mn
In passing we mention that one obtains from (3.1.6)
the classical quasi-primary projection of
the product of two classical quasi-primary fields:
$$\Q(w^{(i)}(z) \de^n w^{(j)}(z)) =
   \lim_{\hbar \to 0} \n(\hat{W}_\hbar^{(i)},\de^n \hat{W}_\hbar^{(j)})(z) .
   \eqno{(\rm 3.1.7)}$$
Applying (3.1.5) and (3.1.6) to the formula for the quasi-primary normal
ordered product $\n$ (eq.\ (1.1.4)) we immediately obtain an explicit formula
for $\Q(w^{(i)}(z) \de^n w^{(j)}(z))$ (under assumptions such as
$\alpha_k < 1 + \alpha_i + \alpha_j$ for all $k$). Let $w^{(i)}(z)$
and $w^{(j)}(z)$ be two classical quasi-primary fields. Then
$$\Q(w^{(i)}(z) \de^{n} w^{(j)}(z)) := \sum_{r=0}^{n} (-1)^r
{{n\choose r} {2d(w^{(j)})+n-1 \choose r} \over
{2(d(w^{(j)})+d(w^{(i)})+n-1) \choose r}}
\de^r (w^{(i)}(z) \de^{n-r} w^{(j)}(z))
   \eqno{(\rm 3.1.8)}$$
is quasi-primary and has dimension $d(w^{(i)})+d(w^{(j)})+n$
\footnote{${}^{8})$}{
This statement can be proven directly (without classical
limit) by taking Poisson brackets with the modes $l_{\pm 1}$ and
$l_0$ of the classical Virasoro generator
on the right hand side of (3.1.8) and verifying that
it does indeed transform like a quasi-primary field.
\par\noindent
Note that finding the projection of the $n$th derivative of the product of
two modular forms onto a modular form (see e.g.\ $\q{\don}$) is analogous
to determining the quasi-primary projection. In fact, the formula (1) of
$\q{\don}$ (called the `$n$th Rankin-Cohen bracket') is more compact but
equivalent to (3.1.8).
}.
\mn
Let us now apply these ideas to the algebras outside the DS class.
Requiring the linear term (in terms of the generators) on the r.h.s.\
of the commutator to be bounded and nonzero for $\hbar \to 0$
we conclude that the exponents $\alpha_i$
in (3.1.5) and (3.1.2) are actually identical. In particular,
for the three examples already discussed above the exponent
for the Virasoro field $L$ is $\alpha_0 = 0$ and all other
$\alpha_i = {1 \over 2}$. In order to have a well-defined limit of the
commutator in (3.1.6) all structure constants connecting the additional
simple fields with quadratic fields not containing $L$ must be at most
of order ${\cal O}(c^{-1})$ as $c \to \infty$. However,
from the discussion of the VPA we conclude that the coupling constants
to the quadratic fields in terms of the generators tend to
a nonzero constant for $c \to \infty$. This means that one
must decouple the quadratic fields from the ring and rescale
them independently, i.e.\ one must introduce further relations
and further generators in order to make sense of the BW classical
limit. This is similar to the VPA, in particular those fields
have to be introduced as new generators whose modes had to be
included into the VPA. This indicates that the BW classical limits
of these three algebras are probably infinitely generated
and satisfy infinitely many constraints.
\mn
At least for the bosonic projection of the $N=1$ Super Virasoro
algebra we can verify in terms of the underlying realization
that one field each at scale dimensions
8 and 10 decouples in the classical limit and gives rise
to a new generator. Recall that the field $G$ in appendix {\appD}
has to be rescaled with $\alpha_1 = 0$ (the exponent $\alpha_0$ for $L$
is also zero). Denote the classical limits corresponding to $L(z),G(z)$
by $l(z),g(z)$. After these substitutions, the coefficient of $g \de^5 g$
in (\appD.3) vanishes in this classical limit and one has to add its primary
projection ${\cal P}(g \de^5 g)$ to the generating set.
This primary projection can be obtained by applying the limiting procedure
to the corresponding primary quantum field and is explicitly given by
$${\cal P}(g \de^5 g) = \Q(g \de^5 g)
-{1380 \over 13 c} \Q(\Q(g \de^3 g) l)
-{182 \over 11 c} \Q(\Q(g \de g) \de^2 l)
+{16524 \over 11 c^2} \Q(\Q(\Q(g \de g) l) l) .
   \eqno{(\rm 3.1.9)}$$
A similar phenomenon happens at scale dimension 10 where the primary
projection of $g \de^7 g$ has to be added to the generating
set (compare also the discussion in $\q{\ajl}$, in particular
appendix A loc.\ cit.).
\mn
For these three examples (the $\w(2,3,4,5)$ and the two algebras of
type $\w(2,4,6)$) the realization in terms of a reduction is
known both on the quantum as well as on the classical level.
Therefore, we can compare the result of the limiting procedure
(3.1.5) and (3.1.6) to the corresponding classical algebras
$\q{\ajl}$. First, we remark that the fields which we had to
introduce as generators in the BW classical limit indeed turn
up as generators in the classical reduction and that the
classical counterparts of these algebras are indeed infinitely
generated $\q{\ajl}$. Rescaling of the generators with exponents
$\alpha_i > 0$ implies a vanishing central term in the limit
$\hbar \to 0$. On the classical level, all generators (with
exception of the Virasoro field $l(z)$) are at least second
order in the fields of the underlying algebra and the Poisson
brackets of such fields do indeed not contain any central term.
For any orbifold (including in particular the bosonic projection
of the $N=1$ Super Virasoro algebra), the energy momentum tensor
$L$ in the projection is noncomposite and the central charge $c$ is
a free parameter on the classical level. Thus, the BW classical
limit of this $\w(2,4,6)$ has a chance to be identical to the $\w$-algebra
obtained from the classical orbifold. The situation is different
for the other $\w(2,4,6)$ and the $\w(2,3,4,5)$ which we realized
in terms of cosets. Note that the classical coset energy momentum
tensor $l(z)$ is composite and therefore has no central term,
i.e.\ $c=0$ on the classical level. However, the BW limiting
procedure (3.1.5) and (3.1.6) gives rise to nonzero Virasoro
centre $c$ showing that the BW classical limit is not identical to
the classical coset. Even more, close inspection of the structure
constants of the $\w(2,4,6)$ arising in the diagonal $\sltr$
coset shows that some structure constants still are proportional
to ${1 \over c^r}$ after taking the BW classical limit. Because
this $c$-dependence cannot be completely scaled away, one cannot
simply set $c=0$ in the BW classical limit. This is not very surprising
because for cosets of Kac-Moody algebras the energy momentum tensor $L$
satisfies
$\lb L_m, L_n \rb = \hbar (n-m) L_{m+n} + \hbar^2 c (n^3 - n) \delta_{m+n,0}$.
This means that one should modify the BW procedure by substituting $c \mapsto
\hat{c}$ instead of (3.1.5) in order to obtain at least the correct classical
form of the Virasoro algebra. However, this substitution does not introduce
any $\hbar$'s in the structure constants and would therefore
leave us with $c$-dependent structure constants in this modified BW
classical limit -- something we do not want either. Furthermore, in the case
of the $\w(2,3,4,5)$, the classical counterpart of the realization in terms
of the coset \sviru\ has a non-vanishing Virasoro centre in contrast to
the classical coset \slu. So, the limiting procedure (3.1.5), (3.1.6) might
correspond to the classical coset \sviru\ but it is not clear how to obtain
the classical counterpart of the other coset realization by a limiting
procedure of this type. Even more, one can check that the classical coset
\sviru\ admits a {\it primary} generating set whereas the classical coset
\slu\ does not. This shows that the Poisson brackets carried by these two
classical cosets are not at all related to each other.
These ambiguities for the classical counterpart
might be related to the possibility that the BW procedure (3.1.5), (3.1.6)
does not always give rise to a classical $\w$-algebra.
\mn
In summary, we have seen that one can introduce the VPA and study the
BW classical limit of any quantum $\w$-algebra in an algorithmic
manner. In doing so one recovers many features of a corresponding
classical counterpart without using any knowledge about the underlying
construction (although the attempt to construct the BW classical limit
could fail). This gives rise to the hope that all $\w$-algebras
outside the DS class which are finitely generated on the quantum
level belong to the same class of $\w$-algebras with infinitely
generated classical counterparts $\q{\ajl}$. From this point of
view coset constructions and orbifolds behave in a very similar manner.
\vfill
\eject
\section{3.2.\ Coset realization of unifying $\w$-algebras and
               level-rank-duality}
\sn
In section 2.1.3.\ we were able to show that the special $\w(2,4,6)$
is realized by the coset \slsl\ and we have studied also its minimal models.
Below we will show that this algebra is the first member of a new
series of unifying algebras - denoted as  ${\cal WD}_{-m}$ - which
are unifying objects for some ${\cal WC}$ minimal models $\q{\letter}$.
In the spirit of $\q{\klausREP}$ one can write down the
minimal series of these algebras. Using character arguments it
is possible to give an explicit coset realization based on the symplectic
Lie algebras $sp(2m)$. Furthermore, we present coset realizations of
unifying algebras of some series of minimal models
of the (Casimir) algebras ${\cal WA, WB}$ and
${\rm Orb}({\cal WD})$ proposed in $\q{\letter}$.
These relationships generalize level-rank-duality of coset pairs.
Furthermore, we study the diagonal cosets
$\hat{g}_\kp \oplus \hat{g}_\mu / \hat{g}_{\kp + \mu}$ for
$g={\cal A}_n,{\cal B}_n,{\cal C}_n,{\cal D}_n$ and some special
values of $\kappa,\mu$ on the level of characters.
\mn
\section{3.2.1.\ Unifying $\w$-algebras for the ${\cal WA}_n$ Casimir
                 algebras}
\sn
Due to the level-rank duality $\q{\bogo,\altschuler}$
$$ {\widehat{sl(n)}_k\oplus \widehat{sl(n)}_1
       \over {\widehat{sl(n)}_{k+1}}} \cong
   {{\widehat{sl(k+1)}_n \over
     {\widehat{sl(k)}_n\oplus \widehat{U(1)}}} = {\cal CP}(k) }
\eqno{\rm(3.2.1)}$$
one expects that the symmetry algebra of the ${\cal CP}(k)$ model
is a unifying $\w$-algebra for the $k$th unitary model of
${\cal WA}_{n-1}$. Note that the l.h.s.\ of (3.2.1) is defined for
integer $n$ and arbitrary $k$, whereas the r.h.s.\ is defined for integer
$k$ and general $n$. The isomorphism in (3.2.1) is valid iff $k$ and $n$
are both positive integers.
\sn
We will calculate the spin content of ${\cal CP}(k)$ using
character techniques \footnote{${}^9)$}{The (unique) simple field
with spin 3 has been calculated in $\q{\koos}$.}.
According to $\q{\ajl}$ we have to count the states in the complement of
$\widehat{sl(k)}_n\oplus \widehat{U(1)}$ in $\widehat{sl(k+1)}_n$
that are invariant under $sl(k) \oplus U(1)$. This will be carried out
for generic level $n$ using the character argument described in
section 2.1.3. Let $\Delta^k$ be the root
system of $sl(k+1)$ and let $\vec\Theta$ be in $h^*$ ($h$ is a Cartan
subalgebra of the horizontal subalgebra $sl(k+1)$ of $\widehat{sl(k+1)}_n$).
We will write $\vec\Theta = (\vec\theta, \Theta_k)$ where $\vec\theta$
corresponds to $sl(k)$ and $\Theta_k$ to $U(1)$. The counting function for
the subspace of the vacuum module of $\widehat{sl(k+1)}_n$ that does not
contain any states generated by $\widehat{sl(k)}_n\oplus \widehat{U(1)}$
is given by:
$${q^{-{2k\over 24}}\over
\prod_{\vec\alpha\in \Delta^k\backslash \Delta^{k-1}}
  \prod_{n \ge 1} (1-e^{2\pi i\,\vec\Theta\cdot\vec\alpha}q^n) }
=\sum_{\vec\lambda,m} C^0_{\vec\lambda,m}(q) \
    e^{2\pi i\, (\vec\theta\cdot\vec\lambda+\Theta_k m)}.
\eqno{\rm(3.2.2)}$$
Since this module carries a representation of $sl(k) \oplus U(1)$ it can
be decomposed into representations of $sl(k) \oplus U(1)$ which is
indicated by the r.h.s.\ of (3.2.2). The $C^0_{\vec\lambda,m}(q)$ are
the string functions where $\vec\lambda$ and $m$ are the weights of $sl(k)$
and $U(1)$ respectively that label the representation. Invariance under
$U(1)$ is equivalent to $m=0$. The part invariant under $sl(k)$ can be
obtained by summation over its Weyl group. Thus, the vacuum character of the
coset under consideration is given by the branching function $B_{0,0}^0(q)$
which can be written as the following sum over string functions (compare
(2.1.55)):
$$B^0_{0,0}(q)=\sum_{w\in W}\epsilon(w)\,C^0_{-w(\rho)+\rho,0}(q)
\eqno{\rm(3.2.3)}$$
where $W$ is the Weyl group of $sl(k)$. The string functions can be obtained
as follows. Inserting $z^2 = e^{2\pi i\,\vec\Theta\cdot\vec\alpha}$ with
$\vec\alpha$ a positive root into (2.1.6), the left hand side of (3.2.2) can
be written in the form
$${q^{-{2k\over 24}}\over
\prod_{\vec\alpha\in \Delta^k\backslash \Delta^{k-1}}
  \prod_{n \ge 1}
(1-e^{2\pi i\,\vec\Theta\cdot\vec\alpha}q^n) }= \!\!\!\!\!\!\!\!\!\!
\sum_{\quad\{n_1,\ldots,n_k\}\in{\BZ^k}} \!\!\!\!\!\!\!\!\!
{\prod_{i=1}^k(\phi_{n_i}(q)-\phi_{n_i-1}(q)) \over \eta^{2k}(q)}
\ e^{2\pi i\ \vec\Theta\cdot
\bigl(\sum_{i=1}^k \!\! n_i\vec\alpha_i\bigr)} .
\eqno{\rm(3.2.4)}$$
Using an embedding of the roots of $sl(k+1)$ into $\BR^k$ one obtains
a map $(\vec\lambda,m)\to(n_1,\ldots,n_k)$. Now, comparison of the r.h.s.\
of eqs.\ (3.2.2) and (3.2.4) gives an explicit representation of the
string functions in terms of the $\phi '$s.
It is now not difficult to calculate the vacuum character of the
${\cal CP}(k)$ model for the first few values of $k$. In table 4 we present
the results for $1\le k \le 4$ and give the spin content of the
coset as well as a conjecture for general $k$ (for $k=1$ compare
section 2.1.1.). The character of the vacuum Verma module freely generated
by fields of dimension $d_1$, $d_2$, ..., $d_n$ is denoted
by $\chi_{d_1,d_2,\dots,d_n}$.
\mn
\centerline{\vbox{
\hbox{\vrule \hskip 1pt\vbox{ \offinterlineskip
\def\tablespace{height2pt&\omit&&\omit&\cr}
\def\tablerule{ \tablespace\noalign{\hrule}\tablespace}
\hrule\halign{&\vrule#&\strut\quad\hfil#\hfil\quad\cr
\tablespace\tablespace
& $k$    && {\it vacuum character} &\cr
\tablerule
& $1$ &&  $B_{0,0}^0(q)=\eta(q)^{-2}\bigl(\phi_0(q)-q\,\phi_1(q)\bigr)$ &\cr
\tablespace\tablespace\tablespace
& \omit && $B_{0,0}^0(q)-\chi_{2,3,4,5}(q)=
-2q^8 - 4q^9 - 9q^{10} + {\cal O}(q^{11})$ &\cr
\tablerule
& $2$ && $B_{0,0}^0(q)=\eta(q)^{-4}\bigl((\phi_0(q)-q\,\phi_1(q))^2-
    (\phi_1(q)-\phi_0(q))^2\bigr)$ &\cr
\tablespace\tablespace\tablespace
& \omit && $B_{0,0}^0(q)-\chi_{2,\ldots,11}(q)=
-2q^{14} - 6q^{15} - 15q^{16} + {\cal O}(q^{17})$ &\cr
\tablerule
& $3$ && $B_{0,0}^0(q)-\chi_{2,\ldots,19}(q)=
  -2q^{22} - 6q^{23} - 17q^{24} + {\cal O}(q^{25})$ &\cr
\tablerule\tablespace
& $4$ && $B_{0,0}^0(q)-\chi_{2,\ldots,29}(q)=
  -2q^{32}-6q^{33} + {\cal O}(q^{34})$&\cr
\tablerule\tablespace
& $k$ && $B_{0,0}^0(q) -\chi_{2,\ldots,{k^2+3k+1}}(q)=
  -2q^{k^2+3k+4}-6q^{k^2+3k+5} +{\cal O}(q^{k^2+3k+6})$ &\cr
\tablespace
}
\hrule}\hskip 1pt \vrule}
\hbox{\quad Table 4: Vacuum character for the ${\cal CP}(k)$-model}}
}
\mn
{}From the vacuum characters given in table 4 and the arguments presented
above we conjecture that
$${\cal CP}(k) =  {\widehat{sl(k+1)}_n \over
                    \widehat{sl(k)}_n\oplus \widehat{U(1)} }
   \cong \w(2,3,\ldots,k^2+3k+1).
   \eqno{(3.2.5)}$$
Note that (3.2.5) is compatible with the truncations predicted in
$\q{\letter}$ for the l.h.s.\ of (3.2.1).
\mn
\subs{Relations to linear $\w_\infty$ algebras}
\sn
At this point a few remarks on the relation of our results and
the various linear $\w_\infty$ algebras are in place. First, we
note that
$$\eqalign{
\w_\infty &\cong \lim_{n \to \infty} {\cal WA}_{n-1} , \cr
\w_{1+\infty} &\cong \lim_{k \to \infty} \widehat{U(1)}\oplus{\cal CP}(k) \cr
}   \eqno{\rm(3.2.6)}$$
where at least the first equality is well-known (see e.g.\ $\q{\bakinf}$).
Note that the limit in the second line of (3.2.6) is defined via the coset
realization of ${\cal CP}(k)$, i.e.\ the limit is taken for fixed level $n$.
Furthermore, the algebras $\w_n^{gl(n)}$ which were used in $\q{\frenkel}$
for the classification of quasifinite representations of $\w_{1+\infty}$
are related to ${\cal WA}_{n-1}$ by $\q{\frenkel}$:
$$\w_n^{gl(n)} \cong \widehat{U(1)} \oplus {\cal WA}_{n-1} .
   \eqno{\rm(3.2.7)}$$
{}From (3.2.6) and (3.2.7) one can immediately derive further identities.
An important one is that $\w_{1+\infty} / \widehat{U(1)} \cong
\lim_{k \to \infty} {\cal CP}(k)$ where the latter non-linear infinitely
generated algebra can {\it formally} identified with ${\cal WA}_{-1}$,
i.e.\ its structure constants can be obtained from those of ${\cal WA}_{n-1}$
by setting $n=0$. In
particular, all truncations presented in $\q{\letter}$ for ${\cal WA}_{n-1}$
can immediately applied to $\w_{1+\infty}$ setting $n=0$ and shifting the
central charge by one. Since the unifying algebras for the unitary
models  of ${\cal WA}_{n-1}$ are the ${\cal CP}(k)$ algebras which we
have just discussed, it is clear that $\widehat{U(1)} \oplus {\cal CP}(k)$
are unifying algebras of $\w_n^{gl(n)}$ and vice versa. In particular,
all known truncations of the linear $\w_{\infty}$ algebras arise
as accumulation points of the minimal series of some unifying $\w$-algebra.
These truncations in turn seem to be in one-to-one correspondence with
the quasifinite representations of the linear $\w_{\infty}$ algebras
(see e.g.\ $\q{\frenkel}$ for a proof in the case of $\w_{1+\infty}$
and $c \in \BN$).
\sn
For example, one can take the limit $n \to \infty$ of
the level $n$ in the ${\cal CP}(k)$ models in order to obtain unitary
representations of $\w_\infty$ at $c=2k$ which has been done already some
time ago $\q{\bakaskiritsis}$. Our previous computations show that
the identity $\w_\infty \cong {\cal CP}(k)$ at $c=2k$ implies a truncation
of $\w_\infty$ to an algebra of type $\w(2,3,\ldots,k^2+3 k + 1)$ at
$c=2k$. Similarly, the truncations of $\w_\infty$ to an algebra of type
$\w(2,3,\ldots,2 r + 1)$ at $c=-2r$ can be understood as accumulation
points of the unifying algebras $\w^{sl(r)}_{r-1,1}/\widehat{U(1)}$.
Above, we already explained the relation of the unifying property of
$\w^{gl(n)}_n$ to the truncations of $\w_{1+\infty}$ to algebras of
type $\w(1,2,\ldots,n)$ at $c=n$. The truncations of $\w_{1+\infty}$
to algebras of type $\w(1,2,\ldots,n^2-1)$ at $c=-n+1$ $\q{\awata}$ also
arise from unifying $\w$-algebras: They are related to algebras of type
$\w(2,3,\ldots,n^2-1) \cong {\cal WA}_{-n}$ that can be considered as
continuations of the ${\cal WA}_l$ series to negative rank. 
The algebras ${\cal WA}_{-n-1}$ can be realized in terms of the coset
$\widehat{sl(n)}_k \oplus \widehat{sl(n)}_{-1} / \widehat{sl(n)}_{k-1}$
$\q{\kneu}$.
\mn
\subs{The general unifying algebra for ${\cal WA}_n$-minimal models}
\sn
{}From the preceding discussion it is natural to expect that the unifying
$\w$-algebras for all ${\cal WA}_{n-1}$ minimal models can be obtained
by cosets of some Drinfeld-Sokolov reductions based on $sl(r)$ that
have a $\widehat{sl(k)} \oplus \widehat{U(1)}$ Kac-Moody subalgebra
\footnote{$^{10})$}{
We would like to thank E.\ Ragoucy for pointing this out to us.}.
Therefore, let us consider the algebras
$$\w^{sl(r)}_{r-k,1^{k}} = \w(1^{k^2},2,3,\ldots,r-k,
\left({\textstyle {r-k+1 \over 2}}\right)^{2 k}) .
\eqno(3.2.8)$$
The $k^2$ currents form a $\widehat{sl(k)} \oplus \widehat{U(1)}$
Kac-Moody algebra, the fields of dimension 2, $\ldots$, $r-k$ are singlets
with respect to this Kac-Moody and the $2 k$ fields of dimension
${r-k+1 \over 2}$ are a Kac-Moody multiplet. More precisely, these $2 k$
fields transform as two $U(1)$-charge conjugate defining representations of
$sl(k)$. The conjecture is that
$${\w^{sl(r)}_{r-k,1^{k}} \over \widehat{sl(k)} \oplus \widehat{U(1)}}
\cong {\cal WA}_{n-1} \qquad \hbox{at} \qquad c_{{\cal A}_{n-1}}(n+k,n+r).
\eqno(3.2.9)$$
Using the truncations of $\q{\letter}$ the conjecture (3.2.9)
implies in particular
$${\w^{sl(r)}_{r-k,1^{k}} \over \widehat{sl(k)} \oplus \widehat{U(1)}}
\cong \w(2,3,\ldots,(k + 1) r + k)
\eqno(3.2.10)$$
with two generic null fields at dimension $(k+1) r + k + 3$. The case
$k=1$ was already discussed at the end of section 2.1.2.\ and the conjectures
(3.2.9) and (3.2.10) were confirmed. Identifying $\w^{sl(r+1)}_{1^{r+1}}$
with the unconstrained $\widehat{sl(r+1)}$ Kac-Moody algebra, the case
$r = k+1$ is identical to the ${\cal CP}(k)$ cosets which we have just
discussed. The results for the ${\cal CP}(k)$ cosets also confirm the
conjectured identities (3.2.9) and (3.2.10).
One can also apply (3.2.9) to the case $k=0$. In this case, the $sl(2)$
embedding is the principal one and there is no Kac-Moody subalgebra.
Thus, the l.h.s.\ of (3.2.9) is just ${\cal WA}_{r-1}$ and we recover
eq.\ (2.4) of ref.\ $\q{\letter}$ (note that this equality was originally
observed in $\q{\nakan}$).
\mn
As a first check we compute the relation of the level $l$ of the underlying
$\widehat{sl(r)}_l$ as a function of $r$, $k$ and $n$ which generalizes
(2.1.37):
$$l = {n + r \over r - k} - r .
\eqno(3.2.11)$$
Like in (2.1.37) the level $l$ is linear in the rank $n$ of ${\cal WA}_{n-1}$
and becomes rational for the identifications in (3.2.9) ($r$, $k$ and $n$
positive integers). Note that the level $l'$ of the Kac-Moody subalgebra
$\widehat{sl(k)}_{l'}$ satisfies $l' = l + r - k - 1$
\footnote{${}^{11})$}{We would like to
acknowledge help for these computations by J.\ de Boer.}.
\sn
In order to perform a further check let us consider the case $k=2$.
In this case we have according to (3.2.8)
$\w^{sl(r)}_{r-2,1^2} = \w(1^4,2,3,\ldots,r-2,
\left({r-1 \over 2}\right)^4)$. Denote the four fields of scale
dimension ${r-1 \over 2}$ by $W_1^{\pm}$ and $W_2^{\pm}$. The upper index
of these fields refers to the $U(1)$-charge. $W_{1,2}^+$ and $W_{1,2}^-$
are two $sl(2)$-doublets. In the spirit of $\q{\ajl}$ one can see that
the {\it classical} invariants under $sl(2) \oplus U(1)$ are generated by
$$S_{m,n} := \de^m W_1^+ \de^n W_1^- + \de^m W_2^- \de^n W_2^+
\eqno({\rm 3.2.12a})$$
and the relations are generated by
$$\det \pmatrix{
S_{m_1,n_1} & S_{m_1,n_2} & S_{m_1,n_3} \cr
S_{m_2,n_1} & S_{m_2,n_2} & S_{m_2,n_3} \cr
S_{m_3,n_1} & S_{m_3,n_2} & S_{m_3,n_3} \cr
} = 0
\eqno({\rm 3.2.12b})$$
with two sets of pairwise distinct integers $\{m_1, m_2, m_3\}$ and
$\{n_1, n_2, n_3\}$. Following the argumentation of $\q{\ajl}$ it is
now simple to determine the field content of the coset (3.2.10) for $k=2$:
$${\w^{sl(r)}_{r-2,1^{2}} \over \widehat{sl(2)} \oplus \widehat{U(1)}}
  \cong \w(2,3,\ldots,3 r + 2)
  \eqno(3.2.13)$$
with two generic null fields at dimension $3 r + 5$.
\mn
In order to perform further checks of (3.2.10) on the level of the characters
we need a generalization of (2.1.7):
$${1 \over \prod_{m\ge\Delta-1} (1-e^{2\pi i\theta}q^m)\prod_{m\ge\Delta}
           (1-e^{-2\pi i\theta}q^m)}=\sum_{n\in\BZ} {\psi^{\Delta}_n(q)\over
           \prod_{m>0} (1-q^m)^2} e^{2\pi i\theta n}
\eqno({\rm 3.2.14a})$$
with
$$\psi^\Delta_n(q)=\sum_{m\ge\Delta-1} (-1)^{m-\Delta+1} q^{ {(m-\Delta+1)
         (m-\Delta+2)\over 2}+nm} \prod_{\nu=m-\Delta+2}^{m+\Delta-1}
         (1-q^\nu).
\eqno({\rm 3.2.14b})$$
Now one can compute the vacuum character of the coset (3.2.9) in
complete analogy to the ${\cal CP}(k)$ models using (3.2.14), i.e.\ the
only modification is that one has to substitute $\psi^\Delta_n(q)$ with
$\Delta = {r - k + 1 \over 2}$ for $\phi_n(q)$. We have checked
for $k=3$, $4 \le r \le 8$ and $k=4$, $5 \le r \le 8$ that this character
argument is in agreement with (3.2.10).
\mn
\section{3.2.2.\ Level-rank-duality for the cosets
        $\soh{n}_k \oplus \soh{n}_1 / \soh{n}_{k+1}$}
\sn
We consider now the impact of level-rank-duality on the coset
$\soh{n}_k \oplus \soh{n}_1 / \soh{n}_{k+1}$. From known results in the
literature $\q{\bogo,\altschuler}$ and arguments to be presented below,
we conjecture that
$$(\hbox{Orb})
\left({\soh{n}_k \oplus \soh{n}_1 \over \soh{n}_{k+1}}\right)
\cong \hbox{Orb} \left({\soh{k+1}_n \over \soh{k}_n}\right).
\eqno{\rm(3.2.15)}$$
For $n$ even, the coset on the l.h.s.\ realizes ${\cal WD}_{n \over 2}$
and one has to take an orbifold (see section 2.2.) whereas for $n$ odd, the
coset already corresponds to the orbifold of ${\cal WB}(0,{{n-1}\over 2})$
and no additional projection has to be taken. Eq.\ (3.2.15) implies that the
orbifold of the coset $\w$-algebra $\soh{k+1}_{n} / \soh{k}_{n}$ is a
unifying $\w$-algebra for the orbifolds of the $k$th unitary minimal models
of ${\cal WD}_{n\over 2}$ and ${\cal WB}(0,{{n-1}\over 2})$ for $n$ even or
odd respectively. It should be clear to the reader that the methods used in
$\q{\bogo,\altschuler}$ to derive level-rank-duality are insensitive to
orbifolds and therefore they originally did not turn up in the equality
(3.2.15). Because of the orbifolding procedure the coset
$\soh{k+1}_{n} / \soh{k}_{n}$ is more difficult to deal with in
full generality than those we discussed before. Therefore we will
just look at the first three examples.
\sn
{\underbar {$\bf k=1$}}: One observes that $\soh{2}_n / \soh{1}_n
\cong \widehat{U(1)}$ and that the first unitary minimal models of
${\cal WD}_{n \over 2}$ and ${\cal WB}(0,{n-1 \over 2})$ all have $c=1$.
Therefore, the RCFTs related to $k=1$ boil down to the classification of
$c=1$ theories $\q{\ginsparg,\kiritsis}$.
The r.h.s.\ of $(3.2.15)$ is given by the orbifold branch
of the $c=1$ RCFTs $\q{\ginsparg}$ leading to a symmetry algebra
of type $\w(2,4,{d\over 2})$ with $d\in \BN$.
Notice that at least for the first members of the series
${\cal WD}_{n\over 2}$ and ${\cal WB}(0,{{n-1} \over 2})$ truncate to an
algebra of type $\w(2,4,{n\over 2})$ for $c=1$ $\q{\howcl}$. The $\BZ_2$
orbifold of these algebras is again a $\w$-algebra of type
$\w(2,4,{d\over2})$. However, the spin content of this algebra depends
on the level $n$ in $(3.2.15)$, i.e.\ $d=d(n)$.
In this respect the case $k=1$ is different from the cases $k > 1$.
\sn
{\underbar {$\bf k=2$}}: For the unifying $\w$-algebra of the second
unitary minimal models we can make use of the fact that
$\soh{3}_n / \soh{2}_n \cong \sltr_{2n} / \widehat{U(1)}$.
This coset has been extensively discussed in section 2.1.1.
In particular, we know that the set of spins of the generators
is not a subset of those of the Casimir algebra ${\cal WD}_{n \over 2}$.
Therefore, we definitely have to take the $\BZ_2$ orbifold on the r.h.s.\
of (3.2.15). This orbifold has been argued in section 2.2.1.\ to lead to
a $\w(2,4,6,8,10)$. Note that the structure constants $C_{4\,4}^4$ in
appendix A and that of ${\cal WD}_{n \over 2}$ are indeed equal for
$k = 2 n$, i.e.
$$c = {2 n -1 \over n +1}
\eqno{\rm(3.2.16)}$$
(compare $\q{\letter}$).
Since for $k=2$ the minimal models on the r.h.s.\ of (3.2.15) are
orbifolds of $\BZ_{2n}$ parafermions, one can easily look at the first
few minimal models on both sides of (3.2.15) (for the relation between
$\BZ_{2n}$ parafermions and ${\cal WD}_{n \over 2}$ and
${\cal WB}(0,{n-1 \over 2})$ see also $\q{\flrep}$). One observes that
already the $\BZ_{2n}$ parafermionic models contain more fields than
the second unitary minimal models of ${\cal WD}_{n \over 2}$. This is
a first argument that we also have to take a $\BZ_2$ orbifold
of the l.h.s.\ of (3.2.15). Furthermore, one can also examine the
structure constants of the $\w(2,4,4,6)$ that corresponds to ${\cal WD}_4$.
{}From their explicit expressions $\q{\kauschpriv}$ one concludes that
for no positive value of the central charge $c$ any of the two fields
with conformal dimension 4 drops out. This is a further strong argument
to take the $\BZ_2$ orbifold also on the l.h.s.\ of (3.2.15).
\sn
{\underbar {$\bf k=3$}}: Finally we look at the third unitary minimal models.
Observing that
$\soh{4}_n / \soh{3}_n$ $\cong$ $\sltr_n \oplus \sltr_n /\sltr_{2 n}$
leads us to a coset which we have already discussed in section
2.1.3. Like for $k=2$ we know that the set of spins of the generators is
not a subset of those of ${\cal WD}_{n \over 2}$
showing once again that a $\BZ_2$ orbifold
has to be taken on the r.h.s.\ of (3.2.15). In section 2.2.1.\ we have argued
that this orbifold leads to an algebra of type $\w(2,4,6,8,10,12,14,16,18)$.
Again, we can check equality of the structure constants
$C_{4\,4}^4$ of ${\cal WD}_{n \over 2}$ with (2.1.52) at
$\kp = \mu = n$, i.e.
$$c = {3 n^2 \over (n+2) (n +1)}.
\eqno{\rm(3.2.17)}$$
{}From these examples we conjecture for the spin content of
Orb($\soh{k+1}_{n} / \soh{k}_{n}$) the following:
$${\rm Orb}\left({{\soh{k+1}_{n}} \over {\soh{k}_{n}}}\right)
         \cong \w(2,4,\dots,k(k+3)).
\eqno{\rm(3.2.18)}$$
This spin content of the coset can also be obtained by looking at the
Kac-determinant $\q{\letter}$.
\eject
\section{3.2.3.\ Realization of ${\cal WD}_{-n}$ as diagonal $sp(2n)$ cosets}
\sn
In $\q{\letter}$ we proposed unifying $\w$-algebras for the minimal
models of ${\cal WC}_n$ with central charge $c_{{\cal C}_n}(n+k+1,2n+2k+1)$.
{}From the study of the Kac-determinant we conjectured the following spin
content for the unifying algebras
$${\cal WD}_{-k} \cong \w(2,4,\dots,2k(k+2)).
\eqno{\rm (3.2.19)}$$
Our aim is to give an explicit coset realization of these algebras.
Therefore, one has to pose the question if one can make sense of
${\cal D}_{-n}= so(-2n)$. Indeed negative-dimensional groups
$SU(-n),SO(-2n),Sp(-2n)$ have been introduced in the context
of representation theory of the classical groups
$SU(n)$, $SO(2n)$, $Sp(2n)$ (see e.g.\ $\q{\dminusn}$).
There exist striking relations in representation theory which can
be explained in a more natural way by `analytic continuations' in $n$.
For example the dimension formula of an irreducible representation
of $SO(2n)$ equals up to a sign the dimension formula of $Sp(2n)$
for the transposed Young tableau upon the substitution $n\to -n$.
Furthermore, the $p$th order Casimir of $SO(2n)$ in the totally
antisymmetric rank-$r$ tensor representation equals up to a sign
the $p$th order Casimir of $Sp(2n)$ in the totally symmetric
rank-$r$ tensor representation upon the substitution $n\to -n$ $\q{\dminusn}$.
These relations arise naturally if one {\it defines} the
negative-dimensional groups via $SO(-2n)\cong \overline{Sp(2n)}$ and
$Sp(-2n)\cong \overline{SO(2n)}$. The overbar means the interchange of
symmetrization and antisymmetrization.
\sn
We conclude that ${\cal D}_{-n}= so(-2n)$ is related to ${\cal C}_n = sp(2n)$.
{}From the Sugawara central charge one can establish the identification:
$\widehat{so(-2n)}_\kp \leftrightarrow \widehat{sp(2n)}_{-{\kp\over 2}}$.
This leads us to the study of the general coset
${\widehat{sp(2n)}}_\kappa\oplus{\widehat{sp(2n)}}_\mu/
{\widehat{sp(2n)}}_{\kappa+\mu}$. The central charges for the minimal models
read
$$c_{\kappa,\mu}(n) = {{n\mu\kappa(2n+1)(2n+2+\mu+\kappa)}
\over{(\mu+\kappa+n+1)(\mu+n+1)(\kappa+n+1)}} .
\eqno{\rm (3.2.20)}$$
Specification to $\mu=-{1\over 2}$ yields the formula
$$c(n,\kappa)=-{\kappa n(2\kappa+4n+3)\over
(\kappa+n+1)(2\kappa+2n+1)} .
\eqno{\rm (3.2.21)}$$
This coincides with the
$c_{{\cal C}_\kappa}(\kappa+n+1,2\kappa+2n+1)$ minimal models
of the ${\cal WC}_\kappa$ Casimir algebras.
Furthermore, looking at formula eq.\ (2.15) of $\q{\letter}$
(this equation describes the truncation
of ${\cal WC}_n$ to ${\cal WD}_m$)
and substituting $m$ by $-m$ we recover eq.\ (3.2.21).
These facts strongly indicate that the ${\cal WD}_{-n}$ algebras
can be realized as the diagonal $\widehat{sp(2n)}$ coset.
Further confirmation for this relationship comes from the comparison
of the coupling constant $C_{4\,4}^4$ of ${\cal WD}_{-n}$ (obtained
by analytic continuation of the coupling constant of ${\cal WD}_n$)
with $C_{4\,4}^4$ of the diagonal coset (see section 3.2.5.).
\sn
The next step to be carried out is the determination of the
vacuum character of the diagonal $\widehat{sp(2n)}$ coset.
As before, it is useful to study the coset
$\widehat{sp(2n)}_{-{1\over 2}}/sp(2n)$ first,
since the former one can be viewed as a deformation of it.
To determine the vacuum character of $\widehat{sp(2n)}_{-{1\over 2}}$ one uses
the realization of $\widehat{sp(2n)}_{-{1\over 2}}$ by $n$
commuting $(\beta,\gamma)$ systems. This is due to the fact that the
vacuum module of $\widehat{sp(2n)}_{-{1\over 2}}$ is freely
generated in terms of the $(\beta,\gamma)$ systems which is not true
for the canonical choice in terms of the currents themselves.
We start with $n$ bosonic ghost-antighost fields $(\beta_i,\gamma_i)$
$$\beta_i(z)\ \gamma_i(w)={1\over z-w}+ reg.
\eqno{\rm(3.2.22)}$$
Define the currents $H_i=\beta_i\, \gamma_i$
satisfying the OPEs
$$H_i(z)\ H_j(w)=-{\delta_{ij}\over (z-w)^2}+ reg.
\eqno{\rm(3.2.23)}$$
The ghost fields $\beta_i,\gamma_i$ have charge $\pm 1$
with respect to the currents $H_i$:
$$H_i(z)\ \gamma_j(w)={\delta_{ij}\gamma_j(w)\over z-w}+ reg.
\quad\quad
H_i(z)\ \beta_j(w)=-{\delta_{ij}\beta_j(w)\over z-w}+ reg.
\eqno{\rm(3.2.24)}$$
The realization of $\widehat{sp(2n)}_{-{1\over 2}}$ in the
Cartan-Weyl basis  $\{ H_i(z),E_{\alpha}(z) \}$ is given by
\footnote{${}^{12}$)}{If $\{e_1,\ldots,e_n\}$ are orthonormal
unit vectors in the standard euclidean $\BR^n$, the root
system of ${\cal C}_n =sp(2n)$ is realized by the vectors $\pm 2e_i$ and
$\pm(e_i\pm e_j)$.}
\sn
\centerline{\vbox{\hbox{\vrule \hskip 1pt
\vbox{\offinterlineskip
\def\tablespace{height2pt&\omit&&\omit&\cr}
\def\tablerule{\tablespace\noalign{\hrule}\tablespace}
\hrule\halign{&\vrule#&\strut\quad\hfil#\hfil\quad\cr
\tablespace
\tablespace
& {\it root vector} $\alpha$ && {\it current} $E_{\alpha}(z)$ &\cr
\tablerule
& $2 e_i$ &&  $\gamma_i \gamma_i$ &\cr
\tablespace
& $- 2 e_i$  &&  $- \beta_i \beta_i$ &\cr
\tablespace
& $e_i+e_j$ && $\gamma_i \gamma_j$ &\cr
\tablespace
& $-e_i-e_j$ && $-\beta_i \beta_j$ &\cr
\tablespace
& $e_i-e_j$ && $\gamma_i \beta_j$ & \cr
\tablespace
& $-e_i+e_j$ && $\beta_i \gamma_j$ & \cr
\tablespace
}
\hrule}\hskip 1pt \vrule}
}}
\mn
The branching functions of the coset $\widehat{sp(2n)}_{-{1\over 2}}/sp(2n)$
are determined by the decomposition of the highest weight modules
$\chi_{\vec\Lambda}(q,\vec\theta)$ of $\widehat{sp(2n)}_{-{1\over 2}}$
into highest weight modules $\chi_{\vec\lambda}(\vec\theta)$ of $sp(2n)$:
$$\chi_{\vec\Lambda}(q,\vec\theta)=\sum_{\lambda}
  b^{\vec\Lambda}_{\vec\lambda}(q)\ \chi_{\vec\lambda}(\vec\theta).
\eqno{\rm(3.2.25)}$$
For the vacuum character one has the identity ($\kappa=-{1\over 2}$)
(see eq.\ (2.1.55)):
$$b_0^{\kappa\La_0}(q)=\sum_{w\in W} \eps(w)
c_{w\star\rho+\kappa\La_0}^{\kappa\La_0}(q)
q^{{1\over{2\kappa}}\vert w\star\rho\vert^2}
\eqno{\rm (3.2.26)}$$
with $\rho=\sum_{i=1}^n (n-i+1) e_i$.
We have to determine the string functions
$c_{w\star\rho+\kappa\La_0}^{\kappa\La_0}(q)$. One obtains from (2.1.7a) with
$z^2 = e^{2\pi i\theta}q^{1\over2}$ the following identity
$${1\over \prod_{n\ge 0}(1-e^{2\pi i\theta}q^{n+{1\over2}})
(1-e^{-2\pi i\theta}q^{n+{1\over2}})}=
\sum_{m\in \BZ} {q^{m\over 2}\phi_m(q)
e^{2\pi i\,\theta m} \over \prod_{n\ge 1}(1-q^n)^2} .
\eqno{\rm(3.2.27)}$$
Using the realization of $\widehat{sp(2n)}_{-{1\over 2}}$ in terms
of the $(\beta,\gamma)$ systems one obtains for the vacuum
character of $\widehat{sp(2n)}_{-{1\over 2}}$:
$${q^{n\over 24}\over \prod\limits_{i=1}^n
\prod\limits_{k\ge 0}(1\!-\!e^{2\pi i\,\theta_i}
q^{k+{1\over2 }}) (1\!-\!e^{-2\pi i\,\theta_i}q^{k+{1\over2 }})}
\!=\!\sum_{\vec\lambda\in \BZ^n}\!\!{q^{3n\over 24}\over \eta^{2n}(q)}
q^{\lambda_1+\ldots+\lambda_n\over 2} \phi_{\lambda_1}(q)\ldots
\phi_{\lambda_n}(q) e^{2\pi i\, \vec{\theta}\cdot\vec{\lambda}}.
\eqno{\rm(3.2.28)}$$
Thus the string functions of $\widehat{sp(2n)}_{-{1\over 2}}$
are given by
$$ c^{-{1\over 2}\La_0}_{\vec\lambda-{1\over 2}\La_0}(q)\
  q^{-|\vec\lambda|^2} = {q^{3n\over 24}\over \eta^{2n}(q)}
  q^{\lambda_1+\ldots+\lambda_n\over 2}
  \phi_{\lambda_1}(q)\ldots \phi_{\lambda_n}(q) .
\eqno{\rm(3.2.29)}$$
Inserting this into eq.\ (3.2.26) yields the explicit form of
the vacuum character of the $\widehat{sp(2n)}_{-{1\over 2}}/sp(2n)$ coset.
In table 5 below we present the first few examples and
give a conjecture for the spin content in the general case.
The spin content of ${\cal WD}_{-n}$ is indeed compatible
with the truncations of ${\cal WC}_m$ Casimir algebras $\q{\letter}$.
\mn
\centerline{\vbox{\hbox{
\vrule \hskip 1pt
\vbox{ \offinterlineskip
\def\tablespace{height2pt&\omit&&\omit&\cr}
\def\tablerule{ \tablespace\noalign{\hrule}\tablespace}
\hrule\halign{&\vrule#&\strut\quad\hfil#\hfil\quad\cr
\tablespace\tablespace
& $n$    && {\it vacuum character} &\cr
\tablerule
& $1$ &&  $b^0_0(q)=\eta(q)^{-2}\bigl(\phi_0(q)-q\,\phi_2(q)\bigr)$ &\cr
\tablespace\tablespace\tablespace
& \omit && $b^0_0(q)-\chi_{2,4,6}(q)=
-q^{11} - 2q^{12} -3q^{13} + {\cal O}(q^{14})$ &\cr
\tablerule
& $2$ && $b^0_0(q)=\eta(q)^{-4}\bigl(\phi_0^2\!-\!
\phi_0\phi_2\!-\!q\phi_1^2\!+\!
2q^2\phi_3\phi_1\!-\!q^2\phi_4\phi_0\!+\!q^3\phi_4\phi_2\!-
\!q^3\phi_3^2\bigr)$ &\cr
\tablespace\tablespace\tablespace
& \omit && $b^0_0(q)-\chi_{2,4,6,\ldots,16}(q)=
-q^{21} - 2q^{22} + {\cal O}(q^{23})$ &\cr
\tablerule
& $3$ && $b^0_0(q)-\chi_{2,4,6,\ldots,30}(q)=
  -q^{35} - 2q^{36} + {\cal O}(q^{38})$ &\cr
\tablerule\tablespace
& $4$ && $b^0_0(q)-\chi_{2,4,6,\ldots,48}(q)=
  -q^{53}-2q^{54} + {\cal O}(q^{55})$&\cr
\tablerule\tablespace
& $n$ && $b^0_0(q) -\chi_{2,\ldots,{2n(n+2)}}(q)=
  -q^{2n(n+2)+5}-2q^{2n(n+2)+6} + {\cal O}(q^{2n(n+2)+7})$ &\cr
\tablespace
}
\hrule}\hskip 1pt \vrule}
\hbox{\quad Table 5: Vacuum character for
           $\widehat{sp(2n)}_{-{1\over 2}}/sp(2n)$}}
}
\mn
\section{3.2.4.\ Minimal models of ${\cal WD}_{-m}$}
\sn
It has been argued in $\q{\klausREP}$ that the coset algebra
\slsl\ $\cong \w(2,4,6)$ $\q{\ajl}$ can be regarded in
a formal way as an algebra ${\cal WD}_{-1}$ in the following sense:
Its structure constants are given by those of the orbifold
of ${\cal WD}_m$ by setting $m = -1$.
Even its minimal models could be deduced with the help of
some known examples $\q{\howcl}$ from those of the
${\cal WD}_m$-algebras.
\sn
We will now try to continue the minimal models of ${\cal WD}_m$ to
negative values of $m$ beyond $m=-1$. The central charges
of the minimal models of the ${\cal WD}_{m}$-algebras are given by
$$c_{{\cal D}_m}(p,q) = m\Bigl(1-2(m-1)(2m-1){{(p-q)^2}\over{pq}}\Bigr)
\eqno{\rm(3.2.30)}$$
and following $\q{\klausREP}$ we have to make certain assumptions
for the allowed values of ${\cal WD}_{-m}$:
$q = p+1; p = 2n + 2x + 1\ \hbox{odd}.$
The value $n=0$ corresponds to the trivial model at $c=0$ implying $x = m$.
This leads to the following ansatz for the central
charge of the minimal models of ${\cal WD}_{-m}$
$$c_{{\cal D}_{-m}}(n) = - {{ mn(3+4m+2n)}\over{(1+m+n)(1+2m+2n)}} .
\eqno{\rm(3.2.31)}$$
This central charge equals the one of the $\widehat{sp(2n)}$ cosets
eq.\ (3.2.21). The dimensions of the highest weights can be obtained
as follows: Starting from the dimensions for the
${\cal WD}_{m}$-algebra
$\q{\luky}$~\footnote{${}^{13}$)}{The formula in $\q{\luky}$ contains a
misprint.}
$$h(p,q=p+1)  = {{\bigl(\sum^m_{r=1}
 (x_r + 1)\omega_r\bigr)^2-\bigl(\sum^m_{r=1} \omega_r\bigr)^2}
\over {2 p  (p+1)} }
\eqno{\rm(3.2.32)}$$
where $x_r = l_r (p+1) - l_r' p - 1$ with positive integers
$l_r$ and $l_r'$. The fundamental weights obey
$$\eqalign{
\omega_r \cdot \omega_s &= r \quad \quad\quad\quad\quad
 \quad \quad      1 \leq r \leq s \leq m-2 \cr
\omega_r \cdot \omega_{m-1} &=
\omega_r \cdot \omega_m =
{\textstyle{r \over 2}} \quad \quad   1 \leq r \leq m-2\cr
\omega_{m-1} \cdot \omega_{m} &=
{\textstyle{{m-2} \over 4}} \quad\quad\quad\quad\quad
\omega_{m-1}^2 = \omega_{m}^2 = {\textstyle{m \over 4}}.\cr
}\eqno{\rm(3.2.33)}$$
The experience with the algebra $\w(2,4,6)$ tells us that we have to
extend the summation in (3.2.32) to infinity and take
$l_r = l_r' = 1$ (i.e.\ $x_r = 0$) for all $r > n$
and therefore we obtain by inserting the fundamental weights eq.\ (3.2.33)
into eq.\ (3.2.32) and replacing $m \rightarrow -m$,\ $p \rightarrow 2n+2m+1$
$$h_{\vec{l},\vec{l'}}(n)  =  { {\sum_{r=1}^{n} r (x_r - 2m - 1 - r) x_r +
  2 \sum_{r<s} r x_r x_s } \over {4  (n+m+1)  (2n+2m+1)}} .
\eqno{\rm(3.2.34)}$$
On the $l_r, l_r'$ we have to impose the additional constraints
$$\sum_{r=1}^{n} (l_r + l_r' - 2)  \leq m+1 \, , \qquad
1 \le l_r , l_r' \le m+1 .
\eqno{\rm(3.2.35)}$$
\mn
Using eq.\ (\appF.2) for the minimal models of the Casimir algebras
we have checked in a few cases that the minimal models
of ${\cal WD}_{-n}$ coincide with those minimal models
of ${\cal WC}_m$ that one expects from  eq.\ (2.15) of~$\q{\letter}$.
\mn
{}From the realization of ${\cal WD}_{-m}$ in terms of the diagonal
$\widehat{sp(2m)}$ coset it should be possible to check these formulae with
representation theory of Kac-Moody algebras in the way outlined in
section 2.1.3.
\vfill
\eject
\section{3.2.5.\ The coset $\hat{g}_k/g$ for a simple Lie algebra $g$}
\sn
In order to confirm the above identifications
we compute a general structure constant for the cosets $\hat{g}_k/g$.
For general simple $g$ we have the following invariants at dimension 2 and~4:
$$\eqalign{
\hbox{dim = 2:}&\quad  L = \sum_{i,j}
           {{g_{i j}} \over {(k + \dcn) }} J^i \, J^j \cr
\hbox{dim = 4:}&\quad   L L, \quad     \de^2 L, \quad
 V_4 = \sum_{i,j} {{g_{ij}}  \over {(k + \dcn)}}  J^i \, \de^2 J^j \cr}
\eqno{\rm(3.2.36)}$$
where $g_{i j}$ is the metric of $g$ and $\dcn$ the dual Coxeter number and
the central charge is given by $c = (k \dim g)/(k + \dcn)$.
These are all independent invariants up to conformal dimension~4
for $sl(2)$ with arbitrary level $k$, for $sl(n),so(2n),
so(2n+1)$ with level $k=1$ and for $sp(2)$ with level $k=-{1\over 2}$
whereas one has to take into account additional invariants in the
general case. In these special cases where $V_4$ is the only new invariant
at scale dimension $4$ one can construct a primary field
$W_4$ out of $V_4$:
$$W_4 = V_4 - {3\over 5}\de^2 L - {{24}\over{5c+22}} L L.
\eqno{\rm (3.2.37)}$$
It is now relatively easy to compute the structure constant $C_{4\,4}^4$ for
this coset $\w$-algebra. For this purpose it is sufficient to consider
the 4th-order pole of $W_4 \star W_4$. Here we find
$$
\hbox{4th-order pole of\ \ } V_4 \star V_4  =
 12 (k + \dcn) \,\dcn\, \de^2 L + 4\, (18 k + 11 \dcn) \, (k + \dcn)\, V_4
\eqno{\rm(3.2.38)}$$
and from eq.\ (3.2.37)
$$\eqalign{
&\hbox{4th-order pole of\ \ } W_4 \star W_4  =
4\,{{(-118 \dcn +19 c \dcn +36 k +54 c k)}\over{(22+5 c)\,(k+\dcn)}}\,W_4 +\cr
&672\,{{(-10\dcn+c\dcn+12 k+6 c k)} \over {(22 + 5 c)^2\,(k+\dcn)}}\,L L +
{{24}\over{5}}\,{{(-10\dcn+c\dcn+12 k+6 c k)} \over {(22+5 c)\,(k+\dcn)}} \,
\de^2 L.}
\eqno{\rm(3.2.39)}$$
Since for a general spin-4 field with OPE
$W_4 \star W_4 = d_{4,4} I + \widetilde{C}_{4\,4}^4 W_4 + ...$
the 4th-order pole has the form
$$
\widetilde{C}_{4\,4}^4 W_4 + {{168 \, d_{4,4}} \over {c \, (22+5c)}}\, L L +
{{6\,d_{4,4}} \over {5\,c}} \, \de^2 L
\eqno{\rm(3.2.40)}$$
we can read off
$$\eqalign{
d_{4,4} &=
 {{4\, c \, (-10\dcn + c \dcn + 12 k+6 c k)} \over {(22+5c) \, (k+\dcn)}}\cr
\widetilde{C}_{4\,4}^4 &=
4\,{{(-118 \dcn +19 c \dcn +36 k +54 c k)} \over {(22+5 c)\,(k+\dcn)}}\cr}
\eqno{\rm(3.2.41)}$$
and for the normalized structure constant
$$(C_{4\,4}^4)^2 = {{(\widetilde{C}_{4\,4}^4)^2 \, c}\over{4 d_{4,4}}} =
{{(-118 \dcn +19 c \dcn +36 k +54 c k)^2} \over {(22+5 c)\,(k+\dcn) \,
 (-10\dcn+c\dcn+12 k+6 c k)}} .
\eqno{\rm(3.2.42)}$$
\sn
\centerline{\vbox{
\hbox{
\vrule \hskip 1pt
\vbox{ \offinterlineskip
\def\tablespace{height2pt&\omit&&\omit&&\omit&&\omit&&\omit&\cr}
\def\tablerule{ \tablespace\noalign{\hrule}\tablespace}
\hrule
\halign{&\vrule#&\strut\quad\hfil#\hfil\quad\cr
\tablespace
\tablespace
& $g$ && $h^\vee $ && $k$ && $(C_{4\,4}^4)^2$ && {\it algebra} &\cr
\tablerule
& $so(2n) $ && $2(n-1)$ && 1 && ${{2(19n-34)^2}\over{(2n-1)(5n+22)}}$ &&
     $[{\cal WD}_n, c=n]$ &\cr\tablerule
& $so(2n+1)$ && $2n-1$ &&  1 && ${{(38n-49)^2}\over{2n(10n+49)}}$ &&
$[Orb\,\bigl({\cal WB}(0,n)\bigr), c = n + {1\over 2}]$ &\cr\tablerule
& $sp(2n)$  && $n+1$ && $-{1\over 2}$ &&
   ${{2(19n+34)^2}\over{(2n+1)(5n-22)}}$ &&
     $ [{\cal WD}_{-n}, c=-n]$ &\cr
\tablespace
}
\hrule}\hskip 1pt \vrule}
\hbox{\quad Table 6: Structure constant $C_{4\,4}^4$ for some $\hat g /g$}}}
\sn
For $g = so(2n)$ and $k=1$ we recover the algebra ${\cal WD}_n$ at $c=n$.
For $g = so(2n+1)$ and $k=1$ we find the structure constant
for the bosonic projection of ${\cal WB}(0,n)$ at $c=n+{1\over 2}$.
Both solutions are in agreement with the general structure constant
for ${\cal WD}_n$ and the bosonic projection of ${\cal WB}(0,n)$
given in $\q{\klausREP}$. We should mention that for $g=sl(2)$
and arbitrary level $k$ we recover eq. (2.1.47).
\sn
For the coset $\widehat{sp(2n)}_{-{1\over 2}}/sp(2n)$
we obtain an identical expression for $(C_{4\,4}^4)^2$ as for
${\cal WD}_n$ if we replace $n$ by $-n$.
As in $\q{\ajl}$ one can deform this algebra
to generic central charges giving rise to the coset
$\widehat{sp(2n)}_{k} \oplus \widehat{sp(2n)}_{-{1\over 2}}
/\widehat{sp(2n)}_{k-{1\over 2}}$
which we formally identified with the algebra ${\cal WD}_{-n}$.
\bn
\section{4.\ Conclusion}
\sn
In this paper we studied various examples of quantum $\w$-algebras
belonging to a new class of deformable $\w$-algebras with infinitely nonfreely
generated classical limits which showed up recently $\q{\ajl}$. By explicit
calculation of the operator product expansions we provided evidence
that the coset algebras \slu\ and \beru\ are finitely generated
on the quantum level as one can infer e.g.\ from character arguments.
The mechanism which prohibits the additional generators in the quantum
case is a cancellation due to normal ordering of the classical relations.
{}From comparison of structure constants we obtained that \slu\ as well as
the isomorphic coset \sviru\ provide a realization of the special
nonfreely generated $\w(2,3,4,5)$ found earlier $\q{\hornfeck}$. We
performed calculations showing that the fourth special
solution of $\w(2,4,6)$ $\q{\kauwatts}$ is realized as the diagonal
coset \slsl. We presented also the self-coupling constant of the unique
primary spin 4 field in the general coset \slslgen.
\sn
Starting from the representation theory of the cosets realizing
$\w(2,3,4,5)$ and $\w(2,4,6)$ we collected further evidence that they are
unifying objects of special series of minimal models of Casimir algebras,
e.g.\ $\w(2,3,4,5)$ `unifies' the first unitary model of ${\cal WA}_n$.
These aspects were already discussed in an earlier
work $\q{\letter}$ where we put the emphasis on the truncations
of the Casimir algebras which they suffer at the particular
values of the central charge $c$. From these truncations infinitely
many unifying algebras can be proposed. For some of them we were
able to present coset realizations thereby generalizing the level-rank-duality
of coset pairs (see e.g.\ $\q{\bouschou,\bogo,\altschuler}$).
For the Casimir algebras ${\cal WA}_{n-1}$ we conjectured
coset realizations for all unifying $\w$-algebras predicted in
table 1 of $\q{\letter}$. Further
important examples are the ${\cal WD}_{-n}$ algebras which arise from
the symplectic cosets
${\widehat{sp(2n)}}_\kappa\oplus{\widehat{sp(2n)}}_{-{1\over 2}}/
{\widehat{sp(2n)}}_{\kappa-{1\over 2}}$. They are unifying
algebras for ${\cal WC}$ minimal models. The coset realizations of
unifying algebras known to us are collected in table 7.
\mn
Finally, we investigated orbifolds of quantum $\w$-algebras in a purely
algebraic manner. They behave similarly to the coset models discussed above,
i.e.\ they belong to the class of finitely nonfreely generated $\w$-algebras.
They occur also naturally in the context of unifying $\w$-algebras. We stress
the fact that in contrast to the orbifold of a classical Casimir $\w$-algebras
the orbifold of a quantum Casimir $\w$-algebra does not contain a Casimir
subalgebra. In examples we discussed some properties of the vacuum preserving
algebra (VPA) and the BW classical limit of nonfreely generated quantum
$\w$-algebras. In contrast to algebras of the Drinfeld-Sokolov class the VPA
does not yield a finite-dimensional Lie-algebra. Analogously, the classical
limits are not finitely generated any more.
\mn
\centerline{\vbox{\hbox{
\vrule \hskip 1pt
\vbox{ \offinterlineskip
\def\tablespace{ height2pt&\omit&&\omit&&\omit&&\omit&&\omit&\cr }
\def\tablerule{ \tablespace\noalign{\hrule}\tablespace }
\hrule\halign{&\vrule#&\strut\hskip1mm\hfil#\hfil\hskip1mm\cr
\tablespace\tablespace
& {\it Casimir}  && {\it central charge} && {\it coset realization} &&
  {\it dimensions of} && {\it dimension of} & \cr
\tablespace
& {\it algebra}  && $c$ && {\it of unifying algebra} &&
  {\it simple fields} && {\it first null field} & \cr
\tablerule
& ${\cal WA}_{n-1}$  && $c_{{\cal A}_{n-1}}(n \pl k,n \pl r)$ &&
  ${\w^{sl(r)}_{r-k,1^{k}} \over \widehat{sl(k)} \oplus \widehat{U(1)}}$ &&
    $2,3,\ldots,kr+r+k$ && $kr+r+k+3$ & \cr
\tablerule
&${\cal WB}_n$   && ${\hbox{$c_{{\cal B}_n}(2n \pl k \hbox{$-$} 1, 2n\pl 1)$}
                    \atop \hbox{$c_{{\cal B}_n}(2n , 2n \pl k)$}}$ &&
 ${\rm (Orb)}\left({\widehat{so(k)}_\kappa \oplus \widehat{so(k)}_1 \over
    \widehat{so(k)}_{\kappa+1}}\right)$ &&
    $2,4,\ldots,2 k$ && $2k + 4$ & \cr
\tablerule
&${\cal WC}_n$   && $c_{{\cal C}_n}(n \pl k \pl 1,2 n \pl 2 k \pl 1)$ &&
 ${\widehat{sp(2k)}_n \oplus \widehat{sp(2k)}_{-{1 \over 2}} \over
    \widehat{sp(2k)}_{n-{1 \over 2}} }$ &&
    $2,4,\ldots,2 k^2 + 4  k$ && $2 k^2 + 4 k +5$ & \cr
\tablerule
& $\orb{{\cal WD}_n}$ && $c_{{\cal D}_n}(2n \pl k \mi 2,2n \pl k \mi 1)$ &&
   $\orb{ {\widehat{so(k+1)}_{2 n} \over \widehat{so(k)}_{2n} } }$ &&
    $2,4,\ldots,k^2+3 k$ && $k^2+3k+4$ & \cr
\tablespace}
\hrule}\hskip 1pt \vrule}
\hbox{\quad Table 7: Coset realizations of unifying $\w$-algebras}
}}
\mn
The coset realization of unifying $\w$-algebras gives a so far unknown
coset realization for some of the minimal models of
the non-simply laced Casimir $\w$-algebras ${\cal WB}_n$ and
${\cal WC}_n$. A further interesting observation concerning
minimal models of diagonal cosets is the following conjecture: At least
one of the levels $\kp$, $\mu$ has to be an integer if
$\hat{g}_\kp \oplus \hat{g}_\mu / \hat{g}_{\kp+\mu}$ has a minimal model.
Note that the only counterexample to this conjecture we know of which was
presented in $\q{\ahn}$ is incorrect (see end of section 2.1.3.).
\sn
There are several interesting open question to answer in the future.
It has been conjectured in $\q{\letter}$ that all minimal models of
Casimir algebras related to the classical Lie algebras can also
be obtained as minimal models of unifying $\w$-algebras. However, the
existence of all these unifying $\w$-algebras is not yet firmly established.
Secondly, one could address the question whether all these algebras can be
realized as cosets of quantum DS reductions.
Finally, the representation theory of unifying $\w$-algebras as well as
the representation theory of quantum DS reductions related to nonprincipal
$sl(2)$ embeddings has not been worked out so far. However, in the case
of \sluk\ and \slsl\ one can see from the considerations in this
paper that all their minimal models also arise from identifications with
Casimir algebras. It would be interesting to know if all minimal models of
all unifying $\w$-algebras are also minimal models of the corresponding
Casimir algebras. If this should be true, one could hope to reconstruct
the representation theory of $\w$-algebras obtained by DS reduction for a
nonprincipal embedding using the fact that the coset by the Kac-Moody
subalgebra that survives the reduction
is a unifying $\w$-algebra for Casimir $\w$-algebras.
\mn
\section{Acknowledgements}
\sn
We are very grateful to L.\ Feh\'er and  W.\ Nahm for
many illuminating discussions and useful comments on the
manuscript.
\sn
W.E.\ thanks the Max-Planck-Institut f\"ur Mathematik
in Beuel for financial support.
K.H.\ is indebted to the University of Torino,
Department of Theoretical Physics, for kind hospitality.
R.H.\ thanks the NRW-Graduierten\-f\"orderung for a research studentship.
\vfill\eject
\section{Appendix A: The simple fields with spin 3, 4, 5 of \slu}
\sn
$$\eqalign{
&\wsd = -6{\ks^2}\Jp\pa\Jm - 12\ks\Jz\Jp\Jm + 4\Jz\Jz\Jz +
  6{\ks^2}\pa\Jp\Jm + 6\ks\pa\Jz\Jz + {\ks^2}\pa^2\Jz \cr
  &       \cr
&\wsv =\, 48\ks\left( 3 + 2\ks \right)(17 + 16\ks)^{-1}
     \bigl( 6\left( 5 - 6\ks \right) {\ks^2} \Jp\Jp\Jm\Jm +
       12{\ks^2}\left( 1 + \ks + {\ks^2} \right) \Jp\pa^2\Jm + \cr
 &12{\ks^2}\left( 11 + 5\ks \right) \Jz\Jp\pa\Jm +
       12\ks\left( 6 + 11\ks \right) \Jz\Jz\Jp\Jm -
       3\left( 6 + 11\ks \right) \Jz\Jz\Jz\Jz - \cr
 &12{\ks^2}\left( 11 + 5\ks \right) \Jz\pa\Jp\Jm +
       12{\ks^2}\left( -8 + 3\ks - 3{\ks^2} \right) \pa\Jp\pa\Jm + \cr
 &12{\ks^2}\left( -5 + 6\ks \right) \pa\Jz\Jp\Jm -
 6\ks\left( 6 + 11\ks \right) \pa\Jz\Jz\Jz +
       3\ks\left( 6 - 5{\ks^2} \right) \pa\Jz\pa\Jz + \cr
 &12\left( -3 + \ks \right) \left( -2 + \ks \right) {\ks^2} \pa^2\Jp\Jm -
 12\ks\left( 1 + \ks + {\ks^2} \right) \pa^2\Jz\Jz -
       {\ks^2}\left( 1 + \ks + {\ks^2} \right) \pa^3\Jz \bigr)\cr
  &       \cr
&\wsf =\, -288{\ks^2}\left( 3 + 2\ks \right) \left( 4 + 3\ks \right)
(107 + 64\ks)^{-1}\bigl( 60\left( 7 - 10\ks \right) {\ks^3}
     \Jp\Jp\pa\Jm\Jm +\cr
 & 20{\ks^3}\left( 5 \!+\! 3\ks \!+\! {\ks^2} \right) \Jp\pa^3\Jm \!+\!
   120\left( 7\! -\! 10\ks \right) {\ks^2} \Jz\Jp\Jp\Jm\Jm \!+\!
   60{\ks^2}\left( 8 \!+\! 19\ks \!+\! 3{\ks^2}\right)\Jz\Jp\pa^2\Jm +\cr
 & 60{\ks^2}\left( 64 + 17\ks \right) \Jz\Jz\Jp\pa\Jm +
   120\ks\left( 12 + 19\ks \right) \Jz\Jz\Jz\Jp\Jm -
   24\left( 12 + 19\ks \right) \Jz\Jz\Jz\Jz\Jz - \cr
 &  60{\ks^2}\!\left( 64 \!+\! 17\ks \right)\! \Jz\!\Jz\pa\Jp\!\Jm \!\!-\!
    480{\ks^2}\!\left( 5 \!+\! 3\ks \!+\!{\ks^2}\right)
      \!\Jz\pa\Jp\pa\Jm \!\!+\!
    180\!\left( \ks-4 \right)\! \left(\ks-3 \right)\!{\ks^2}
      \Jz\pa^2\Jp\Jm \!+\! \cr
 &  60{\ks^3}\!\left( 10\ks -7  \right) \pa\Jp\Jp\Jm\Jm +
    60{\ks^3}\!\left( 4\ks - 2{\ks^2} -17 \right) \pa\Jp\pa^2\Jm +
   180{\ks^2}\!\left(  3{\ks^2}\! - 4 \right) \pa\Jz\Jp\pa\Jm + \cr
 & 240{\ks^2}\!\left( 10\ks -7 \right) \pa\Jz\Jz\Jp\Jm -
    60\ks\left( 12 + 19\ks \right) \pa\Jz\Jz\Jz\Jz +
    60{\ks^2}\!\left( 14\ks \!-\! 11{\ks^2}\! - \! 12\right)
      \pa\Jz\pa\Jp\Jm + \cr
 & 180\ks\left( 4 - 3{\ks^2} \right) \pa\Jz\pa\Jz\Jz +
   120\left( \ks-4 \right) \left( \ks-3 \right){\ks^3}\pa^2\Jp\pa\Jm +
   60{\ks^2}\left( 8 - 7\ks + 4{\ks^2} \right) \pa^2\Jz\Jp\Jm - \cr
 & 30\ks\left( 16 + 12\ks + 7{\ks^2} \right) \pa^2\Jz\Jz\Jz +
   30{\ks^2}\left( 4 - 3{\ks^2} \right) \pa^2\Jz\pa\Jz +
   20\left( 3 - \ks \right) \left( \ks-4 \right) {\ks^3}\pa^3\Jp\Jm - \cr
 & 20{\ks^2}\left( 5 + 3\ks + {\ks^2} \right) \pa^3\Jz\Jz -
   {\ks^3}\left( 5 + 3\ks + {\ks^2} \right) \pa^4\Jz \bigr)\cr
&\cr
&d_{3,3} = 48\left( \ks-2 \right) \left(\ks-1 \right) {\ks^3}
       \left( 4 + 3\ks \right)  \cr
&d_{4,4} = 331776\left(\ks-3 \right) \left( \ks-2 \right)
     \left(  \ks-1 \right){\ks^6}\left( 1 + 2\ks \right)
 {{\left( 3 + 2\ks \right)}^2}\left( 4 + 3\ks \right) ({17 + 16\ks})^{-1} \cr
&d_{5,5} = 477757440(\ks\!-\!4)(\ks\!-\!3)(\ks\!-\!2)(\ks\!-\!1)\ks^9
     (1 \!+\! 2\ks )(3 \!+\! 2\ks)^2 (4 \!+\! 3\ks)^2 (8 \! +\! 5\ks)
    (107\! +\! 64\ks)^{-1} \cr
&\cr
&\SCu_{4\,4}^{4} =
     {10368{\ks^3}\left( 3 + 2\ks \right)
     \left( 4{\ks^3}- 15{\ks^2}- 33\ks -4 \right) }
     ({17 + 16\ks})^{-1} \cr
&\SCu_{4\,5}^5 = {25920{{\ks}^3}\left( 3 + 2\ks \right)
     \left( 4 + 3\ks \right)
     \left( 32{{\ks}^3}- 236{{\ks}^2}- 535\ks -125  \right) }
 \bigl({\left( 17 + 16\ks \right)\left( 107 + 64\ks \right)}\bigr)^{-1}. \cr
}$$
\mn
\section{Appendix B: Some structure constants of \beru}
\sn\sn
$$\eqalign{
&d_{3,3} = -{{{\left( 1 + \kb \right) }^2}\left( 3 + \kb \right)
      \left( 1 + 2\kb \right) \left( 9 + 4\kb \right) }({3 + 2\kb})^{-1}\cr
&d_{4,4} = {48{\kb^2}\!{{\left( 1\! +\! \kb \right)}^2}\!
     {{\left( 3\! +\! \kb \right) }^4}\!
     \left( 1\! +\! 2\kb \right)\! \left( 5\! +\! 3\kb \right)\!
     \left( 9\! + \!4\kb \right)\! \left( 12 \!+\! 5\kb \right)}\!
   {{{\left( 3\! +\! 2\kb \right)
}^{-2}}\!\left( 18 \!- \!19\kb \!- \! 15{\kb^2} \right)^{-1} }\cr
&d_{5,5} = {{480{\kb^2}{{\left( 1 \! + \! \kb \right) }^2}
  {{\left( 3 \! + \! \kb \right) }^6}
     \!\left( 2\kb\! - \!1 \right)\! \left( 1 \! + \! 2\kb \right)\!
     {{\left( 5 \! + \! 2\kb \right) }^2}\!\left( 5 \! + \! 3\kb \right)\!
     \left( 9 \! + \! 4\kb \right)\! \left( 12 \! + \! 5\kb \right) }
   {{{\left( 3 \! + \! 2\kb \right) }^{-3}}\!\left( 50 \! + \! 5\kb \!
    - \! 7{\kb^2} \right)^{-1} }}\cr
&d_{6,6} =
{{1600{\kb^2}{{\left( 1 + \kb \right) }^2}{{\left( 3 + \kb \right) }^8}
     \left( -1 + 2\kb \right) \left( 1 + 2\kb \right)
     {{\left( 5 + 2\kb \right) }^2}\left( 5 + 3\kb \right)
     \left( 9 + 4\kb \right) \left( 12 + 5\kb \right) }\over
   {{{\left( 3 + 2\kb \right) }^4}\left( 11 + 2\kb - {\kb^2} \right)
     {{\left( -50 - 5\kb + 7{\kb^2} \right) }^2}}}   \times   \cr
& \hskip 1cm \left( -57600 + 25260\kb + 67829{\kb^2} - 3738{\kb^3}
       - 19182{\kb^4} +
       2540{\kb^5} + 4809{\kb^6} + 882{\kb^7} \right) \cr
}$$
$$\eqalign{
d_6 =&
{{80{\kb^2}{{\left( 1 + \kb \right) }^2}{{\left( 3 + \kb \right) }^6}
     \left( -1 + 2\kb \right) \left( 1 + 2\kb \right)
     {{\left( 5 + 2\kb \right) }^2}\left( 5 + 3\kb \right)
     \left( 9 + 4\kb \right) \left( 12 + 5\kb \right) }\over
   {3{{\left( 3 + 2\kb \right) }^3}\left( 11 + 2\kb - {\kb^2} \right)
\left( 7{\kb^2}- 5\kb-50 \right)\left( 33 + 46\kb + 21{\kb^2} \right)^{-1}}}
\cr
\cr
\SCu_{4\,4}^4 =&{24{{\left( 3\! +\! \kb \right)}^2}\!
     \left(216\! +\! 186\kb\! +\! 823{\kb^2}\! +\! 1184{\kb^3}\!
       +\! 573{\kb^4}\! + \! 90{\kb^5} \right) }\!{\bigl(
   \left( 3\! +\! 2\kb \right)\!
      \left( -18\! +\! 19\kb\! +\! 15{\kb^2} \right)\bigr)^{-1}} \cr
\SCu_{4\,4}^6 =&        {\textstyle{4\over 5}} \cr
\SCu_{4\,5}^5 =&        {{60{{\left( 3 + \kb \right) }^2}
     \left( -10800 - 1740\kb - 2408{\kb^2} - 14301{\kb^3} -
5041{\kb^4} +
       4947{\kb^5} + 3233{\kb^6} + 510{\kb^7} \right) }\over
   {\left( 3 + 2\kb \right) \left( -50 - 5\kb + 7{\kb^2} \right)
     \left( -18 + 19\kb + 15{\kb^2} \right) }} \cr
\SCu_{4\,5}^7 =& {\textstyle{2\over 3}}\cr
\SCu_{4\,6}^4 =&{\textstyle{160\over 3}}
 \left( 3 + \kb \right)^4 \left( 2\kb\! - \!1  \right)
 \left( 5 + 2\kb \right)^2 \left( 15\kb^2+ 19\kb \! - \!18  \right)
 \left( 3 + 2\kb \right)^{-2}
 \!\bigl(\left(\kb^2 \! - \! 2\kb \! - \!11 \right)\!
\left( 7\kb^2\! - \! 5\kb\! - \!50 \right)\bigr)^{-1} \cr
& \times\left( 576 - 132\kb - 143{\kb^2} + 485{\kb^3} + 351{\kb^4}
   + 63{\kb^5} \right)\cr
\SCu_{4\,6}^6 =&        {\textstyle{4\over 3}}
{{{{\left( 3 + \kb \right) }^2}
 }{\Bigl(\left( 3 + 2\kb \right) \left( {\kb^2}- 2\kb-11 \right)
     \left( 7{\kb^2}- 5\kb-50 \right)
     \left( 15{\kb^2}+9\kb - 18 \right)\Bigr)^{-1} }} \cr
& ( 10254600 + 3254580\kb - 4772094{\kb^2} + 5168399{\kb^3} +
       6444501{\kb^4} -  \cr
&   267576{\kb^5} - 1247560{\kb^6} + 114411{\kb^7}+
       233577{\kb^8} + 39690{\kb^9} ) .  \cr
}$$
\mn
\section{Appendix C: The primary spin 4 generator of \slslgen}
\sn\sn
$$\eqalign{
&\Phi^{(4)} := {(3 \mu + 11) \mu}\SS{1,1}{0,0}\ \SS{1,1}{0,0}
 - {4 (3 \mu + 11) (\kp + 2)}\ \SS{1,2}{0,0}\SS{1,1}{0,0} \cr
& +(2 \mu \kp - 11 \mu - 11 \kp - 22) (3 \kp + 4) (\mu - 1)^{-1}
   \ \SS{2,2}{0,0}\SS{1,1}{0,0} \cr
& + (4 \mu \kp + 23 \mu + 23 \kp + 76) (3 \kp + 4) (\mu - 1)^{-1}
   \ \SS{1,2}{0,0}\SS{1,2}{0,0} \cr
&  - 4 (\mu + 2) (3 \kp + 11) (3 \kp + 4) (\kp - 1)
 \bigl((3 \mu + 4) (\mu - 1)\bigr)^{-1} \SS{2,2}{0,0}\SS{1,2}{0,0} \cr
& + (3 \kp + 11) (3 \kp + 4) (\kp - 1) \kp \bigl((3 \mu + 4) (\mu - 1)
   \bigr)^{-1}
   \SS{2,2}{0,0}\SS{2,2}{0,0} \cr
& \! +\! (37 \mu^2 \kp \! +\! 44 \mu^2 \! +\! 37 \mu \kp^2 \! +\! 192 \mu
 \kp\! +\! 176 \mu \! +\! 44 \kp^2 \! +\! 176 \kp \! +\! 176) (3 \kp \! +\! 4)
 \bigl((3 \mu \! +\! 4) (\mu \!-\! 1)\bigr)^{-1} \SS{1,2,2}{0,0,1}\cr
& + (37 \mu^2 \kp + 44 \mu^2 + 37 \mu \kp^2 + 192 \mu \kp + 176 \mu +
 44 \kp^2 + 176 \kp + 176) (\mu - 1)^{-1} \SS{2,1,1}{0,0,1} \cr
& + {\textstyle {3 \over 2}} (5 \mu \kp + 4 \mu + 5 \kp^2 + 20 \kp + 8)
 \mu \ \SS{1,1}{1,1}\cr
& - (22 \mu^2 \kp + 32 \mu^2 + 22 \mu \kp^2 + 147 \mu
\kp + 164 \mu + 59 \kp^2 + 236 \kp + 200) \mu (2(\mu - 1))^{-1}
   \SS{1,1}{0,2} \cr
&+ (5 \mu \kp^2 - 23 \mu \kp - 36 \mu + 5 \kp^3 + 5 \kp^2 - 96 \kp - 112) \
   \SS{1,2}{2,0} \cr
&  - (5 \mu^2 \kp - 5 \mu^2 + 5 \mu \kp^2 + 6 \mu \kp -
38 \mu - 5 \kp^2 - 38 \kp - 56) (3 \kp + 4) (\mu - 1)^{-1} \SS{1,2}{1,1} \cr
& + \bigl( 5 (3 \mu + 4) (\mu - 1)\bigr)^{-1}
 \bigl( {(4 \mu \kp + 23 \mu + 23 \kp + 76) (3 \mu + 4)
(3 \mu \kp^2 + \mu \kp + 16 \mu - 9 \kp^2 - 23 \kp + 12)} \cr
&\quad + {(\mu + 2) (3 \kp + 11) (\kp - 1)
(- 12 \mu^2 \kp + 44 \mu^2 + 60 \mu \kp + 280 \mu
 + 168 \kp + 384)} \bigr) \ \SS{1,2}{0,2} \cr
&  \! - \! (22 \mu^2 \kp \! +\! 59 \mu^2 \! +\! 22 \mu \kp^2
\! +\! 147 \mu \kp \! +\! 236 \mu \! +\!
 32 \kp^2 \! +\! 164 \kp \! +\! 200) (3 \kp \! +\! 4) \kp
\bigl( 2 (3 \mu \! +\! 4) (\mu \! - \! 1)\bigr)^{\! - \!1}\SS{2,2}{0,2} \cr
&+ 3 (5 \mu^2 + 5 \mu \kp + 20 \mu + 4 \kp + 8) (3 \kp + 4) (\kp - 1) \kp
\bigl( 2 (3 \mu + 4) (\mu -1) \bigr)^{-1} \SS{2,2}{1,1} . \cr
}$$
\vfill
\eject
\section{Appendix \appF: Minimal models of Casimir $\w$-algebras}
\sn
Let ${\cal K}$ be a simple Lie algebra of rank $l$ over $\BC$.
The rational models of the Casimir $\w$-algebra
related to this Lie algebra have central charge
$$c_{\cal K}(p,q) = l -
{\textstyle{12\over pq}}(q\,\rho - p\,{\rho^\vee})^2
\quad p,q\ \hbox{coprime}, \quad {h^\vee} \le p \quad h\le q
\eqno{\rm(\appF.1)}$$
where $p$ and $q$ have to be chosen minimal,
$h$ ($h^\vee$) denotes the (dual) Coxeter number of
${\cal K}$ and $\rho$ ($\rho^\vee$) denotes the sum of
its (dual) fundamental weights $\lambda_i$ ($\lambda_i^\vee$).
The conformal dimensions of the minimal model are given by
$\q{\kfw}$:
$$h_{\lambda,\nu^\vee} = {1\over{2pq}}
     \left( (q\lambda-p\nu^\vee)^2-(q\rho-p\rho^\vee)^2\right)
\eqno{\rm(\appF.2)}$$
where $\lambda$ ($\nu^\vee$) lies in the (dual) weight lattice
so that
$\lambda = \sum_{i=1}^l l_i \lambda_i,
\ \nu^\vee = \sum_{i=1}^l l_i^\vee \lambda_i^\vee.$
$\lambda$ and $\nu^\vee$ have to satisfy
$\sum_{i=1}^l l_i m_i \le p-1,
\ \sum_{i=1}^l l_i^\vee m_i^\vee \le q-1$
where $m_i$ are the normalized components of the highest root $\psi$ in
the directions of the simple roots $\alpha_i$, i.e.\
${{\psi}\over{\psi^2}} = \sum_{i=1}^l m_i {{\alpha_i}\over{\alpha_i^2}}$.
$m_i^\vee$ is given by $m_i^\vee  = {2\over{\alpha_i^2}} m_i$.
Note that the set of conformal
dimensions given by this condition has a symmetry so that all conformal
dimensions of the minimal model occur with the same multiplicity in it
(in the nonsimply laced cases the multiplicity is just 2). For more details
see $\q{\kfw,\bouwknegt,\flrep,\god}$.
\sn
\centerline{\vbox{\hbox{\vrule \hskip 1pt\vbox{\offinterlineskip
\def\tablespace{height2pt&\omit&&\omit&&\omit&\cr}
\def\tablerule{\tablespace\noalign{\hrule}\tablespace}
\hrule\halign{&\vrule#&\strut\quad\hfil#\hfil\quad\cr
\tablespace\tablespace
& {\it Lie algebra}  && $(m_i)$ && $(m_i^\vee)$ &\cr
\tablerule
& ${\cal A}_l$ && $(1,\dots,1)$ && $(1,\dots,1)$ &\cr
\tablespace
& ${\cal B}_l$  && $(1,2,\dots,2,1)$ && $(1,2,\dots,2)$ &\cr
\tablespace
& ${\cal C}_l$  && $(1,\dots,1)$ && $(2,\dots,2,1)$ &\cr
\tablespace
& ${\cal D}_l$ && $(1,2,\dots,2,1,1)$ && $(1,2,\dots,2,1,1)$ &\cr
\tablespace
& ${\cal E}_6$ && $(1,2,2,3,2,1)$     && $(1,2,2,3,2,1)$ &\cr
\tablespace
& ${\cal E}_7$ && $(2,2,3,4,3,2,1)$   && $(2,2,3,4,3,2,1)$ &\cr
\tablespace
& ${\cal E}_8$ && $(2,3,4,6,5,4,3,2)$ && $(2,3,4,6,5,4,3,2)$ &\cr
\tablespace
& ${\cal F}_4$ && $(1,2,3,2)$         && $(2,4,3,2)$ &\cr
\tablespace
& ${\cal G}_2$ && $(2,1)$             && $(2,3)$ &\cr
\tablespace}\hrule}\hskip 1pt \vrule}
\hbox{Table 8:
      Values of $m_i,m_i^\vee$ for all simple Lie algebras $\q{\god}$.}
}}
\mn
\section{Appendix \appD: The orbifold of the $N=1$ Super Virasoro algebra}
\sn
The orbifold of this algebra has been proposed by P.\ Bouwknegt who also
determined the field content $\q{\bouwknegt}$. Explicit constructions were
carried out before in $\q{\horst}$ and $\q{\commute}$ and the classical
version of this orbifold was discussed in $\q{\ajl}$. In $\q{\ajl}$
it was also shown how two normal ordered analogues of
classical relations ensure that the orbifold contains
at least a closed $\w(2,4,6)$ as subalgebra and how a
third relation gives rise to a first generic null field
at scale dimension 10. However, neither a {\it primary} basis
of generating fields nor the corresponding structure constants
were computed in $\q{\ajl}$. These calculations will
be presented in this appendix.
\sn
For our calculations we adopt the (noncovariant) conventions
for the extension of the Virasoro algebra by a spin ${3 \over 2}$
fermion $G$ used e.g.\ in $\q{\supwir}$. Orthogonalizing
the $\BZ_2$ invariant normal ordered products $\n(G,\de G)$ and
$\n(G,\de^3 G)$ with respect to other normal ordered products
one obtains the composite primary fields in the projection as:
$$\eqalign{
\Phi^{(4)} =& (5 c + 22) \n(G,\de G) - 17 \n(L,L) \cr
\Phi^{(6)} =& 5 (7 c + 68) (2 c - 1) (c + 24) \n(G,\de^3 G)
- 130 (7 c + 68) (2 c - 1) \n(\n(G,\de G),L) \cr
&+ 20 (218 c - 293) \n(\n(L,L),L)
- 6 (11 c - 86) (c + 24) \n(L,\de^2 L). \cr
} \eqno({\rm \appD.1})$$
The two point functions of these fields can easily
computed to be
$$\eqalign{
d_{4,4} =& {\textstyle{1 \over 6}} c (10c -7) (5c+22) (4c+21) \cr
d_{6,6} =& 50 c (14 c + 11) (10c -7)
(7 c + 68) (4c+21) (2 c -1) (c+24) (c+11) . \cr
} \eqno({\rm \appD.2})$$
For the structure constants one obtains exactly those of the
first solution (Set 1) in $\q{\kauwatts}$.
\sn
In order to demonstrate the notational advantages of quasi-primary fields
we also present a quasi-primary analogue of the relations presented in
appendix A of $\q{\ajl}$. The quasi-primary version of (A.2) in
$\q{\ajl}$ is:
$$\eqalign{
&(192 - 31 c) \ \n(G, \de^5 G)
+ 90 \ \n(\n(G, \de G), \n(G, \de G))
+ {\textstyle{81 \over 10}} \ \n(L, \de^4 L) \cr
&+ 154 \ \n(\n(G, \de G), \de^2 L)
- 420 \ \n(\n(G,\de^3 G), L)
= 0 . \cr
}   \eqno{(\rm \appD.3)}$$
Clearly, $\n(G, \de^5 G)$ can be expressed in terms of normal ordered
products of invariant fields with lower dimension.
\sn
The commutators of the simple fields of an algebra
of type $\w(2,4,6)$ involve fields up to dimension 10. Therefore,
closure of the algebra is ensured if in addition to eq.\ (\appD.3) also
$\n(G, \de^7 G)$ can be expressed in terms of $L$, $\n(G, \de G)$
and $\n(G,\de^3 G)$. A suitable relation in the spirit of
(A.4) in $\q{\ajl}$ can easily be established but is omitted.
\sn
Finally, we recall the discussion of singularities for special $c$-values
that has been carried out in $\q{\commute}$.
For $c \in \{ -{11 \over 14}, -{68 \over 7}, {1 \over 2}, -24, -11 \}$
the field $\Phi^{(6)}$ is a null field before normalization and one obtains
a $\w(2,4)$. For $c \in \{ {7 \over 10}, -{21 \over 4}
\}$ also the field $\Phi^{(4)}$ is a null field
before normalization and the bosonic
sector of the Super Virasoro algebra coincides with the Virasoro algebra.
It might seem that for $c=-{22 \over 5}$ we obtain a $\w(2,6)$ which is not
the case.
Here, singularities in the structure constants $C_{6 \, 6}^{X}$ forces
one to normalize $\Phi^{(6)}$ to zero.
This shows that there is no consistent
$\w$-algebra in the bosonic sector at $c=-{22 \over 5}$ and one is left
with the Virasoro algebra again.
\sn
Note that it is straightforward to derive the representations of this
orbifold $\w$-algebra from the well-known representations of the $N=1$ Super
Virasoro algebra. In particular,
it has been observed in $\q{\wirrep}$ that the representation theory of
$\w(2,4)$ and the $N=1$ Super Virasoro algebra at $c=-11$ is much the same
which is clear keeping the above truncations of the orbifold in mind
(see also $\q{\rwofus}$).
\vfill
\eject
\section{Appendix \appE: Generators and structure constants of the orbifold
                   of $\w(2,3)$}
\sn
Using the procedure described in section 2.2.1.\ one calculates
the composite primary fields of dimension 8 and 10 in the
orbifold of $\w(2,3)$ to be:
$$\eqalign{
\Phi^{(8)} =& 262080 (1919 c \!-\! 642) \n(\n(\n(L,L),L),L) \cr
& +1512 (3965 c^2\! -\! 168232 c \!+\! 1940316) \n(\n(L,\de^2 L),L) \cr
&- 205 (13 c^2 \!-\! 1096 c \!+\! 14556) (13 c \!+\! 516) \n(L, \de^4 L) \cr
&- 589680 (5 c \!+\! 22) (5 c \!+\! 3) (3 c \!+\! 46)
   \n(\n(W^{(3)},W^{(3)}),L) \cr
&+2730 (13 c \!+\! 516) (5 c \!+\! 22) (5 c\! +\! 3) (3 c \!+\! 46)
\n(W^{(3)},\de^2 W^{(3)}) \cr
\Phi^{(10)} =& 22522500 (17 c + 944) (11 c + 232) (5 c + 22) (c + 47)
 (c + 2) \n(W^{(3)},\de^4 W^{(3)}) \cr
& - 42702660000 (11 c + 232) (5 c + 22) (c + 47) (c + 2)
 \n(\n(W^{(3)},\de^2 W^{(3)}),L) \cr
& - 18711000 (17 c + 944) (11 c + 232) (7 c - 130) (5 c + 22)
 \n(\n(W^{(3)},W^{(3)}),\de^2 L) \cr
& + 11675664000 (32 c + 25) (11 c + 232) (5 c + 22)
 \n(\n(\n(W^{(3)},W^{(3)}),L),L) \cr
& - 2431 (3325 c^3 - 642870 c^2 + 39648336 c - 320267008) (5 c + 22)
 \n(L, \de^6 L) \cr
& + 26520 (522225 c^3 - 42458420 c^2 + 1123770804 c - 8347445152)
 \n(\n(L,L),\de^4 L) \cr
& + 1108800 (708305 c^2 + 132859254 c - 2814883952)
 \n(\n(\n(L,L),L),\de^2 L) \cr
& - 20756736000 (9421 c - 13918) \n(\n(\n(\n(L,L),L),L),L) \ . \cr
} \eqno({\rm \appE.1})$$
The two point functions of these fields turn out to be
$$\eqalign{
d_{8,8} =& 178869600 (13 c + 516)
(7 c + 114) (5 c + 186) (5 c + 22) \cr
&(5 c + 3) (5 c - 4)
(3 c + 46) (c + 2)^2 c \cr
d_{10,10} =& 116873396640000000 (17 c + 944) (11 c + 490) (11 c +
 232) (7 c + 114) \cr
&(7 c + 40) (5 c + 22) (5 c - 4) (c + 47) (c + 23) (c + 2)^2 c \ . \cr
} \eqno({\rm \appE.2})$$
After rescaling to standard normalization one obtains the following
structure constants:
$$\eqalign{
\left( C_{6 \, 6}^8 \right)^2 =& {300 (7 c + 68)^2 (5 c + 186) (2 c - 1)^2
 (c + 30)^2 \over (
13 c + 516) (7 c + 114) (5 c + 22) (5 c + 3) (5 c - 4) (3 c +
46)} \cr
C_{6 \, 6}^6 C_{8 \, 8}^6 =& {60 (14 c^3 + 915 c^2 + 14758 c - 22344)
 (21 c^2 + 754 c - 1176) (5 c + 3) (3 c + 46) \over (13 c + 516) (7 c +
 114) (7 c + 68) (5 c + 22) (5 c - 4) (2 c - 1)} \cr
C_{6 \, 6}^8 C_{8 \, 8}^8 =& {70 \ \Q \ (7 c + 68) (2 c - 1) (c + 30) \over
 (13 c + 516)^2 (7 c + 114) (5 c + 22) (5 c + 3) (5 c - 4) (3 c + 46)
(c + 2)} \cr
} \eqno({\rm \appE.3a})$$
with
$$\eqalign{
\Q :=\  & 54525 c^6 + 5031245 c^5 +
148843726 c^4 + 1411010708 c^3 +\cr
&  1061946744 c^2 - 1038009888 c - 12099262464 .\cr}
\eqno({\rm \appE.3b})$$
We omit the structure constants involving fields of scale dimension 10 or
higher because they are complicated but not very illuminating.
\vfill
\eject
\hsize = 15cm
\section{References}
\sn
\bibitem{\bouschou} P.~Bouwknegt, K.~Schoutens, {\it $\w$-Symmetry in
    Conformal Field Theory}, Phys.~Rep.~{\bf 223} (1993) p.~183
\bibitem{\flrep} V.A.\ Fateev, S.L.\ Luk'yanov,
    {\it Additional Symmetries and Exactly-Soluble Models in Two-Dimensional
    Conformal Field Theory}, Sov.\ Sci.\ Rev.\ A.\ Phys.\ {\bf 15/2} (1990)
\bibitem{\zam} A.B.\ Zamolodchikov,
    {\it Infinite Additional Symmetries in Two-Dimensional
    Conformal Quantum Field Theory}, Theor.\ Math.\ Phys.\ 65 (1986) p.\ 1205
\bibitem{\diealten} R.\ Blumenhagen, M.\ Flohr, A.\ Kliem,
    W.\ Nahm, A.\ Recknagel, R.\ Varnhagen,
    {\it $\w$-Algebras with Two and Three Generators},
    Nucl.\ Phys.\ {\bf B361} (1991) p.\ 255
\bibitem{\kauwatts} H.G.\ Kausch, G.M.T.\ Watts,
    {\it A Study of $\w$-Algebras Using Jacobi Identities},
    Nucl.\ Phys.\ {\bf B354} (1991) p.\ 740
\bibitem{\wirrep} W.\ Eholzer, M.\ Flohr, A.\ Honecker,
    R.\ H{\"u}bel, W.\ Nahm, R.\ Varnhagen,
    {\it Representations of $\w$-Algebras with Two Generators
    and New Rational Models}, Nucl.\ Phys.\ {\bf B383} (1992) p.\ 249
\bibitem{\hornfeck} K.~Hornfeck, {\it $\w$-Algebras with Set of Primary
    Fields of Dimensions (3,4,5) and (3,4,5,6)},
    Nucl.~Phys.~{\bf B407} (1993) p.~237
\bibitem{\klausREP} K.~Hornfeck, {\it Classification of
    Structure Constants for $\w$-Algebras from Highest Weights},
    Nucl.~Phys.~{\bf B411} (1994) p.~307
\bibitem{\ajl} J.~de Boer, L.~Feh\'{e}r, A.~Honecker,
    {\it A Class of $\w$-Algebras with Infinitely Generated
    Classical Limit}, Nucl.~Phys.~{\bf B420} (1994) p.\ 409
\bibitem{\FORT} L.\ Feh\'er, L.\ O'Raifeartaigh, P.\ Ruelle, I.\ Tsutsui,
    {\it On the Completeness of the Set of Classical $\w$-Algebras
    Obtained from DS Reductions}, Commun.\ Math.\ Phys.\ {\bf 162} (1994)
    p.\ 399
\bibitem{\deboer} J.\ de Boer, T.\ Tjin, {\it The Relation between Quantum
    $\w$ Algebras and Lie Algebras},
    Commun.\ Math.\ Phys.\ {\bf 160} (1994) p.\ 317
\bibitem{\laszlorep} L.\ Feh\'er,  L.\ O'Raifeartaigh, P.\ Ruelle,
    I.\ Tsutsui, A.\ Wipf, {\it On Hamiltonian Reductions of the
    Wess-Zumino-Novikov-Witten Theories}, Phys.\ Rep.\ {\bf 222} (1992) p.\ 1
\bibitem{\bowwatts} P.\ Bowcock, G.M.T.\ Watts,
    {\it On the Classification of Quantum $\w$-Algebras},
    Nucl.\ Phys.\ {\bf B379} (1992) p.\ 63
\bibitem{\fehort} L.\ Feh\'{e}r, L.\ O'Raifeartaigh, I.\ Tsutsui,
    {\it The Vacuum Preserving Lie Algebra of a Classical
    $\w$-Algebra}, Phys.\ Lett.\ {\bf B316} (1993) p.\ 275
\bibitem{\bbss} F.A.\ Bais, P.\ Bouwknegt, M.\ Surridge, K.\ Schoutens,
    {\it Extensions of the Virasoro Algebra Constructed
    from Kac-Moody Algebras Using Higher Order Casimir Invariants};
    {\it Coset Construction for Extended Virasoro Algebras},
    Nucl.\ Phys.\ {\bf B304} (1988) p.\ 348; p.\ 371
\bibitem{\howcl} W.~Eholzer, A.~Honecker, R.~H\"ubel,
    {\it How Complete is the Classification of $\w$-Sym\-me\-tries ?},
    Phys.~Lett.~{\bf B308} (1993) p.~42,
\bibitem{\letter} R.\ Blumenhagen, W.\ Eholzer, A.\ Honecker,
    K.\ Hornfeck, R.\ H\"ubel, {\it Unifying $\w$-Algebras},
    Phys.\ Lett.\ {\bf B332} (1994) p.\ 51
\bibitem{\bogo}   P.\ Bowcock, P.\ Goddard, {\it Coset Constructions
    and Extended Conformal Algebras}, Nucl.\ Phys.\ {\bf B305} (1988) p.\ 685
\bibitem{\altschuler} D.\ Altschuler, {\it Quantum Equivalence of Coset
    Space Models}, Nucl.\ Phys.\ {\bf B313} (1989) p.\ 293
\bibitem{\fatzam} V.A.\ Fateev, A.B.\ Zamolodchikov,
    {\it Nonlocal (Parafermion) Currents in Two-Dimen\-sional
    Conformal Quantum Field Theory and Self-Dual
    Critical Points in $\BZ_N$-Sym\-metric Statistical Systems},
    Sov.\ Phys.\ JETP {\bf 62} (1985) p.\ 215
\bibitem{\dminusn} G.V.\ Dunne, {\it Negative-Dimensional Groups in
    Quantum Physics}, J.\ Phys.\ A: Math.\ Gen.\ {\bf 22} (1989) p.\ 1719
\bibitem{\dijkgraaf} R.\ Dijkgraaf, C.\ Vafa, E.\ Verlinde, H.\ Verlinde,
    {\it The Operator Algebra of Orbifold Models},
    Commun.\ Math.\ Phys.\ {\bf 123} (1989) p.\ 485
\bibitem{\twists} A.~Honecker, {\it Automorphisms of $\w$-Algebras
    and Extended Rational Conformal Field Theories},
    Nucl.~Phys.~{\bf B400} (1993) p.~574
\bibitem{\frfolding} L.~Frappat, E.~Ragoucy, P.~Sorba,
    {\it Folding the $\w$-algebras}, Nucl.~Phys.~{\bf B404} (1993) p.~805
\bibitem{\commute} A.\ Honecker, {\it A Note on the Algebraic
    Evaluation of Correlators in Local Chiral Conformal Field Theory},
    preprint BONN-HE-92-25 (1992), hep-th/9209029
\bibitem{\thielemans} K.\ Thielemans, {\it A Mathematica Package for Computing
    Operator Product Expansions},
    Int.\ Jour.\ of Mod.\ Phys.\ {\bf C2} (1991) p.\ 787
\bibitem{\poly} A.M.\ Polyakov, {\it Gauge transformations and
    diffeomorphisms},
    Int.\ Jour.\ of Mod.\ Phys.\ {\bf A5} (1990) p.\ 833
\bibitem{\bersh} M.\ Bershadsky, {\it Conformal Field Theories via Hamiltonian
    Reduction}, Commun.\ Math.\ Phys.\ {\bf 139} (1991) p.~71
\bibitem{\bakas} I.\ Bakas, E.\ Kiritsis, {\it Beyond the Large $N$ Limit:
    Non-Linear $\w_\infty$ as Symmetry of the $SL(2,\BR)/U(1)$ Coset Model},
    Int.~Jour.~of Mod.~Phys.~{\bf A7}, Suppl.~1A (1992) p.\ 55
\bibitem{\bakpriv} I.~Bakas, E.~Kiritsis, private communication
\bibitem{\narganes} F.J.\ Narganes-Quijano, {\it On the Parafermionic $\w_N$
    Algebra}, Int.\ Jour.\ of Mod.\ Phys.\ {\bf A6} (1991) p.\ 2611
\bibitem{\peiwang}X.-M.\ Ding, P.\ Wang, {\it Nahm's Product and $\w$-Algebra},
    Phys.\ Lett.\ {\bf B335} (1994) p.\ 56
\bibitem{\bmcp} P.\ Bouwknegt, J.\ McCarthy, K.\ Pilch, {\it On the
    Free Field Resolutions for Coset Conformal Field Theories},
    Nucl.\ Phys.\ {\bf B352} (1991) p.\ 139
\bibitem{\yuwu} F.\ Yu, Y.-S.\ Wu, {\it On the Kadomtsev-Petviashvili
    Hierarchy, $\hat W_\infty$ Algebra, and Conformal $SL(2,\BR)/U(1)$ Model,
    I.\ The Classical Case,
    II.\ The Quantum Case}, J.\ Math.\ Phys.\ {\bf 34} (1993)
    p.\ 5851; p.\ 5872
\bibitem{\yuwuC} F.\ Yu, \ Y.-S.\ Wu, \ {\it An \ Infinite \ Number \ of \
    Commuting \ Quantum \ $\hat{W}_\infty$ \ Charges \ in \ the \
    $SL(2,\BR)/U(1)$ Coset Model},
    Phys.\ Lett.\ {\bf B294} (1992) p.\ 177
\bibitem{\toppan} F.\ Toppan, {\it Generalized NLS Hierarchies from
    Rational $\w$-Algebras},
    Phys.\ Lett.\ {\bf B327} (1994) p.\ 249
\bibitem{\ralphnpb} R.\ Blumenhagen, {\it $N=2$ Supersymmetric
    $\w$-Algebras}, Nucl.\ Phys.\ {\bf B405} (1993) p.\ 744
\bibitem{\nemesch} D.\ Nemeschansky, {\it Feigin-Fuchs Representation
    of $\widehat{su(2)}_k$ Kac-Moody Algebra},
    Phys.\ Lett.\ {\bf B224} (1989) p.\ 121
\bibitem{\qiu} Z.\ Qiu, {\it Nonlocal Current Algebra and $N=2$
    Superconformal Field Theory in Two Dimensions},
    Phys.\ Lett.\ {\bf B188} (1987) p.\ 207
\bibitem{\klausWAN} K.~Hornfeck, {\it The Minimal
    Supersymmetric Extension of ${\cal WA}_{n-1}$},
    Phys.~Lett.~{\bf B275} (1992) p.~355
\bibitem{\delduc} F.\ Delduc, L.\ Frappat, E.\ Ragoucy, P.\ Sorba, F.\ Toppan,
    {\it Rational $\w$-algebras from  Composite Operators},
    Phys.\ Lett.\ {\bf B318} (1993) p.\ 457
\bibitem{\gosch}  P.\ Goddard, A.\ Schwimmer, {\it Unitary Construction
    of Extended Conformal Algebras}, Phys.\ Lett.\ {\bf B206} (1988) p.\ 62
\bibitem{\ralph} R.\ Blumenhagen, {\it $\w$-Algebren in Konformer
    Quantenfeldtheorie}, Diplom\-arbeit BONN-IR-91-06 (1991)
\bibitem{\bouwknegt} P.~Bouwknegt, {\it Extended Conformal
    Algebras from Kac-Moody Algebras},
    Proceedings of the meeting `Infinite dimensional Lie algebras and Groups'
    CIRM, Luminy, Marseille (1988) p.~527
\bibitem{\kacpeter} V.G.\ Kac, D.H.\ Peterson, {\it Infinite-Dimensional
    Lie Algebras, Theta Functions and Modular Forms},
    Adv.\ Math.\ 53 (1984) p.\ 125
\bibitem{\kacwaki} V.G.\ Kac, M.\ Wakimoto, {\it Classification of
    Modular Invariant Representations of Affine Algebras},
    Infinite dimensional Lie algebras and groups,
    Advanced Series in Mathematical Physics Vol.\ 7, p.\ 138
\bibitem{\ahn} C.\ Ahn, S.\ Chung, S.-H.H.\ Tye, {\it New
    Parafermion, $SU(2)$ Coset and $N=2$ Superconformal Field Theories},
    Nucl.~Phys.~{\bf B365} (1991) p.~191
\bibitem{\kiritsis} E.\ Kiritsis, {\it Proof of the Completeness of the
    Classification of Rational Conformal Theories with c=1},
    Phys.\ Lett.\ {\bf B217} (1989) p.\ 427
\bibitem{\wolfgang} W.\ Eholzer, {\it Fusion Algebras Induced by
    Representations of the Modular Group},
    Int.\ Jour.\ of Mod.\ Phys.\ {\bf A8} (1993) p.\ 3495
\bibitem{\rwofus} W.~Eholzer, R.~H\"ubel, {\it Fusion Algebras of Fermionic
    Rational Conformal Field Theories via a Generalized Verlinde Formula},
    Nucl.~Phys.~{\bf B414} (1994) p.~348
\bibitem{\gso} F.\ Gliozzi, J.\ Scherk, D.\ Olive, {\it Supersymmetry,
    Supergravity Theories  and the Dual Spinor Model},
    Nucl.\  Phys.\ {\bf B122}  (1977) p.\ 253
\bibitem{\mfl} M.~Flohr, {\it $\w$-Algebras, New Rational Models and
    Completeness of the c=1 Classification},
    Commun.~Math.~Phys.~{\bf 157} (1993) p.~179
\bibitem{\andrdipl} A.\ Honecker, {\it Darstellungstheorie von
    $\w$-Algebren und Rationale Modelle in der Konformen Feldtheorie},
    Diplomarbeit BONN-IR-92-09 (1992)
\bibitem{\don} D.\ Zagier, {\it Modular Forms and Differential Operators},
    preprint Max-Planck-Institut f\"ur Mathematik MPI/94-13
\bibitem{\koos} K.~de Vos, {\it Coset Algebras, Integrable Hierarchies
    and Matrix Models}, Ph.D.~thesis, University of Amsterdam, September 1992
\bibitem{\bakinf} I.\ Bakas, {\it The Large-$N$ Limit of Extended
    Conformal Symmetries}, Phys.~Lett.~{\bf B228} (1989) p.\ 57
\bibitem{\frenkel} E.\ Frenkel, V.\ Kac, A.\ Radul, W.\ Wang,
    {\it $\w_{1+\infty}$ and $\w(gl_N)$ with Central Charge $N$},
    preprint hep-th/9405121
\bibitem{\bakaskiritsis} I.\ Bakas, E.\ Kiritsis, {\it Grassmanian Coset
    Models and Unitary Representations of $\w_\infty$},
    Mod.\ Phys.\ Lett.\ {\bf A5}, (1990) p.\ 2039
\bibitem{\awata} H.\ Awata, M.\ Fukuma, Y.\ Matsuo, S.\ Odake,
    {\it Character and Determinant Formulae of Quasifinite
    Representation of the $\w_{1+\infty}$ Algebra},
    preprint YITP/K-1060, YITP/U-94-17, SULDP-1994-3, hep-th/9405093
\bibitem{\kneu} K.\ Hornfeck, {\it $\w$-Algebras of Negative Rank},
    Phys.\ Lett.\ {\bf B343} (1995) p.\ 94
\bibitem{\nakan} A.\ Kuniba, T.\ Nakanishi, J.\ Suzuki, {\it Ferro and
    Antiferro-Magnetizations in RSOS Models}, Nucl.\ Phys.\ {\bf B356} (1991)
    p.\ 750
\bibitem{\ginsparg} P.\ Ginsparg, {\it Curiosities at c=1},
    Nucl.\ Phys.\ {\bf B295} (1988) p.\ 153
\bibitem{\kauschpriv} H.G.\ Kausch, private communication
\bibitem{\luky} S.L.\ Luk'yanov, V.A.\ Fateev, {\it Exactly soluble models
    of conformal quantum field theory associated with the simple Lie
    algebra $D_n$}, Sov.\ J.\ Nucl.\ Phys.\ {\bf 49} (1989) p.\ 925
\bibitem{\kfw} E.\ Frenkel, V.\ Kac, M.\ Wakimoto, {\it Characters and Fusion
    Rules for $\w$-Algebras via Quantized Drinfeld-Sokolov Reduction},
    Commun.\ Math.\ Phys.\ {\bf 147} (1992) p.\ 295
\bibitem{\god} P.\ Goddard, D.\ Olive, {\it Kac-Moody
    and Virasoro Algebras in Relation to Quantum Physics},
    Int.\ Jour.\ of Mod.\ Phys.\ {\bf A1} (1986), p.\ 303
\bibitem{\horst} H.G.~Kausch, {\it Chiral Algebras in Conformal Field
    Theory}, Ph.D.~thesis, Cambridge University, September 1991
\bibitem{\supwir} R.\ Blumenhagen, W.\ Eholzer, A.\ Honecker, R.\ H{\"u}bel,
    {\it New $N=1$ Extended Superconformal Algebras with Two and Three
    Generators}, Int.\ Jour.\ of Mod.\ Phys.\ {\bf A7} (1992) p.\ 7841
\vfill
\end